\documentclass[12pt, a4paper]{book}
\usepackage{etex}
\usepackage{thesis}
\usepackage{graphicx,tikz-cd}
\usepackage{booktabs}
\usepackage{array}
\usepackage{tabularx}
\usepackage{pdfpages}
\usepackage{comment}
\usepackage[english]{babel}
\usepackage[nottoc]{tocbibind}
\usepackage{dcolumn}

\usepackage{enumitem}

\usepackage{tikz}
\usetikzlibrary{positioning}

\tikzset{>=stealth}

\definecolor{green2}{HTML}{00CC18}
\definecolor{green3}{HTML}{03888C}

\usepackage[backend=bibtex,natbib=true,style=ieee,eprint=true]{biblatex}

\AtEveryBibitem{%
  \ifentrytype{book}{%
    \clearfield{month} 
    \clearfield{pages} 
    \clearfield{volume}  
  }{}
 \ifentrytype{inbook}{%
    \clearfield{month}
    \clearfield{volume}
  }{}
 \ifentrytype{incollection}{%
    \clearfield{month}
    \clearfield{volume}
  }{}
  \ifentrytype{misc}{%
    \clearfield{month}
  }{}
}

\usepackage{hyperref}
\hypersetup{
    colorlinks=true,      
    linkcolor=blue,       
    citecolor=green3,       
    urlcolor=blue,        
    pdfborder={0 0 0}     
}

\newcommand{\Eqref}[1]{(\ref{#1})}
\newcommand{\half}{\frac{1}{2}}
\newcommand{\expo}[1]{\mathrm{e}^{#1}}
\newcommand{\im}{\mathrm{i}}
\newcommand{\outpr}[2]{\vert{#1}\rangle\langle{#2}\vert}
\newcommand{\proj}[1]{\vert{#1}\rangle \langle{#1}\vert}
\newcommand{\Ket}[1]{\vert{#1}\rangle}
\newcommand{\Bra}[1]{\langle{#1}\vert}
\newcommand{\inpr}[2]{\langle{#1},{#2}\rangle}
\newcommand{\Inpr}[2]{\langle{#1} \vert{#2}\rangle}
\newcommand{\abs}[1]{\vert{#1}\vert}
\newcommand{\norm}[1]{\vert{#1}\vert}
\newcommand{\dnorm}[1]{\vert\vert{#1}\vert\vert}

\usepackage{bm}
\usepackage{epsf}
\usepackage{braket}
\usepackage{tensor}
\usepackage{soul}
\usepackage{float}
\usepackage{bbm}

\usepackage{mathrsfs}
\usepackage{mathtools}
\usepackage{multirow}

\usepackage{amsfonts}
\usepackage{appendix}
\usepackage[utf8]{inputenc}
\usepackage[]{qcircuit}

\usepackage{apptools}

\numberwithin{equation}{section}

\makeatletter
\newcommand*\rel@kern[1]{\kern#1\dimexpr\macc@kerna}
\newcommand*\widebar[1]{%
  \begingroup
  \def\mathaccent##1##2{%
    \rel@kern{0.8}%
    \overline{\rel@kern{-0.8}\macc@nucleus\rel@kern{0.2}}%
    \rel@kern{-0.2}%
  }%
  \macc@depth\@ne
  \let\math@bgroup\@empty \let\math@egroup\macc@set@skewchar
  \mathsurround\z@ \frozen@everymath{\mathgroup\macc@group\relax}%
  \macc@set@skewchar\relax
  \let\mathaccentV\macc@nested@a
  \macc@nested@a\relax111{#1}%
  \endgroup
}
\makeatother

\usepackage{etoolbox}
\makeatletter
\patchcmd{\chapter}{\if@openright\cleardoublepage\else\clearpage\fi}{}{}{}
\makeatother

\usepackage{filecontents}
\makeatletter
\def\thickhrulefill{\leavevmode \leaders \hrule height 1ex \hfill \kern \z@}
\def\@makechapterhead#1{%
    \reset@font
    \vspace*{10\p@}%
    {\parindent \z@ 
        \begin{flushleft}
            \normalfont \scshape \Huge \thechapter \par 
        \end{flushleft}
        \hrule
        \begin{flushleft}
            \reset@font \LARGE \strut #1\strut \par
        \end{flushleft}
        \vskip 100\p@
}}
\def\@makeschapterhead#1{%
    \reset@font
    \vspace*{-30\p@}%
    {\parindent \z@ 
        \begin{flushleft}
            \normalfont \scshape \Huge \vphantom{\thechapter} \par 
        \end{flushleft}
        \begin{flushleft}
            \reset@font \LARGE \strut #1\strut \par
        \end{flushleft}
        \hrule
        
        \vskip 40\p@
}}
\makeatother

\makeatletter
\def\section{\@startsection{section}{1}{\z@}%
  {2ex plus .2ex minus .2ex}%
  {1ex plus .2ex}%
  {\normalfont\bfseries\Large}} 

\def\subsection{\@startsection{subsection}{2}{\z@}%
  {2ex plus .2ex minus .2ex}%
  {1ex plus .2ex}%
  {\normalfont\bfseries\large}} 
\makeatother

\usepackage{mdframed}

\newmdenv[
  backgroundcolor=gray!9,
  linecolor=gray!9, 
  linewidth=.1pt, 
  roundcorner=5pt, 
  innerleftmargin=10pt,
  innerrightmargin=10pt,
  innertopmargin=10pt,
  innerbottommargin=10pt
]{custombox}

\newcommand{\be}{\begin{equation}}
\newcommand{\ee}{\end{equation}}
\newcommand{\bea}{\begin{eqnarray}}
\newcommand{\eea}{\end{eqnarray}}

\def\l{\lambda}

\def\bra{\big\rangle}

\allowdisplaybreaks

\begin{document}

\includepdf[pages=1]{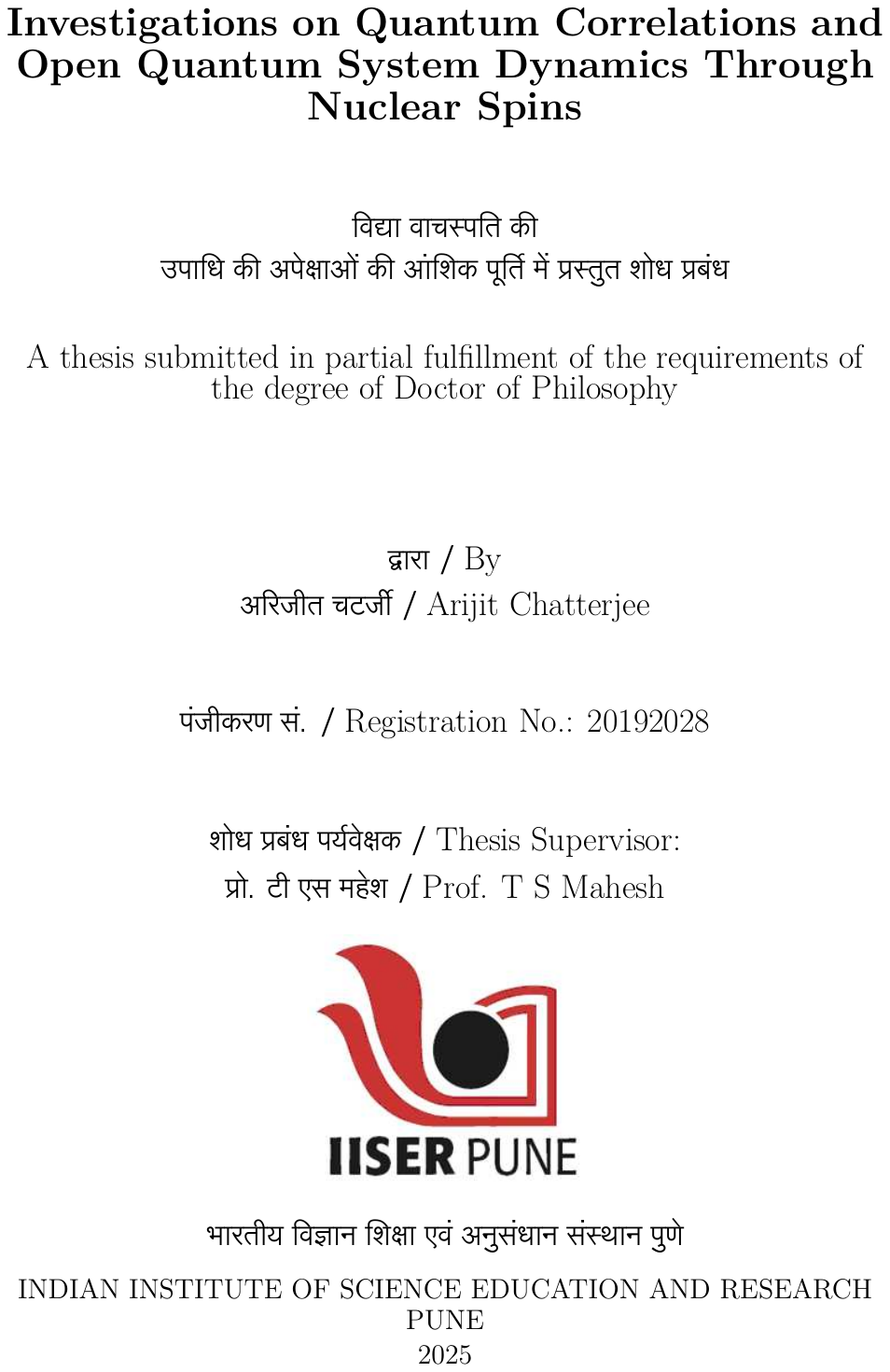}

\setlength{\parindent}{15pt}

\pagenumbering{gobble}


\vspace*{\fill}
\begingroup
\begin{center}
\textbf{\fontsize{15pt}{0mm}{\it  Dedicated to  Ma, Baba, and Archi,} \\ {\it{ and to anyone who is trying to push the boundaries that he, or she, is born with ...}}}
\end{center}
\vskip10cm
\endgroup
\vspace*{\fill}
\cleardoublepage

\chapter*{Declaration}

\noindent
{Name of Student:} Arijit Chatterjee\\
{Reg. No.:} 20192028\\
{Thesis Supervisor:} Prof. T S Mahesh\\
{Department:} Physics\\
{Date of joining program:} August 1, 2019\\
{Date of Pre-Synopsis Seminar:} July 24, 2025\\
{Title of Thesis:} Investigations on Quantum Correlations and Open Quantum System Dynamics Through Nuclear Spins\\[0mm]

I declare that this written submission represents my ideas in my own words and where others’ ideas have been included, I have adequately cited and referenced the original sources. I declare that I have acknowledged collaborative work and discussions wherever such work has been included. I also declare that I have adhered to all principles of academic honesty and integrity and have not misrepresented or fabricated or falsified any idea/data/fact/source in my submission. I understand that violation of the above will be cause for disciplinary action by the Institute and can also evoke penal action from the sources which have thus not been properly cited or from whom proper permission has not been taken when needed.

The work reported in this thesis is the original work done by me under the guidance of Prof. T S Mahesh.\\[18mm]

{\em Date: September 24, 2025 } \hfill {\em Arijit Chatterjee}\\ \vspace{-10mm}
\begin{flushright}
{\em Reg. No. 20192028}
\end{flushright}


\cleardoublepage

\chapter*{Certificate}
Certified that the work incorporated in the thesis entitled ``\textit{Investigations on Quantum Correlations and Open Quantum System Dynamics Through Nuclear Spins}'',
submitted by \textit{ Arijit Chatterjee} was carried out by the candidate, under my supervision. The work presented here or any part of it has not been included in any other thesis submitted previously for the award of any degree or diploma from any other university or institution.\\[25mm]
{\em Date: September 24, 2025 } \hfill {\em Prof. T S Mahesh}\\ \vspace{-10mm}
\begin{flushright}
{\em Thesis Supervisor}
\end{flushright}
\cleardoublepage

\chapter*{Acknowledgements}
\begin{center}
``\textit{Happiness is an elusive state of mind - not to be gained by clumsy pursuit. It is given to those who do not sue for it: to be unconcerned about a desired good is probably the only way to possess it.}'' \\
~~~~~~~~~~~~~~~~~~~~-- Ruskin Bond. \textit{`The Lamp is Lit', p $149$}
\end{center}

The above line is perhaps one of the key lighthouses that kept me moving in my PhD journey over the last few years. There are some people whose contributions in keeping me `\textit{unconcerned about a desired good}' so that I can enjoy the journey can only be understated. First and foremost, I extend my heartfelt acknowledgment to my supervisor Prof. T S Mahesh, whom I joined in September 2019. After two years of my master's, I officially began my doctoral work under him in September 2021. Over all these years, not even once did he restrict me from exploring and working on the topics that I found interesting. Even though this free exploration resulted in a somewhat gentle start of the PhD, he never asked me to rush. Looking back I can safely say, it were those initial months that shaped me as a researcher more than ever, and I could hardly amount the debt that I owe to him for letting me freely do this exploration. It was always a great pleasure to discuss physics with him, especially because in every conversation he made me feel equal and let me express my ideas freely irrespective of how impractical and bogus they were at times. I am truly grateful to my other RAC members, Prof. M.S. Santhanam and Dr. Sreejith G.J. for their constant guidance and support. The works reported here would not be possible to do without my collaborators : Prof. Yen-Kheng Lim, Prof. Mounir Nisse, Prof. Tomsaz Paterek, Dr. Sreejith G.J, Dr. Sachin Jain, Sakil Khan, Dr. Krithika, Prof. Usha Devi, Dr. Karthik. I extend a great sense of regards and greatfulness to each one of them. It is very hard to forget the support and help I got from Dr. Sandeep Mishra, and Nitin Dalvi during the NMR experiments. I would always remember the joyful company they provided during long experiments, especially calling me `\textit{Kaviji}' and teasingly asking me to narrate some couplets each time we sat together. I also like acknowledging our physics office staff, Mr. Prabhakar Angare, Ms. Dhanashree Sheth, and Mr. Palash, for their prompt action anytime I went to them for any logistic issues. I thank IISER Pune for providing the financial support during my PhD, and the central facility of NMR Research center for allowing me to carry out all the experimental works reported here.    

It would be impossible for me to consider doing a PhD and then completing it without the support and love of my parents. They always made me feel everything is going well so that I can freely focus on my work. Even when things got a little difficult at the home, they took care of it, and never let me feel anything. I can hardly forget how hard it was for them to live without me over the last years, but in any conversations they never let me sense that they were sad, instead they filled them with utmost joy and lightness of which I was at dire need at times. A thesis is never got written alone, it is you and your very own support system that co-write it. I could hardly express how deeply fortunate I am to have them as my own `support system' over all these years. I would also like to thank kaku (Abhijit Roy) and kakima (Prof. Aparna Roy) for their constant enthusiasm and understanding.

I would always be grateful to my PhD years as they lead me to meet \textit{Archi}, who is not only my best friend and a supporting partner, but also a very inspiring peer. She sustained me at times that seems unsustainable, kept me sane, and always fueled me with a constant energy and inspiration. Throughout these four years, we loved, teased, laughed, cried, got lost and found, but above all, cherished the grand monotony of waking up and coming to the main building, working hard throughout the day and stealing small `chai breaks' at the MDP, and then walking back to the hostel together at late nights. Those days, like most beautiful things in life, had now come to a halt, but I believe we have packed them enough in our memories to sustain ourselves for coming years. I would also like to mention my friends at IISER : Hari, Debesh da, Kaustav, Saikat, Debjyoti da, who never let me feel away from home. I could hardly forget the childish demands and adult-like scolding upon not fulfilling them that I received from  `Supritha', especially her coming to each Friday and saying ``dada, let's go to a cafe and see some lights," which would long outlast those weekends and stay forever in my memory.  I owe a great deal of thanks to all my hostel neighbors, especially Akash, Digvijay, Aman, Suraj, Iktesh, Dhurv, Shushant and Ajay for the memories they help me create from playing IISER premier league together with our very own team to have those crazy Sunday night parties.

The initial days of my PhD were full of a lot of self-doubts and uncertainties, and I would really like to thank my lab seniors, especially Dr. Soham Pal, Dr. Priya Batra, and Dr. Krithika, for calming me down each time and fueling me with self-belief. Among peers and juniors, I greatly enjoyed the physics discussions I had with Vishal and Conan. I would never forget Jiten's teasings each time I ate non veg and our long debates about environment and spirituality. Staying and working for long hours in the office would have been very boring, without the joyful leg-pulling and departmental gossip I had with Rishika, Rinku, and Vishal, along with the occasional fights over the air conditioning. I enjoyed the working experience with Aniket, Ashay, and Deepika, who always brought a fresh air in the discussions and taught me to think in new, unconventional ways. 

Finally, I wish to thank the city of Pune for the motherly shelter she provided throughout all these years. With the glittering skyline of Balewadi high streets and the whimsical monologues of post-monsoon drizzles, this city has been like that elderly woman, who has lost her everything many years ago, but always styles so nicely just to greet the mornings each day. In a few days from now, I will not be able to call IISER Pune my `\textit{home}' anymore, but this city and this campus, with all these people, will continue to moisten my heart as the most beautiful days of my life forever. Thank you all ...

\begin{flushright}{\em -- Arijit Chatterjee}\end{flushright}
\cleardoublepage


\pagestyle{fancy}
\fancyhf{}
\fancyhead[LO]{\nouppercase{\rightmark}}
\fancyhead[RE]{\nouppercase{\rightmark}} 
\fancyhead[LE,RO]{\thepage} 

\tableofcontents

\chapter*{Abstract}
\addcontentsline{toc}{chapter}{Abstract}
In the presence of a high magnetic field, nuclear spins serve as an ideal testbed for studying quantum correlations and open quantum system dynamics in various physical phenomena ranging from quantum information processing and quantum foundations to quantum many-body physics. This is primarily due to the long classical (T1) and quantum (T2) memories of nuclear spins and robust control via radio frequency pulses. In this thesis, I will discuss my works on using nuclear spins to explore these areas. First, I will address our investigations of temporal correlations, quantified by the Leggett-Garg Inequality (LGI), of a qubit evolving under the superposition of unitary operators. Using a three-qubit quantum register, we experimentally realized superposed unitaries and demonstrated the violation of LGI beyond its maximal quantum value of 1.5, indicating enhanced non-classicality. Interestingly, we found that this superposed unitary dynamics improves robustness against decoherence. Next, I will discuss our work on Lee Yang (LY) zeros, points in the complex plane where the partition function goes to zero, which provide insights into a system's thermodynamics and its behaviour near critical points. We propose a method to determine the full set of LY zeros of an asymmetric Ising system using a single quantum probe and demonstrate it experimentally with a three-qubit nuclear spin register. Interestingly, here also correlation plays an important role as the mutual information between the system and the quantum probe becomes maximum at the time points corresponding to LY zeros. Next, I will report our investigations of the quantum Mpemba effect in nuclear spin relaxation, which is an open system dynamics naturally occurring in nuclear spins at long times. We theoretically showed that a system far from steady state can relax faster than one nearer to equilibrium and demonstrated it experimentally. Finally, I will discuss my other work the localization and delocalization of entanglement due to local interactions, which causes and the apparent violation of quantum data processing inequality. Apart from experimentally realizing the respective dynamics using NMR methods, it was shown that the violation and corresponding non-complete positivity was only apparent by constructing a completely positive and trace-preserving map that describes the process.
\chapter*{List of Publications}
\addcontentsline{toc}{chapter}{List of Publications}

\begin{enumerate}[label=(\roman*)]
    \item\label{pub1} \underline{Chatterjee Arijit}, T. S. Mahesh, Mounir Nisse, and Yen-Kheng Lim. \textit{Observing algebraic variety of Lee-Yang zeros in asymmetrical systems via a quantum probe} \href{https://doi.org/10.1103/PhysRevA.109.062601}{Physical Review A 109, no. 6 (2024): 062601}

    \item\label{pub2} \underline{Chatterjee Arijit}, H. S. Karthik, T. S. Mahesh, and A. R. Devi. \textit{Extreme violations of Leggett-Garg inequalities for a system evolving under superposition of unitaries}. \href{https://doi.org/10.1103/vydp-9qqq}{Phys. Rev. Lett. 135, 220202 (2025)}.

    \item\label{pub3} Sengupta Aniket, \underline{Chatterjee Arijit}, Sreejith G. J. and T. S. Mahesh. \textit{Partial Shadow Tomography for structured operators and its experimental demonstration using NMR.} \href{https://arxiv.org/abs/2503.14491}{https://arxiv.org/abs/2503.14491} (2025). (Under review).

    \item\label{pub4} \underline{Chatterjee Arijit}, Sakil Khan, Sachin Jain, and T. S. Mahesh. \textit{Direct Experimental Observation of Quantum Mpemba Effect without Bath Engineering} .\href{https://arxiv.org/abs/2509.13451}{https://arxiv.org/abs/2509.13451}  (2025). (Under Review).

    \item\label{pub4} Pitambar Bagui, \underline{Chatterjee Arijit}, Bijay Kumar Agarwala. "Accelerated relaxation and Mpemba like effect  for operators in Open quantum systems" .\href{https://doi.org/10.48550/arXiv.2510.24630}{https://arxiv.org/abs/2510.24630} (2025). (Under Review).

    \item\label{pub4} Pitambar Bagui, \underline{Chatterjee Arijit}, Bijay Kumar Agarwala. "Detection of Mpemba Effect through good observables in Open Quantum System" .\href{https://arxiv.org/abs/2512.02709}{https://arxiv.org/abs/2512.02709} (2025). (Under Review).
\end{enumerate}
\textbf{Manuscripts under preparation :}
\begin{enumerate}[label=(\roman*)]
   \item \underline{Chatterjee Arijit}, Krithika V R, Tanjung Krisnanda, Tomasz Paterek and T. S. Mahesh. “Experimental Localization of Entanglement onto Noninteracting Systems and apparent Violation of Quantum Data Processing Inequality”.

   \item Deepika Bhargava, \underline{Chatterjee Arijit}, Vishal Varma, T S Mahesh. "Experimental Realization of Superposition of Channels using NMR"
\end{enumerate}

\chapter*{Synopsis}
\addcontentsline{toc}{chapter}{Synopsis}

This thesis reports the use of solution state nuclear magnetic resonance (NMR) architecture for experimental and theoretical exploration of open quantum systems, quantum correlations and as an application quantum simulations. The four research projects, which constitutes the thesis, are briefed below 

\begin{figure}
\centering
\includegraphics[width=12.5cm, clip = true, trim={0cm 2cm 0cm 2.9cm}]
{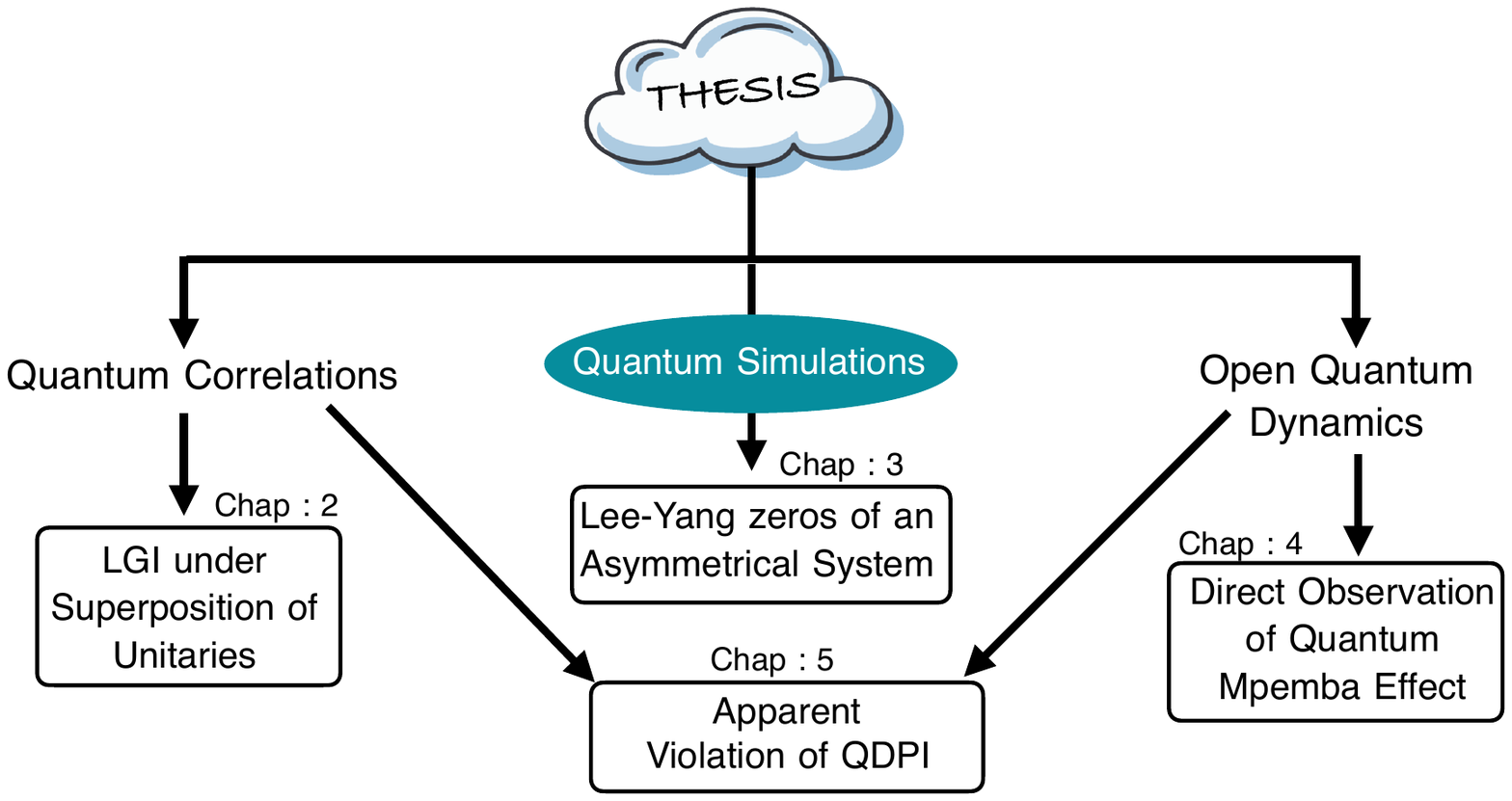}
\caption{Pictorial representation of the contents of this thesis.}
\label{fig:ludwig1}
\end{figure}

\begin{itemize}
\item[\textbf{Project 1:}] \textbf{Enhanced non-macrorealism: Extreme violations of Leggett-Garg inequalities for a system evolving under superposition of unitaries \hyperlink{LGI}{(Chapter 2)}}:
Quantum theory contravenes classical macrorealism by allowing a system to be in a superposition of two or more physically distinct states, producing physical consequences radically different from that of classical physics. We show that a system, upon subjecting to transform under superposition of unitary operators, exhibits enhanced non-macrorealistic feature - as quantified by violation of the Leggett-Garg inequality (LGI) beyond the temporal Tsirelson bound. Moreover, this superposition of unitaries also provides robustness against decoherence by allowing the system to violate LGI and thereby retain its non-macrorealistic behavior for a strikingly longer duration. Using an NMR register, we experimentally demonstrate the superposition of unitaries with the help of an ancillary qubit and verify these theoretical predictions.

\item[\textbf{Project : 2}]{\textbf{Observing algebraic variety of Lee-Yang zeros in asymmetrical systems via a quantum probe \hyperlink{chap_LY}{(Chapter 3)}:}}  
Lee-Yang (LY) zeros, points on the complex plane of physical parameters where the partition function goes to zero, have found diverse applications across multiple disciplines like statistical physics, protein folding, percolation, and complex networks. However, experimental extraction of the complete set of LY zeros for general asymmetrical classical systems remains a crucial challenge to put those applications into practice. Here, we propose a qubit-based method to simulate an asymmetrical classical Ising system, enabling the exploration of LY zeros at arbitrary values of physical parameters like temperature and internal couplings. Without assuming system symmetry, the full set of LY zeros forms an algebraic variety in a higher-dimensional complex plane. To determine this variety, we project it into sets representing magnitudes (amoeba) and phases (coamoeba) of LY zeros. Our approach uses a probe qubit to initialize the system and to extract LY zeros without assuming any control over the system qubits. This is particularly important as controlling system qubits can get intractable with the increasing complexity of the system. Initializing the system at an amoeba point, coamoeba points are sampled by measuring probe qubit dynamics. Iterative sampling yields the entire algebraic variety. Experimental demonstration of the protocol is achieved through a three-qubit NMR register. This paper expands the horizon of quantum simulation to domains where identifying LY zeros in general classical systems is pivotal. Moreover, by extracting abstract mathematical objects like amoeba and coamoeba for a given polynomial, our paper integrates pure mathematical concepts into the realm of quantum simulations.

\item[\textbf{Project 3:}] \textbf{Direct Experimental Observation of Quantum Mpemba Effect without Bath Engineering \hyperlink{chap_Mpemba}{(Chapter 4):}  }
The quantum Mpemba effect refers to the phenomenon of a quantum system in an initial state, far away from equilibrium, relaxing much faster than a state comparatively nearer to equilibrium. We experimentally demonstrate that this highly counterintuitive effect can occur naturally during the thermalization of quantum systems. Considering dipolar relaxation as the dominant decoherence process, we theoretically derive the conditions that can lead to the Mpemba effect in nuclear spins. After experimentally preparing nuclear spin states dictated by those conditions, we observe the occurrence of the Mpemba effect when they are left to thermalize without any external control. We also experimentally observe the genuine quantum Mpemba effect during thermalization of nuclear spins. Our results establish that both these effects are natural in thermalization of quantum systems, and may show up without the need for any bath engineering.

\item[\textbf{Project 4:}] \textbf{Experimental Localization of Entanglement onto Noninteracting Systems and apparent Violation of Quantum Data Processing Inequality \hyperlink{chap_qdpi}{(Chapter 5):}}  
As quantum technologies continue to advance, the ability to preserve and harness shared entanglement over extended durations is becoming increasingly vital. Here, we address the fundamental problem of localizing shared entanglement onto a chosen bipartition, thereby rendering it a practical resource for quantum information processing. At first sight, such localization seems to conflict with the quantum data-processing inequality, hinting at the presence of non–completely positive
reduced dynamics. We show, however, that this apparent violation is only superficial: the entire
process can be rigorously described by a completely positive and trace-preserving quantum channel.
We provide an experimental demonstration of this phenomenon using a three-qubit NMR register.
We further investigate the reverse problem of delocalizing an initially localized entanglement into
higher-dimensional subspaces. Remarkably, we find that delocalization substantially enhances the robustness of entanglement against dephasing noise. Using a four-qubit NMR system, we experimentally establish that this approach enables a dramatic improvement in entanglement storage—by up to a factor of forty. These results not only resolve a fundamental conceptual puzzle but also
introduce a powerful strategy for extending the lifetime of entanglement, a key requirement for
scalable quantum technologies.

\end{itemize}
\fancyhead[LO]{\nouppercase{\leftmark}}
\fancyhead[RE]{\nouppercase{\rightmark}}
\fancyhead[LE,RO]{\thepage}

\pagenumbering{arabic}

 \newcommand\marginal[1]{\marginpar[\raggedleft\tiny #1]{\raggedright\tiny #1}}

\chapter{Introduction}\label{chap_intro}
\begin{center}
``\textit{The weird thing about quantum mechanics is that unless everything you’re working with is quantum mechanical, you’re not taking advantage of the weirdness that gives you an advantage}" \\
~~~~~~~~~~~~~~~~~~ ---  William Phillips
\end{center}

\section{Basics of Quantum Information Processing} \label{intro_QI}
Quantum information science is the story of how we write numbers onto a quantum system, instruct it to perform arithmetic with those numbers, and finally read off the results. “Writing numbers” is a metaphor for storing information in the system. At any given time, a quantum system is mathematically described by an object called the density operator, commonly denoted by $\rho$. I will discuss operators in more detail in later sections; for now, it suffices to know that specifying $\rho$ is equivalent to specifying all the features about the system that determines the outcome of an arbitrary measurement on it.

The first step of any quantum experiment is preparation, where the system is initialized in a desired quantum state $\rho$. In our metaphor, this is the act of writing onto the system. It is worth noting that different preparation procedures may, and often do, lead to the same final state $\rho$. In other words, the protocol to prepare a system in state $\rho$ is not unique.

\begin{figure}
\centering
\includegraphics[width=12cm, clip = true, trim={7.4cm 9.5cm 2.8cm 5.2cm}]{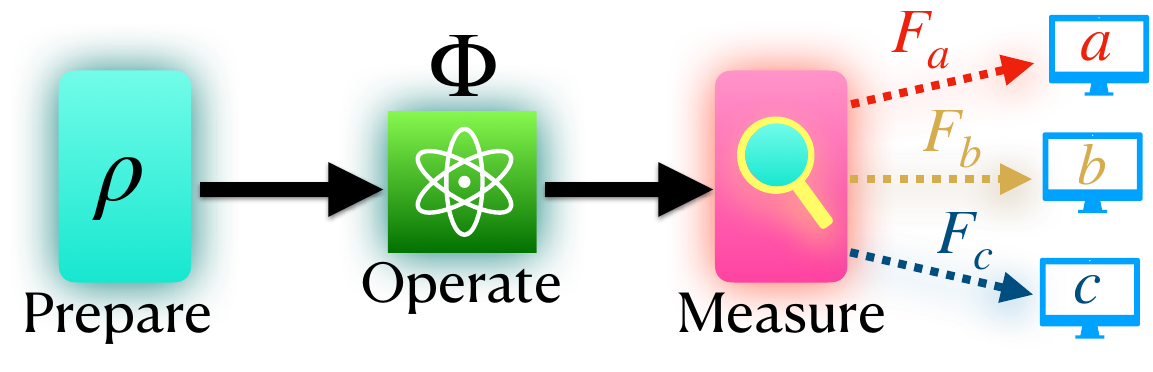}
\caption{Any quantum experiment can be described in terms of three steps: preparation (which corresponds to fixing the initial state $\rho$ of the system), operation (which can be any CPTP quantum channel $\Phi$, in general) and finally measurement (described by a POVM $\{F_n \}$). After the experiment, only one of the possible outcomes occurs in a given run of the experiment.}
\label{fig:ludwig1}
\end{figure}

Once the system is prepared, the natural question is: What can we do with it? This step corresponds to processing the information stored in the system by preparation. Mathematically, the most general physical operation that can be performed on a quantum system is represented by a completely positive, trace-preserving (CPTP) map, often abbreviated as a `quantum channel' and denoted as $\Phi$.

The final step of the experiment, in this narrative, is to read out the processed information. This is done by performing a measurement on the quantum system. The possible outcomes of a measurement can be labeled, for instance, as $\{a,b,c\}$. Each outcome $n$ is associated with a positive semidefinite operator $F_n \geq 0$,  and these operators satisfy the completeness condition $\sum_n F_n = \mathbbm{1}$, where 
$\mathbbm{1}$ is the identity operator. The full set $\{F_n\}$ defines a positive operator-valued measure (POVM) \cite{Nielsen_Chuang_2010}.

Before giving a detailed mathematical description of each of the steps (preparation, operate and measure) shown in Fig~\ref{fig:ludwig1}, I will start with some basics of quantum mechanics. This is done only for the purpose of completion and setting up the notations. For a more comprehensive review of quantum mechanics, the reader may consider \cite{10.1063/1.2815083,von2018mathematical}, or chapter 2 and 3 of \cite{preskill1998lecture}.

\subsection{Mathematical Preliminaries} \label{subsec:mathQM1} 
To mathematically describe a quantum system, the first thing to do is to identify a corresponding Hilbert space $\mathbb{H}$ for it. The Hilbert space $\mathbb{H}$ would be a linear vector space (LVS) over the field of complex numbers $\mathbb{C}$, with a scalar product $\langle \cdot , \cdot \rangle : \mathbb{H} \times \mathbb{H} \rightarrow \mathbb{C}$
\footnote{When necessary, I will mention the Hilbert space in the subscript, for example, $\langle \psi,\phi\rangle_{\mathbb{H}}$ for $\phi,\psi \in \mathbb{H}$.}. The scalar product is such that for any three vectors $\forall \phi, \psi, \xi \in \mathbb{H}$ and for any scalar $\lambda \in \mathbb{C}$ \cite{Nielsen_Chuang_2010} : \\
1. the norm of $\psi$ is defined as $\norm{\psi} := \sqrt{\inpr{\psi}{\psi}},\mbox{with}~\inpr{\psi}{\psi} \geq 0$, and $\norm{\psi}=0 \iff \psi = 0$. \\
2. they are anti-symmetric under complex conjugation : $\inpr{\phi}{\psi}^{*} = \inpr{\psi}{\phi}$ \\
3. the map is linear in its second argument : $\inpr{\phi}{\l(\psi+\xi)} = \l(\inpr{\phi}{\psi} + \inpr{\phi}{\xi})$. \\
 

Given such a Hilbert space $\mathbb{H}$, we can define an orthonormal basis (by basis, I will only mean orthonormal basis in this thesis unless otherwise stated) for $\mathbb{H}$ as a labeled list of vectors $\{ e_\mu \} \subset \mathbb{H}$ such that \\
1. $\inpr{e_{\mu}}{e_{\nu}} = \delta_{\mu,\nu}$, and \\
2. The set of all linear combinations of vectors from $\{e_{\mu}\}$, or the linear hull \cite{lang1987linear} of the basis is $\mathbb{H}$ itself, i.e lin$\{ e_{\mu}\} = \mathbb{H}$. 

Therefore any vector $\psi \in \mathbb{H}$ can be expanded as a linear combination of the basis vectors: $\psi = \sum_{\mu} \inpr{e_{\mu}}{\psi} \, e_{\mu}$, where $\inpr{e_{\mu}}{\psi}$ is called the component of $\psi$ along $e_{\mu}$.

The total number of elements, or cardinality \cite{enderton1977elements} of such a basis, is called the dimension of $\mathbb{H}$. Even though we can have infinite-dimensional Hilbert spaces, in this thesis I will only consider quantum systems that correspond to only finite-dimensional $\mathbb{H}$. Therefore, from now on, by $\mathbb{H}$, I will only mean a finite-dimensional Hilbert space, unless otherwise stated.

Given two such Hilbert spaces $\mathbb{H}_1$ and $\mathbb{H}_2$, one can define a linear operator $A : \mathbb{H}_1 \rightarrow \mathbb{H}_2$, such that $A (\lambda_1 \phi + \lambda_2 \psi) = \lambda_1 A(\phi) + \lambda_2 A(\psi) \in \mathbb{H}_2$, for two vectors $\phi,\psi \in \mathbb{H}_1$, and  two scalars $\lambda_1,\lambda_2 \in \mathbb{C}$. Such a linear operator $A$ is called bounded \cite{kreyszig1991introductory} when there exists a non-negative number $\lambda \geq0$, such that $|A \psi| \leq \lambda |\psi|$ for any $\psi \in \mathbb{H}_1$. The norm of the bounded operator $A$ is defined \cite{conway2019course} by taking an  infimum of $\lambda$ such that the inequality holds : $\dnorm{A} = \inf \{\lambda \geq 0\,\text{such~that}~ |A\psi| \leq \lambda|\psi|,\forall \psi \in \mathbb{H}_1 \}$.     
The set of all such bounded linear operators is denoted as $\mathcal{B}(\mathbb{H}_1,\mathbb{H}_2)$, or by $\mathcal{B}(\mathbb{H})$ when input and output Hilbert spaces $\mathbb{H}$ are the same.

For any such $A \in \mathcal{B}(\mathbb{H}_1,\mathbb{H}_2)$, there exists an adjoint \cite{miller2008quantum} of $A$ denoted as $A^{\dagger} :\mathbb{H}_2 \rightarrow \mathbb{H}_1$ such that 
\begin{gather}
\inpr{\psi}{A^{\dagger} \phi}_{\mathbb{H}_1} = \inpr{A \psi}{\phi}_{\mathbb{H}_2},~~\forall \phi \in \mathbb{H}_2 ~\text{and}~\forall \psi \in \mathbb{H}_1. \label{eq1:adjoint} 
\end{gather}
The adjoint of $A$ is also sometimes called the `hermitian conjugate' of $A$ \cite{Nielsen_Chuang_2010}.

\textbf{The Ket and the Bra : } In 1939, Paul Dirac introduced a notation \cite{Dirac_1939} that eventually became the universal notation of quantum mechanics. Corresponding to every vector $\psi \in \mathbb{H}$, he introduced a `ket' $\Ket{\psi}$ and a `bra' $\Bra{\psi}$. A mathematically nice way of understanding this notation \cite{wernerQI} is to first note that $\mathbb{C}$ itself is a Hilbert space with respect to the inner product $\inpr{z_1}{z_2} = z_{1}^{*}\,z_2$ for $\forall z_1, z_2 \in \mathbb{C}$. For any vector $\psi \in \mathbb{H}$, the `Ket' can then be defined as an operator between two Hilbert spaces $\Ket{\psi}: \mathbb{C} \rightarrow \mathbb{H}$ and it acts on some $z \in \mathbb{C}$ as $\Ket{\psi}z = z\,\psi \in \mathbb{H}$. The `Bra' for that same vector $\psi \in \mathbb{H}$ is then defined as the adjoint of the `Ket' $\Bra{\psi}:=\Ket{\psi}^{\dagger}$. It is therefore an operator that maps $\mathbb{H} \rightarrow \mathbb{C}$. By the definition of adjoint operators in Eq.~\Eqref{eq1:adjoint}, we can read $\forall z \in \mathbb{C}$ and $\forall \psi,\phi \in \mathbb{H}$ ;
\begin{gather}
\inpr{z}{\Ket{\psi}^{\dagger} \phi}_{\mathbb{C}} = \inpr{\Ket{\psi}z}{\phi}_{\mathbb{H}} 
\implies \inpr{z}{\Bra{\psi}\phi}_{\mathbb{C}} =  \inpr{z \psi}{\phi}_{\mathbb{H}} \nonumber \\
\implies z^{*} \Inpr{\psi}{\phi} = z^{*} \inpr{\psi}{\phi}_{\mathbb{H}} \implies \inpr{\psi}{\phi}_{\mathbb{H}} = \Inpr{\psi}{\phi}, \label{eq1:braket} 
\end{gather}
which allows us to write inner products $\inpr{\phi}{\psi}$ as $\langle \phi | \psi \rangle$. Moreover, we can use $\Ket{\psi}$ and $\psi$ interchangeably with the implicit assumption that the operator $\Ket{\psi}$ has acted on complex number $1$ and produced the vector $\psi \in \mathbb{H}$ (which happened in our mind silently). For the basis $\{e_{\mu}\}$, we can now use only the labels as a shorthand and write $\{\Ket{\mu} \}$. In this notation, the expansion of a vector $\psi \in \mathbb{H}$ reads
\begin{gather}
\psi = \sum_{\mu} \inpr{e_{\mu}}{\psi}\,e_{\mu} \implies \Ket{\psi} = \sum_{\mu} \Inpr{\mu}{\psi}\,\Ket{\mu} \nonumber \\ \implies \Ket{\psi} = \sum_{\mu} \left(\proj{\mu}\right) \Ket{\psi} \implies \sum_{\mu} \proj{\mu} = \mathbbm{1}. \label{eq1:basis_complete}
\end{gather}
The `Ket-Bra' operators $\proj{\mu}$ are of prime importance in quantum mechanics. As $\Bra{\mu}:\mathbb{H}\rightarrow \mathbb{C}$ and $\Ket{\psi}:\mathbb{C}\rightarrow \mathbb{H}$, the operators of the form $\proj{\mu}$ maps vectors of the Hilbert spaces to the same space $\proj{\mu}:\mathbb{H}\rightarrow \mathbb{H}$. However, by construction the output spaces of such operators are only one dimensional (spanned by $e_{\mu}$ only). So they are called rank-1 projectors in mathematics. Their action on any vector $\psi$ can be interpreted as projecting that vector onto its direction.

Any operator $A\in\mathcal{B}(\mathbb{H}_1,\mathbb{H}_2)$ can be expanded in a basis of $\mathbbm{H}_1$ ($\{\Ket{\mu}\}$) and $\mathbbm{H}_2$ ($\{\Ket{i}\}$) as : $A = \mathbbm{1}_{\mathbbm{H}_2}\,A\,\mathbbm{1}_{\mathbbm{H}_1} = \sum_{\mu,i} \proj{i}A\proj{\mu} = A_{i\mu}\outpr{i}{\mu}$. Thus the operator could interchangeably denoted by the matrix $A_{i\mu} = \Bra{i}A\Ket{\mu}$. There are some very special classes of operators that we often use :
\begin{itemize}
\item{\textbf{Self-Adjoint Operators:}} $A\in\mathcal{B}(\mathbb{H})$ is self adjoint when  $A = A^{\dagger}$. Self-adjoint operators have real spectrum and their eigenvectors form a basis of $\mathbb{H}$. 
    
\item{\textbf{Isometries:}} An operator $U:\mathbb{H}_1 \rightarrow \mathbb{H}_2$ is called an isometry when it preserves the inner product structure, i.e $\forall \phi,\psi \in \mathbbm{H}_1 $, we have $\inpr{U\psi}{U\phi}_{\mathbb{H}_2} = \inpr{\phi}{\psi}_{\mathbb{H}_1}$. This requires $U^{\dagger}U = \mathbbm{1}_{\mathbb{H}_1}$ (which implies $U U^{\dagger} = \mathbbm{1}_{\mathbb{H}_2}$). When the input and output Hilbert spaces are same, isometry is called an unitary operator.  
\end{itemize}

\textbf{Positive Semi-definite Operators :} I will conclude by stating the very useful concept of positivity \cite{725cb2ad}. The statement that an operator $A \in \mathcal{B}(\mathbb{H})$ is positive semi-definite, $A \geq 0,$ is equivalent to the following three statements :
\begin{enumerate}
\item{For any $\psi \in \mathbb{H}$, $\inpr{\psi}{A\psi} \geq 0$}. \label{pos1} 

\item{$A = A^{\dagger}$, and the spectrum of $A$ contains only non-negative reals : $spec(A) \subset \mathbb{R}_{+}$}.

\item{There exists $B \in \mathcal{B}(\mathbb{H})$, such that $A = B^{\dagger}B$}.
\end{enumerate}

The notion of positive semi-definiteness allows us to introduce partial ordering between operators. For example, $A-B \geq 0$ implies $A \geq B$. For any two positive semi-definite operators $A,B \in \mathcal{B}(\mathbb{H})$, $A+B$ is also positive semi-definite (by condition \ref{pos1}) and so is $\lambda A$, or $\lambda B$, for any non-negative number $\lambda \geq0$. Therefore, set of all positive semi-definite operators forms a cone $\mathcal{C}_{+}(\mathbb{H}) \subset \mathcal{B}(\mathbb{H})$ \cite{725cb2ad,wernerQI}, which is often called the positive cone of $\mathcal{B}(\mathbb{H})$. An operator $A$ is said to be on the `mantle' of the cone, or on an `extremal ray of the cone', if $\forall B \in \mathcal{C}_{+}(\mathbb{H})$, the condition $0 \leq B \leq A$ implies $B = \lambda A$ for some non-negative number $0 \leq \lambda \leq 1$. Therefore, from the eigen-decomposition of positive semi-definite operators, we could easily see that even when $A$ has only two degenerate eigenvectors ($\phi$ and $\psi$) with non-zero eigenvalues, we could write  $0 < \proj{\phi} <A$, with $\proj{\phi}$ being non-proportional to $A$, and hence conclude that $A$ is not on an extremal ray. Therefore, only rank-1 projectors of the form $\proj{\phi}$ for $\phi \in \mathbb{H}$ are the ones that lie on the extremal rays of $\mathcal{C}_{+}(\mathbb{H})$ \cite{wernerQI}.

\subsection{States and Observables (Prepare and Measure)}
\textbf{States :} Consider a system $S$ with a corresponding Hilbert space $\mathbb{H}$. The quantum state of the system is described by a trace $1$, positive semi-definite operator $\rho \in \mathcal{C}_{+}(\mathbb{H})$, called the density operator. Specifying $\rho$ is equivalent to specifying the outcome statistics of any arbitrary measurement on $S$. The set of all density operators forms a convex set, and is often denoted as $\mathcal{T}(\mathbb{H})$. We could find the eigen-decomposition of $\rho$ as $\rho = \sum_{\mu=1}^{n} p_{\mu} \proj{\mu}$, where the trace $1$ condition imposes an extra constrain on the spectrum forcing the eigenvalues to sum to one: $\sum_{\mu=1}^{n} p_{\mu} = 1$. One way to interpret this is to say an ensemble of $S$ can be prepared in state $\rho$ by randomly mixing copies of $S$ prepared in state $\rho_{\mu}=\proj{\mu}$ with probabilities $p_{\mu}$. However, such a preparation protocol is not unique because of the following equation:
\begin{gather}
\rho = \sum_{i=1}^{n} p_i \proj{\psi_i} = \sum_{j=1}^{m} q_j \proj{\phi_j}, \nonumber \\
\text{as long as there exists an isometry $U$ such that :} \nonumber \\
\sqrt{p_i} \Ket{\psi_i} = U_{ij}\sqrt{q_j}\Ket{\phi_j}.
\label{eq1:do_amb}
\end{gather}
Since, in general, there exist infinitely many such isometries, there are infinitely many ways of preparing the system in state $\rho$. Once prepared, all such preparations (or decompositions) of $\rho$ will be statistically indistinguishable with respect to any arbitrary measurement. There is only one exception: when the density operator itself happens to be a rank-1 projector $\rho = \proj{\psi}$. These are the states that are the external points of the convex set $\mathcal{T}(\mathbb{H})$, and they can not be decomposed as a convex sum of other states. They are called `pure states'. All other states are called `mixed state' since they can be prepared by mixing pure states ( in infinitely many ways).

If $S$ is prepared in a pure state, we know more about it than if $S$ is in some mixed state. One way to quantify our ignorance (or information) about some random variable was introduced by Claude Shannon in his landmark paper \cite{6773024} in 1948. Consider a random variable $\mathbb{X}$ which can take $n$ values $(x_1,x_2,...,x_n)$ with respective probabilities $(p_1,p_2,...,p_n)$. If we measure $\mathbb{X}$ and find a highly improbable outcome, we will learn more about it than if a more probable outcome occurs. The information gained by learning $\mathbb{X}=x_k$ can thus be quantified as $-\log(p_k)$ \footnote{Although it may seem that there is a singularity at $p_k=0$, but that outcome never occurs so the statement of information gained by learning that outcome contains no meaning.}. Therefore, if we randomly measure $\mathbb{X}$, the information we can get on average reads $\mathscr{S}(\mathbb{X}) = -\sum_{i=1}^{n}p_i \log(p_i)$, which is called the Shannon entropy of $\mathbb{X}$. It measures our ignorance about $\mathbb{X}$. For example, if we know with certainty the definite outcome of $\mathbb{X}$ to be $x_3$ (with $p_3=1$ and $p_i=0$ for $i=1,2,4,..,n$), then $\mathscr{S}(\mathbb{X})=0$, and on the other hand when the ignorance is maximum and each outcome remains equally probable ($\forall i,\,  p_i=1/n$), $\mathscr{S}(\mathbb{X})$ attains the maximum value of $\log(n)$. This notion is generalized to quantum systems by John von Neumann \cite{von2018mathematical} by defining an entropy for quantum state $\rho$ as $S(\rho) = - \rm{tr}[\rho \log \rho ]$, which we now call the von-Neumann entropy \footnote{In this definition von-Nuemann entropy of a quantum state is a dimensionless positive number lying between $0$ to $\log d$, where $d$ is the dimension of the corresponding Hilbert Space.}. If the eigen decomposition of $\rho$ is $\sum_{\mu} p_{\mu} \proj{\mu}$, we can write $\rho = U\Lambda U^{\dagger}$, where $U$ is a unitary operator having columns as eigenvectors of $\rho$, and $\Lambda$ is a diagonal matrix containing the corresponding eigenvalues $\{p_{\mu}\}$. The von-Neumann entropy of the state reads 
\begin{gather}
S(\rho) = -\rm{tr}[\rho \log \rho] = -\rm{tr}[\rho \log(U\Lambda U^{\dagger})] = -\rm{tr}[\rho U\log(\Lambda)U^{\dagger}] \nonumber \\
= -\rm{tr}[U^{\dagger}(U\Lambda U^{\dagger})U\log \Lambda] = -\rm{tr}[\Lambda \log \Lambda] = -\sum_{\mu}p_{\mu} \log p_{\mu}. \label{eq1:entropy}
\end{gather}
Thus, one way to interpret the von-Neumann entropy is to say it is the Shannon entropy of the probability distribution  $\{p_{\mu}\}$ obtained in the eigen-decomposition of the density matrix. Clearly, for a pure state $\rho = \proj{\psi}$, we will have $S(\rho) = 0$. On the other hand, if the density operator reads $\rho = \sum_{{\mu}=1}^{d}(1/d)\, \proj{\mu}$, the Von-Neumann entropy attains its maximum value $S(\rho) = \log d$, where $d$ is the dimension of the respective Hilbert space. This state is called the `maximally mixed state' and it describes maximal ignorance about the system.

\textbf{Observables:} At the end of any experiment, we perform a measurement on the quantum system (the rightmost block in Fig.~\ref{fig:ludwig1}). We are going to see how to mathematically describe quantum measurement. First thing to do is to identify the set of possible outcomes of the measurement $\mathbb{X} = \{x_1,x_2,...,x_n\}$. In this thesis, we will only consider the cases where $\mathbb{X}$ is a finite set. If we measure the system, one of the $x_i$'s will be read as the measurement outcome, and the system will `collapse' to a corresponding post-measurement state $\rho_i$. All that quantum mechanics could predict is the probability $p_i$ of getting outcome $x_i$. After specifying the outcome space $\mathbb{X}$, we associate a positive operator $F_{i} \in \mathcal{C}_{+}(\mathbb{H})$ for each possible outcome $x_i$, such that $\sum_{i}F_{i} = \mathbbm{1}$. The full set of these operators corresponding to the outcome set $\mathbb{X}$ is called POVM (positive operator valued measure) and denoted as $\mathcal{M}_{\mathbb{X}}$ :
\begin{gather}
\mathcal{M}_{\mathbb{X}} := \{ \mathbbm{1}\geq F_{i} \geq 0, \, \forall x_i \in \mathbb{X}, ~\text{such~that~} \sum_{i}F_{i} = \mathbbm{1}  \}.
\end{gather}
If the state of the system is $\rho$, then the probability of getting outcome $x_i \in \mathbb{X}$ is given as $p_{i} = \rm{tr}[F_i\,\rho]$. Hence the set $\mathcal{M}_{\mathbb{X}}$ completely determines the outcome probabilities $\{p_i\}_{i=1}^{n}$ of the measurement on a given state $\rho$. However, the POVM elements $F_i$ do not uniquely determine the post-measurement state, i.e. the state to which the system would `collapse' after the outcome $x_i$ is recorded in the measuring device. We define a new set of operators $\{M_i\}$ as $M_i = U_{i}\,\sqrt{F_i}$, where $U_i$ is some arbitrary unitary operator. The post measurement state after an outcome $x_i$ is observed is given by $\rho_i = M_i \rho M_{i}^{\dagger}/p_i$, where $p_i = \text{tr}(F_i \rho) = \text{tr}(M_i \rho M_i^{\dagger})$ is the probability of getting that outcome, as before. If we do not keep track of the particular outcome that occurred, then the post measurement state reads $\rho \rightarrow \sum_{i=1}^{n}p_i\,\rho_i = \sum_{i=1}^{n} M_i \rho M_{i}^{\dagger}$. Often this is the state we refer to when we use the phrase `post measurement state', whereas $\rho_i$ is called the `post-selected' state corresponding to outcome $x_i$. The set $\{M_i\}$ determines both the outcome probabilities $\{p_i\}$ of the measurement, and the post measurement states $\{\rho_i\}$, and thus it describes a measurement process. However, as there could be infinitely many unitary operators $U_i$, there exist infinitely many measurement processes $\{M_i\}$ corresponding to a given POVM $(\mathcal{M}_{\mathbb{X}})$.

One particularly interesting measurement scenario happens when, corresponding to an outcome set $\mathbb{X}=\{x_\mu\}$, we have POVM $\mathcal{P}_{\mathbb{X}} = \{F_{\mu} = \proj{\mu} \}$, such that the elements are orthogonal to each other, i.e $\text{tr}[F_{\mu}^{\dagger}F_{\nu}] = \delta_{\mu \nu}$. Such set of operators $\mathcal{P}_{\mathbb{X}}$ are called a Projective Valued Measure and often abbreviated as PVM. 
For a given PVM $\mathcal{P}_{\mathbb{X}}$, the expression $\sum_{\mu=1}^{d}x_{\mu}\proj{\mu}$ represents eigen decomposition of some self-adjoint bounded operator $P\in\mathcal{B}(\mathbb{H})$, which is often called the `quantum mechanical observable'. Given such $P$, the average value of the outcome, which is called the `expectation value of $P$', and is denoted as $\langle P \rangle$,  can be directly computed from $P$ as 
\begin{gather}
\langle P \rangle = \sum_{\mu=1}^{d} x_{\mu}\,p_{\mu} = \sum_{\mu=1}^{d} x_{\mu} \text{tr}\left[\proj{\mu}\rho \right] = \text{tr}\left[\left( \sum_{\mu=1}^{d}  x_{\mu} \proj{\mu} \right)\rho\right] = \text{tr} (P \rho).
\end{gather}

\textbf{An Example (The Qubit) :} The simplest non-trivial quantum system is called a `qubit' and has a two-dimensional Hilbert space $\mathbb{H}_2$ associated with it. A generic basis of $\mathbb{H}_2$ contains only two vectors that span the space, traditionally they are denoted as $\{\Ket{0},\Ket{1}\}$, and called the `computational basis' in analogy with the $\{0,1\}$ symbols used in the logic of classical computers. The vector space of linear bounded self-adjoint operators $\mathcal{B}(\mathbb{H}_2)$ are spanned by four $2 \times 2$ matrices, which can be written in computational basis as :
\begin{gather}
\sigma_0 = \mathbbm{1} = \begin{bmatrix}
    1 & 0 \\ 0 & 1
\end{bmatrix},  ~~
\sigma_1 = \begin{bmatrix}
    0 & 1 \\ 1 & 0
\end{bmatrix},  ~~
\sigma_2 = \begin{bmatrix}
    0 & -i \\ i & 0
\end{bmatrix},  ~~
\sigma_3 = \begin{bmatrix}
    1 & 0 \\ 0 & -1
\end{bmatrix}.  \label{eq1:pauli}
\end{gather}
A generic self-adjoint operator $A \in \mathcal{B}(\mathbb{H}_2)$ can thus be written as 
\begin{gather}
A = r_0 \,\mathbbm{1} + r_1 \,\sigma_1 + r_2 \,\sigma_2 + r_3 \,\sigma_3, ~~\text{with}~~ (r_0,r_1,r_2,r_3) \in \mathbb{R}^4. \label{eq1:qubit_isom}
\end{gather}
The above Eq.~\Eqref{eq1:qubit_isom} establishes an isomorphism between the set of self adjoint operators $\mathcal{B}(\mathbb{H}_2)$ and $\mathbb{R}^4$ which us to geometrically discuss the $\mathcal{B}(\mathbb{H}_2)$ space by studying $\mathbb{R}^4$. To show how this works, let us discuss the positive cone $\mathcal{C}_{+} \subset \mathcal{B}(\mathbb{H}_2)$. The eigenvalues of the operator $A$ of Eq.~\Eqref{eq1:qubit_isom} reads $\lambda_{\pm} = r_0 \pm \norm{\vec{r}}$, where we define $\vec{r} = (r_1,r_2,r_3)$ as a vector in $\mathbb{R}^3$, often called the  Bloch vector, and $|r| = \sqrt{r_{1}^2 + r_{2}^2 + r_{3}^2}$. Clearly, from the conditions of positive semi-definiteness we can state that for $A \in \mathcal{C}_+(\mathbb{H}_2)$ we need :
\begin{enumerate}
\item{$r_0 \geq 0$, and}  

\item{$r_0 \geq \norm{\vec{r}}$}.
\end{enumerate}

\begin{figure}
\centering
\includegraphics[width=13cm, clip = true, trim={3.9cm 4cm 0cm 3cm}]{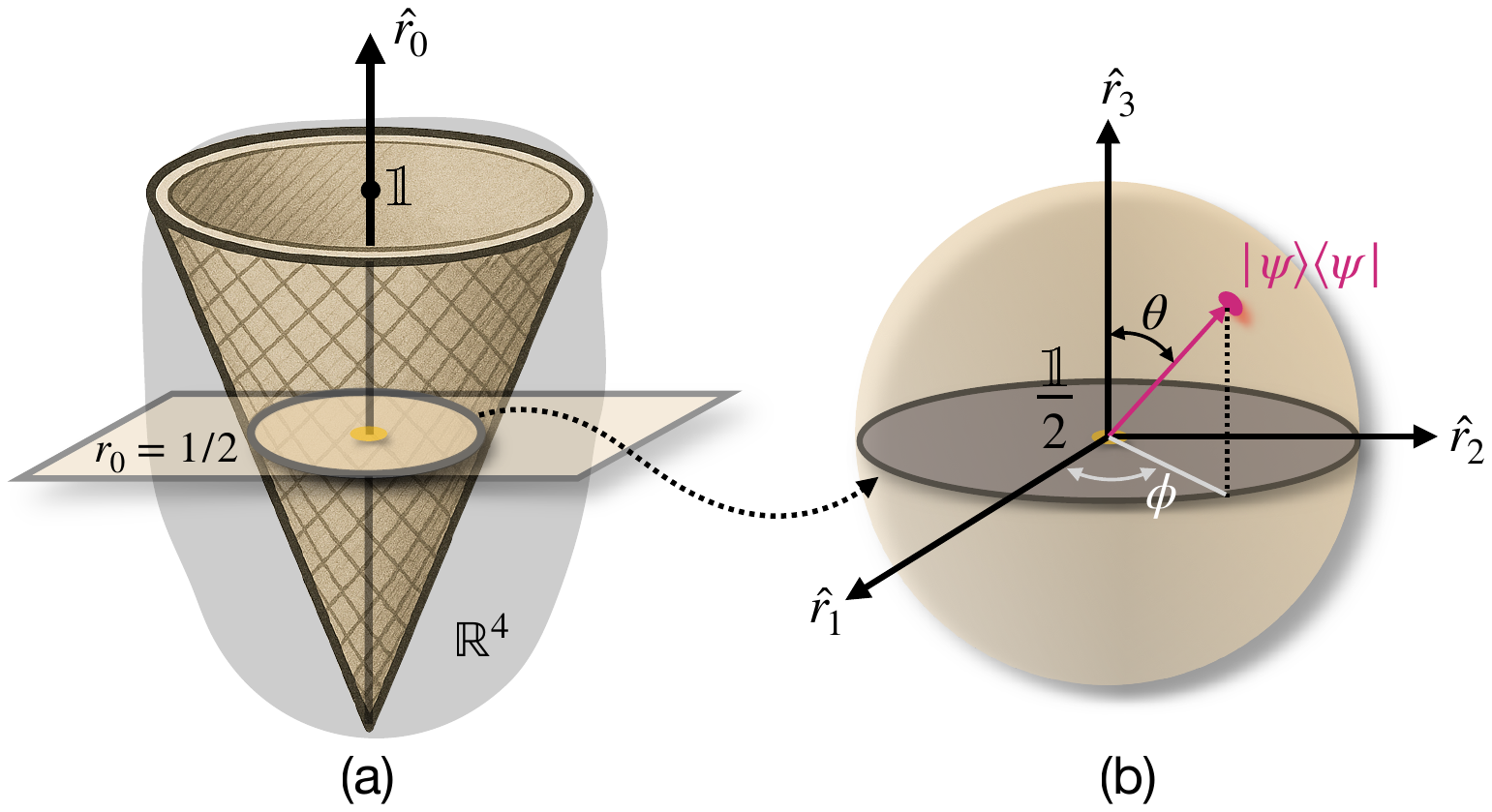} 
\caption{(a) A cartoon representation of the cone $\mathcal{C}_{+}(\mathbb{H}_2)$ in $\mathbb{R}^4$, which is cut by a hypersurface $r_0=1/2$ to obtain the Bloch sphere shown in (b)}.  
\label{fig:cone_sphere}
\end{figure}

Since each point $(r_0,r_1,r_2,r_3)$ in $\mathbb{R}^4$ corresponds to a self adjoint operator in $\mathcal{B}(\mathbb{H}_2)$ by Eq.~\Eqref{eq1:qubit_isom}, we are going to investigate the geometry of the positive cone $\mathcal{C}_{+}(\mathbb{H}_2)$ in $\mathbb{R}^4$. From the above conditions, it is clear that the origin $(0,0,0,0)$, which corresponds to the zero operator, is inside the positive cone : basically it is the origin of the cone. Starting from the origin, we move along the positive $r_0$ axis, abbreviated as $\hat{r}_0$, which always remains inside the positive cone since it contains all the operators of the form $r_0 \mathbbm{1}$ \footnote{This is because throughout this axis $\hat{r}_0$, we have $|\vec{r}|=0$ and thus the corresponding positive operators will be of the form $r_0\mathbbm{1}$}. According to the condition 2 : at any given point $(r_0,0,0,0)$ on the positive $\hat{r}_0$ axis, there exists a sphere of radius $r_0$, having that point as the origin, which lies inside the positive cone. This is similar to how we see an ordinary cone in 3D, there exists a central axis, and at any given point on that axis, we can find a circle that lies inside the cone. Only now the circle is replaced by the sphere, since we are dealing with a four-dimensional cone. As we move up along the positive $\hat{r}_0$, the sphere gets bigger and bigger with its surface defining the `outer layer' or `mantle' of the cone. Therefore, each point of the surface corresponds to rank-1 operators. The $\text{tr}{A}= 1$ condition corresponds to the equation $r_0=1/2$, which defines a hypersurface in $\mathbb{R}^4$ perpendicular to $\hat{r}_0$ . This intersection of this hypersurface and the cone is a sphere of radius $1/2$ centered at the point $r_0 = 1/2$ (which corresponds to the operator $\mathbbm{1}/2$). This sphere, called the `Bloch Sphere' is of prime importance in quantum information since it contains all possible quantum states of the qubit. All the points on the surface of the sphere corresponds to the rank 1 projectors of the form $\proj{\psi}$ with $\Inpr{\psi}{\psi}=1$. 

From the state perspective, each point on the surface of the Bloch Sphere corresponds to a pure state. These states can be parametrized by two angles $\theta \in [0,\pi]$ which is the co-latitude of the Bloch vector, and $\phi \in [0,2\pi)$ which is the longitude \cite{Nielsen_Chuang_2010}. The state can be written as : $\Ket{\psi}=\Ket{\theta,\phi}=\cos (\theta/2) \Ket{0} + \sin(\theta/2) \exp(i\phi)\Ket{1}$.

\subsection{Quantum Channels (Operate)}
A quantum channel $\Phi$, in Schrodinger representation, is a map that transforms the density operators of an `input' Hilbert space $\mathbb{H}_{\text{in}}$ to the density operators of an `output' Hilbert space $\mathbb{H}_{\text{out}}$, $\Phi : \mathcal{T}(\mathbb{H}_{\text{in}}) \rightarrow \mathcal{T}(\mathbb{H}_{\text{out}})$, such that the following property holds for any $\rho_1,\rho_2 \in \mathcal{T}(\mathbb{H}_{\text{in}})$:
\begin{itemize}
\item $\Phi$ is a linear map : $\Phi(\lambda_1\rho_1+\lambda_2\rho_2) = \lambda_1\Phi(\rho_1) + \lambda_2\Phi(\rho_2)$, for $\lambda_1,\lambda_2 \in \mathbb{C}$ 
\item $\Phi$ is trace-preserving (TP) : $\text{tr}[\Phi(\rho)]=\text{tr}[\rho]$. 
\item $\Phi$ is completely positive (CP) : $\Phi$ is a positive map if it maps positive semidefinite operators to positive semidefinite operators : $\Phi(\rho) \geq 0,~\forall \rho \geq 0$. $\Phi$ is completely positive if $\Phi \otimes \mathcal{I}_R^{d}$ is a positive map for all $d$, where $d$ is the dimension of an ancillary Hilbert space $\mathbb{H}_{R}$, and $\mathcal{I}_{R}^{d}$ is an identity map on $\mathcal{B}(\mathbb{H}_R)$. 
\end{itemize}

For any linear map $\Phi : \mathcal{T}(\mathbb{H}_{\text{in}}) \rightarrow \mathcal{T}(\mathbb{H}_{\text{out}})$, the following statements are equivalent : 
\begin{enumerate}
\item $\Phi$ is CPTP, in other words, $\Phi$ is a quantum channel 
\item $\Phi$ admits Kraus decomposition, i.e
\begin{gather}
\Phi(\cdot) = \sum_{a=0}^{n^2-1} K_a(\cdot)K_{a}^{\dagger},~\text{with}~\sum_{a=0}^{n^2-1}K_{a}^{\dagger}K_a = \mathbbm{1},
\end{gather}
where $n$ is the dimension of $\mathbb{H}_{\text{in}}$, and $K_{a}$'s are called the Krauss operators. 
\end{enumerate}

\subsection{Putting in a continuous time}
Up to now, we have only discussed about quantum channels, which are discrete time transformations on the state of a quantum system. However, in most of the experiments, it becomes important to describe the system as it evolves continuously in time. To do so, consider the state of the quantum system at time $t$ be described by a density operator $\rho(t)$, which is a continuous function of $t$. Therefore, by Taylor expansion \cite{lidar2020lecturenotestheoryopen} we get for an infinitesimal time increment $dt$ : 
\begin{gather}
\rho(dt) = \rho(0) + [\dot{\rho}]_{t=0}dt + \mathcal{O}(dt^2) = \sum_{a=0}^{n^2-1}K_a \rho(0) K_a^{\dagger}, \label{eq:cont1}
\end{gather}
where the last equality is written considering the complete positivity and trace preservation of the density matrix in any general transformation from $t=0 \rightarrow dt$. In order to satisfy the left equality of Eq.~\Eqref{eq:cont1}, at least one of the Krauss operator must be of the form \cite{lidar2020lecturenotestheoryopen} $K_{0} = \mathbbm{1} + L_0 dt$, with rest of them being $K_a = L_a \sqrt{dt}$, for $a > 0$. The trace preservation condition implies 
\begin{gather}
K_0^{\dagger}K_0 + \sum_{a>0}^{n^2-1} K_a^{\dagger}K_a = \mathbbm{1} + \left( L_0 + L_0^{\dagger} + \sum_{a>0}^{n^2-1} L_a^{\dagger}L_a\right)dt + \mathcal{O}(dt^2) = \mathbbm{1} \nonumber \\
\implies A = -\frac{1}{2} \sum_{a>0}^{n^2-1} L_{a}^{\dagger}L_a,~\mbox{where}~L_0=A- iH,
\end{gather}
since we can always write $L_0$ (any operator) as linear combination of Hermitian ($A$) and anti-Hermitian operator ($-iH$). Now we can rewrite Eq.~\Eqref{eq:cont1} as 
\begin{gather}
\rho(dt) = \rho(0) + \left[\left(A-iH\right)\rho(0) + \rho(0)\left(A+iH\right) + \sum_{a>0}^{n^2-1}L_a \rho(0)L_a^{\dagger} \right]dt + \mathcal{O}(dt^2) \nonumber \\
\implies \lim_{dt\rightarrow0}  \frac{\rho(dt)-\rho(0)}{dt} = \left[ \frac{d\rho(t)}{dt} \right]_{t=0} = -i\left[H,\rho(0) \right] + \sum_{a>0}^{n^2-1} \left( L_a\rho(0)L_a^{\dagger} -
\frac{1}{2}\{L_a^{\dagger}L_a,\rho(0)\} \right). \label{eq:temp}
\end{gather}
We note that the $L_a$ operators for $a>0$ in Eq.~\Eqref{eq:temp} have the dimension of $1/\sqrt{\text{time}}$. To make them dimensionless, we pull out a factor $\sqrt{\gamma_a}$ from each $L_a$, where $\gamma_a$ has the dimension of $1/\text{time}$. Note that, we demand $\gamma \geq 0$ by construction. With this modification, Eq.~\Eqref{eq:temp} can be written as 
\begin{gather}
\left[ \frac{d\rho(t)}{dt} \right]_{t=0} = -i\left[H,\rho(0) \right] + \sum_{a>0}^{n^2-1} \gamma_a \left( L_a\rho(0)L_a^{\dagger} -
\frac{1}{2}\{L_a^{\dagger}L_a,\rho(0)\} \right). \label{eq:temp}
\end{gather}
This equation is valid only at $t=0$. Now we assume that the equation is valid for all $t\geq0$. In other words, we are assuming that the evolution has a memory that is much smaller than $dt$, so that it ``resets'' itself in every $dt$. This assumption, known as `Markovian approximation', leads us from Eq.~\Eqref{eq:temp} to the GSKL master equation \cite{chruscinski2017brief,lindblad1976generators,gorini1976completely} : 
\begin{gather}
\frac{d\rho(t)}{dt} = \mathcal{L}\,[\rho(t)] =  -i\left[H,\rho(t) \right] + \sum_{a>0}^{n^2-1} \gamma_a \left( L_a\rho(t)L_a^{\dagger} -
\frac{1}{2}\{L_a^{\dagger}L_a,\rho(t)\} \right), \label{eq:def_gskl}
\end{gather}
which describes a continuous time evolution of a quantum system such that between any two arbitrarily chosen times $t_2 \geq t_1 \geq 0$, the quantum state transforms as $\rho(t_2) = \Phi(t_2,t_1)\rho(t_1) =  \exp[{\mathcal{L}(t_2-t_1)}]\rho(t_1)$, where $\Phi(t_2,t_1)$ is guaranteed to be a CPTP map as long as we ensure $\gamma_a \geq0$ in Eq.~\Eqref{eq:def_gskl}.

\section{Basics of Quantum Correlations :} \label{intro_corr}

\begin{center}

``\textit{Correlations cry out for explanation}'' \\
~~~~~~~~~~~ John Bell, `\textit{Speakable and Unspeakable in Quantum Mechanics}'
\end{center}

Correlations are one of the key notions in which quantum systems radically depart from their classical counterparts. In fact, this departure was exploited to point out the bizarreness of quantum mechanical predictions in early days of the theory \cite{PhysRev.47.777,Beller1994}. Spatial correlation can exist between two, or multiple, systems that are physically distant from each other. On the other hand, a single quantum system can be correlated with itself at two, or multiple, instants of time, which is known as `temporal correlations'. It has been shown that in both spatial \cite{PhysicsPhysiqueFizika.1.195,RevModPhys.38.447,PhysRevLett.23.880,PhysRevLett.28.938,PhysRevLett.49.1804} and temporal \cite{PhysRevLett.54.857,Leggett_2008,palacios2010experimental,PhysRevLett.107.130402,PhysRevLett.107.090401,PhysRevA.105.042613} correlations, quantum systems can surpass the bounds that classical systems obey. Therefore, correlation measures can act as a witness of non classicality. Here I present some of the basic concepts of quantum correlations that will be used throughout this thesis. For a more comprehensive review of the topic, the reader is encouraged to consider \cite{Adesso_2016,Leggett_2008,fanchini2017lectures}.   
\subsection{Spatial Correlation :}
\subsubsection{Classical Correlation :}
Consider two quantum systems $A$ and $B$, corresponding to finite dimensional Hilbert spaces $\mathbb{H}_A$ and $\mathbb{H}_B$, respectively. The two systems are said to be in a correlated state $\rho^{AB}$, if the sum of their von-Neumann entropies ($S(\rho^A) + S(\rho^B)$) is more that the von-Neumann entropy of the combined system $S(\rho^{AB})$, where $\rho^{A(B)}=\text{tr}_{B(A)}[\rho^{AB}]$. When $A$ and $B$ are in a correlated state, then knowledge about $A$ provides information about $B$, and therefore our combined ignorance about $A$ and $B$ remains more than our ignorance about $A$ and $B$ combined. This difference, very aptly abbreviated as the `mutual information ($I_{A:B}$) between $A$ and $B$', is used to measure the total correlation between them :
\begin{gather}
I_{A:B}(\rho^{\text{AB}}) = S(\rho^A) + S(\rho^B) - S(\rho^{AB}). \label{eq1:mi}
\end{gather}
As an example, supposing both $A$ and $B$ to be qubits in a state 
\begin{gather}
\sigma^{AB}=p\proj{00} + (1-p)\proj{11}, \label{eq1:mi_eg}
\end{gather}
their mutual information reads $I_{A:B}(\sigma^{AB})=-p\log (p) - (1-p)\log(1-p)$, which is maximum at $p=1/2$ and minimum at $p=0,1$. We can understand this by noting that at $p=0$ (or $1$) we know the exact state of $A$, $B$ and $AB$, and therefore our combined ignorance about them is zero, which is same as our ignorance about them combined. On the other hand, at $p=1/2$, although both $A$ and $B$ are in a maximally mixed state $\mathbbm{1}/2$, the combined system $\sigma^{AB}$ is not maximally mixed. This means even though we are completely ignorant about the individual systems $A$ and $B$, we are not in maximal ignorance about $AB$.

\subsubsection{Quantum Correlation :}
The mutual information is  a `measure' of the total correlation present in the system's state. To measure the amount of quantum correlation present in state $\rho^{AB}$, we need the notion of quantum discord \cite{PhysRevLett.88.017901}. We say the state $\rho^{AB}$ to be classically correlated with respect to (w.r.t) $A$, or having zero quantum discord ($\mathcal{D}_{A:B}(\rho^{AB})=0$) w.r.t to $A$, if there exists a complete measurement basis $\mathcal{P}:=\{\Pi_i=\proj{i}\}$, such that if a projective measurement of $\mathcal{P}$ is performed on $A$, it will leave the state of $AB$ invariant. In other words, the post measurement state of $AB$ would be $\rho^{AB}$ itself. For the choice of the state $\sigma^{AB}$ of Eq.~\Eqref{eq1:mi_eg}, it can be seen that such measurement basis for $A$ would be $\{\proj{0},\proj{1}\}$. Note that $\mathcal{D}_{A:B}(\rho^{AB})=0$ does not imply $\mathcal{D}_{B:A}(\rho^{AB})=0$, and depending on whether such a measurement basis exists for only $A$, only $B$ or both $A$ and $B$, $\rho^{AB}$ can be classically correlated w.r.t $A$, $B$ or both $A$ and $B$, respectively.

To quantify quantum correlation in terms of quantum discord $\mathcal{D}_{A:B}(\rho^{AB})$, we first focus on the post measurement state of $AB$ after a measurement of $\mathcal{P}$ is being done on $A$:
\begin{gather}
\rho^{AB} \rightarrow \mathcal{P}(\rho^{AB}) = \sum_i p_i\, \proj{i} \otimes \rho^{B}_i, \label{eq1:pm} 
\end{gather}
where the probability $p_i=\text{tr}[\Pi_i\,\rho^{AB}\,\Pi_i],  
\mbox{and}~\rho_{B}^{i}=\text{tr}_A[\Pi_i\,\rho^{AB}\,\Pi_i]/p_i$. From the above discussion, we can say that the post measurement state of Eq.~\Eqref{eq1:pm} contains zero discord w.r.t $A$. The conditional entropy of $B$, after a measurement of $\mathcal{P}$ is done on $A$, reads $S(B|A)_{\mathcal P}=\sum_{i}p_{i}\,S(\rho^B_i)$. If $\rho^{AB}$ is such that $A$ is classically correlated with $B$, then it always takes the form $\rho^{AB}=\sum_{q}p_q \proj{q}\otimes \rho^{B}_q$, and thus there always exist a measurement basis $\mathcal{Q}=\{\proj{q}\}$ which leaves the state invariant : $\rho^{AB} = \mathcal{Q}(\rho^{AB})=\sum_{q}p_q \proj{q}\otimes \rho^{B}_q$. In these cases, we can alternatively define mutual information as $I_{A:B}(\rho^{AB})=J_{A:B}^{\mathcal Q}(\rho^{AB}):=S(B)-S(B|A)_{\mathcal{Q}}$. However, if $\rho^{AB}$ is such that $A$ is not classically correlated with $B$, then no such basis exists, measurement on which would leave $\rho^{AB}$ invariant. In those cases, we have $J^{\cal Q}_{A:B}(\rho^{AB}) \neq I_{A:B}(\rho^{AB})$\footnote{omitting the trivial case where $A$ and $B$ are uncorrelated, which would imply $J^{\cal Q}_{A:B}(\rho^{AB}) = I_{A:B}(\rho^{AB})=0$}. Thus a measure of quantum discord of state $\rho^{AB}$ w.r.t $A$ can be defined as 
\begin{gather}
\mathcal{D}_{A:B}(\rho^{AB}) = \min_{\mathcal{P}}\left[I_{A:B}(\rho^{AB}) - J_{A:B}^{\mathcal{P}}(\rho^{AB}) \right], \label{eq:disc}
\end{gather}
where the minimization is performed over all possible measurement basis. Note that in general $\mathcal{D}_{A:B} \neq \mathcal{D}_{B:A}$, which displays that discord is an asymmetric correlation.

\subsubsection{Quantum Entanglement :} 
\begin{center}
``\textit{The world appears quantum mechanical because entanglement is fundamental, not because particles are mysterious}'' \\
~~~~~~~~~~~~~~~~~~~~ -- Lee Smolin
\end{center}

A special type of quantum correlation that deserves a separate mention is `Quantum Entanglement' \cite{horodecki2009quantum,schrodinger1935mathematical}. Even though entanglement can exist between multiple systems \cite{walter2016multipartite,PhysRevA.92.042329}, I will only discuss bipartite entanglement that can exist between two systems, which suffices the requirement of this thesis. Two systems, $A$ and $B$, that are physically apart are said to be in a separable state if their density operator can be written as 
\begin{gather}
\rho^{AB} = \sum_{k} p_k \, \rho^{A}_k \otimes \rho^{B}_k. \label{eq1:sep}
\end{gather}
Clearly, all such separable states form a convex set $\mathcal{S}(\mathbb{H}_{AB})\subset \mathcal{T}(\mathbb{H}_{AB})$, where $\mathbb{H}_{AB}=\mathbb{H}_A \otimes \mathbb{H}_{B}$ is the Hilbert space of the combined system $AB$. Any state $\rho_{AB} \notin \mathcal{S}(\mathbb{H}_{AB})$, which is not separable, is entangled. A function $E:\mathcal{T}(\mathbb{H}_{AB})\rightarrow \mathbb{R}$ is called an `Entanglement measure', if, for a given state $\rho$, the number $E(\rho^{AB})$ quantifies the amount of entanglement between $A$ and $B$. For that to happen, $E$ must satisfy the following conditions :
\begin{itemize}
\item The entanglement between $A$ and $B$ can not increase under any local operations and classical communication (LOCC) \cite{Chitambar2014} : $E(\rho^{AB}) \geq E(\Lambda_{LOCC}[E(\rho^{AB})])$, where $\Lambda_{LOCC}$ represents a CPTP quantum channel representing LOCC.

\item The entanglement between $A$ and $B$ should remain invariant under the action of local unitaries : $E(\rho^{AB})=E((U_{A}\otimes U_{B})\rho^{AB}(U_{A}^{\dagger}\otimes U_{B}^{\dagger}))$, where $U_A$ and $U_B$ are two arbitrary local unitary operators on $A$ and $B$.

\item $E(\rho^{AB})=0$ for any separable states $\rho^{AB} \in \mathcal{S}(\mathbb{H}_{AB})$.
\end{itemize}

\subsection{Temporal Correlation :}
Apart from correlations between spatially separated systems, a single quantum system can be correlated at two different instances of time. To understand temporal correlations within the requirement of this thesis, consider the time evolution of a two-level system while a dichotomous ($\pm 1$ valued) observable $Q$ is measured at different time points $t_n$ to obtain outcomes $q_n\in\{0,1\}$. The two-time correlation between time $t_i$ and $t_j$ in the measurement outcome of observable $Q$ is defined as $C_{ij}=\langle q_i\,q_j \rangle$, where $\langle \cdot \rangle$ represents an average over multiple runs of the experiment. Repeating this for the three different intervals of time : $(t_1,t_2),(t_2,t_3)$ and $(t_1,t_3)$, we obtain $C_{12},C_{23}$ and $C_{13}$. We focus on a particular linear combination of these three two-time correlations $K_3 = C_{12} + C_{23} - C_{13}$. In their landmark paper \cite{PhysRevLett.54.857}, Leggett and Garg had argued that if we assume the classical worldview, abbreviated as the assumptions of `macrorealism', then it can always be shown that $-3 \leq K_3 \leq 1$, which is known as the Leggett-Garg inequality (LGI). However, experiments performed on various systems \cite{palacios2010experimental,PhysRevLett.107.130402,PhysRevLett.107.090401,PhysRevA.105.042613,PhysRevLett.106.040402} have shown $K_3$ to surpass the bound of macrorealism put by the LGI and thereby confirming the quantum nature of the temporal correlations present in those systems.




\section{Basics of Nuclear Magnetic Resonance (NMR)} 
\begin{center}
``\textit{The spinning nuclei in a sample are like spies inside an unmapped and unexplored continent. To learn about the interior of that continent, you don’t have to destroy it, you don’t even have to invade it, all you have to do is learn the spies’ language and listen to them talk.}" \\
~~~~~~~~~~~~~~~~~~~~~~~~~~~~~~~--- Alex Pines
\end{center}

\label{intro_nmr}
\begin{figure}
\centering
\includegraphics[width=9.3cm, clip = true, trim={10.4cm 3.4cm 0.3cm 2.4cm}]
{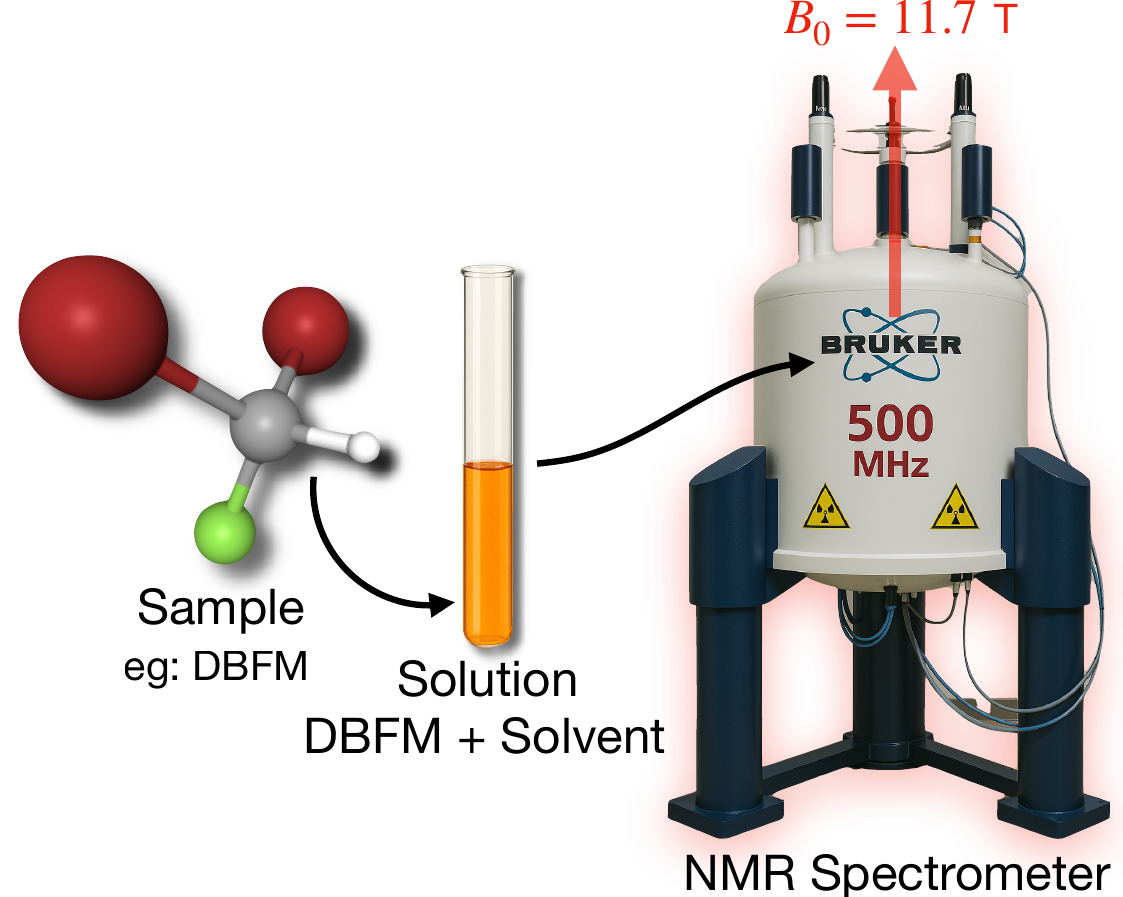} 
\caption{Typical steps involve in the initiation of a solution state NMR experiment. The sample is first dissolved in an appropriately chosen solvent, and then the solution is put in an NMR tube where it stretches to a height of $4-6$ cm. The tube is then put inside the NMR spectrometer where it stands vertically (along $\hat{z}$ axis) in alignment with the strong magnetic field ($B_0 = 11.7$ T).}
\label{fig:Exp_Cartoon}
\end{figure}
We, and all that surrounds us, are built from molecules. These molecules are formed from atoms, and if we look deeper, inside each atom, we discover a dense nucleus, shrouded in swirling clouds of electrons. Each nucleus have an intrinsic property called the nuclear spin, denoted by a quantum number $I$. In nature, we find spinless nuclei ($I=0$) like $^{12}$C,\, $^{16}$O etc, spin half nuclei $I=1/2$ like $^{1}$H, $^{19}$F, $^{13}$C etc and higher spin nuclei like $I = 1$ for $^{14}$N, $^{2}$H; $I=3/2$ of $^{35}$Cl and so on. We can associate a spin angular momentum vector operator $\vec{I}:=(I_x,I_y,I_z)$ to describe nuclear spin. Every nucleus with a non-zero nuclear spin possesses a magnetic dipole moment, represented by the vector operator $\vec{\mu}$, which is proportional to the spin angular momentum $\vec{\mu} = \gamma \, \vec{I}$. The proportionality constant $\gamma$ is called the gyromagnetic ratio and it's value is an intrinsic property of a given isotope, for example, $\gamma/(2\pi) = 42.57$ MHz T$^{-1}$ for $^{1}$H, $40.078$ MHz T$^{-1}$ for $^{19}$F, $10.705$ MHz T$^{-1}$ for $^{13}$C and so on. The presence of this tiny (compared to the electron) yet non-zero magnetic moment for nuclei with non-zero nuclear spins is the source of nuclear magnetism. 

To do experiments, we consider samples whose molecule contain one or more magnetic spin half nuclei \footnote{ NMR experiment involving higher than spin half nuclei is called Quadrupolar NMR, and is a rich and fascinating field of research. However, they interact with the electric field gradient very strongly and suffer extremely rapid relaxation in the kind of experiments I will be talking about in this thesis. Therefore, I will only consider spin half nuclei and safely ignore any higher spin nuclei from now on.}, and dissolve them in any non-reacting, non-magnetic solvent (commonly used solvents are DMSO, D$_2$O, CDCl$_3$ etc). Usually it is recommended to prepare dilute solutions ( typically $50~\mu$L sample in $600-700~\mu$L solvent) to avoid intermolecular dipolar relaxation \cite{ODELIUS1993289,doi:10.1021/acs.jpcb.0c06258}. Typical NMR tubes used are of radius $5$mm and the solution (sample plus solvent) takes a height up to $4$ to $6$ cm in them. This tube is then put inside an NMR spectrometer which is having a very high magnetic field $B_0\,\hat{z}$ (see Fig.~\ref{fig:Exp_Cartoon} ) generated by huge current flowing resistance-free through superconducting coils.  The value of $B_0$ is $11.7$ T in the $500$ MHz Bruker Avance spectrometer, which is used to do all the experiments reported in this thesis. Apart from such solution state, samples can be prepared also in partially ordered or liquid crystal state \cite{emsley2018nmr,DONG200467}, or even in solid state \cite{reif2021solid,Nishiyama2023}; however in this thesis I will only describe the methodology and experiments regarding solution state NMR.   

\subsection{The Spin Hamiltonian} \label{subsec1:spinHam}
Consider a sample molecule with $N$ number of spin half nuclei with respective gyromagnetic ratio $\gamma_i, \gamma_2, ..., \gamma_N$. We will denote the spin angular momentum operators corresponding to $i'$th nucleus as $(I_{ix},I_{iy},I_{iz})$. The amount ($\approx 50~\mu$L) of sample dissolved in the solvent contains an astronomical number of sample molecules, each of them is treated as identical nuclear spin systems. Thus we mathematically describe the dissolved sample in the NMR tube as an ensemble of nuclear spin systems. The spin Hamiltonian $\mathcal{H}(t)$ is generally decomposed into two parts : $\mathcal{H}(t) = \mathcal{H}_{\text{coh}}(t) + \mathcal{H}_{\text{fluc}}(t)$, which are listed in detail below :
\begin{enumerate}
\item $\mathbf{\mathcal{H}_{\text{coh}}(t):}$ This is the coherent part of the spin Hamiltonian, i.e this part of the Hamiltonian is same for all the members of the ensemble at any given time $t$. It can be divided into three parts :
\begin{gather}
    \mathcal{H}_{\text{coh}}(t) = \mathcal{H}_A + \mathcal{H}_B + \mathcal{H}_C(t), ~ \text{with}~ \dnorm{\mathcal{H}_A}\gg \dnorm{\mathcal{H}_B},\dnorm{\mathcal{H}_C (t)},\dnorm{\mathcal{H}_{\text{fluc}}(t)}. \label{eq1:HA_decomp}  
\end{gather}
Thus we can treat $\mathcal{H}_B,~\mathcal{H}_C(t)$ and $\mathcal{H}_{\text{fluc}}(t)$ as perturbations to $\mathcal{H}_A$, whose eigen decomposition is denoted as $\mathcal{H}_A \Ket{r}^A = \omega_{r}^{A}\,\hbar \Ket{r}^A$. We assume that the spacing between non-degenerate eigenvalues of $\mathcal{H}_A$ are large compared to the corresponding matrix elements of $\mathcal{H}_B$ and $\mathcal{H}_C$ \cite{BENGS2020106645} :
\begin{gather}
\Bra{r}^A \mathcal{H}_{B} \Ket{s}^A,~\Bra{r}^A \mathcal{H}_{C}(t) \Ket{s}^A \ll \abs{\omega_{r}^A - \omega_{s}^A}\,\hbar ~~\forall \,\omega_{r}^A \neq \omega_{s}^A, \label{eq1:secassume}
\end{gather}
which is crucial assumption for the validity of the secular approximation in the relaxation theory \cite{burgarth2024taming}, and in this thesis we will always consider the cases where the above assumption is satisfied.
\begin{itemize}
    \item $\mathbf{\mathcal{H}_A}:$ The dominant part of the spin Hamiltonian is the Zeeman interaction, i.e. the interaction of nuclear magnetic moment with the static magnetic field $B_0\,\hat{z}$ of the spectrometer :
    \begin{gather}
    \mathcal{H}_A = \sum_{i=1}^N (-\gamma_i B_0)\hbar \, I_{iz} = \sum_{i=1}^{N} \omega_{0}^{i}\, I_{iz} \,\hbar,
    \end{gather}
    where $\omega_{0}^{i}$ is the Larmor frequency of the $i'$th nucleus. At a static field of $11.7$ T, $\omega_{0}^i/(2\pi)$ typically takes value around few $\approx 100$ MHz (eg: $500.2$ MHz for $^{1}$H, $470$ MHz for $^{19}$F, $125$ MHz for $^{13}$C etc.) Note that we have made a great claim tacitly, by considering $\mathcal{H}_A$ to be part of the coherent Hamiltonian $\mathcal{H}_{\text{coh}}(t)$, that the static field is homogeneous across the macroscopic length of $4-6$ cm of the sample, which is not entirely true and we will come back to this in chapter \ref{chap_Mpemba}.

    \item $\mathbf{\mathcal{H}_B:}$ This part contains two terms : the isotropic chemical shift $\mathcal{H}_{CS}$ and the scalar coupling $\mathcal{H}_{J}$. The presence of electron clouds around the nucleus shields the external magnetic field $B_0$ slightly (parts per million or ppm) and thereby adds a correction to the Larmor frequency of each nuclei. This correction term depends on the local chemical environment of each nuclei in the molecule. It contains an isotropic term, called the isotropic chemical shift $\mathcal{H}_{CS}$, which does not depend on molecular orientation and hence remains same for all the members of the ensemble at all time. Moreover, the magnetic moments of the spin half nuclei within a molecule interact with each other via a passive interaction mediated by the electron clouds between them, which is known as the scalar coupling and denoted as $\mathcal{H}_J$. Therefore we write
    \begin{gather}
    \mathcal{H}_B = \mathcal{H}_{CS} + \mathcal{H}_J = \left(\sum_{i=1}^{N} \,\Delta_i I_{iz} + \sum_{i > j} 2 \pi J\, \vec{I}_{i} \cdot \vec{I}_{j}\right)\,\hbar.
    \end{gather}
    Typically, the values of isotropic chemical shift $\Delta/(2\pi)$ ranges around few kHz, whereas $J$ ranges around few hundreds of Hz.

    \item $\mathbf{\mathcal{H}_C (t):}$ To control and manipulate the spin states of the sample, we irradiate the tube with linearly polarized (the direction of the magnetic field is aligned along an axis $\hat{x}$ perpendicular to the static field) radio frequency (rf) pulse. The magnetic field of the pulse is time dependent and reads $B_{\text{rf}} = 2B_1 \sum_{i=1}^{N} \cos(\omega_{\text{rf}}^i\, t + \phi_i)$, where the frequencies $\omega_{\text{rf}}^i$ are set equal to the respective Larmor frequencies $\omega_{0}^i$ for on-resonant pulsing. The corresponding Hamiltonian reads 
    \begin{gather}
    \mathcal{H}_{C}(t) = 2\,\hbar \, \sum_{i=1}^{N}\Omega_{i} I_{ix}\, \cos\left(\omega_{\text{rf}}^{i}\, t + \phi_i \right), 
    \end{gather}
    whereas $\Omega_i/(2\pi) = -\gamma_i\,B_1/(2\pi)$ typically ranges between $5-20$ kHz for hard pulses and even lower for continuous long time shaped pulses. 
\end{itemize}
\item $\mathbf{\mathcal{H}_{fluc}(t):}$ There are primarily two terms : the direct interaction between the neighboring nuclear magnetic dipole moments $\mathcal{H}_{DD}$, and the anisotropic chemical shift $\mathcal{H}_{CSA}$ which depends on the orientation of the molecule \footnote{Apart from these two, other major molecular orientation dependent interaction is Quadrupolar interaction \cite{doi:10.1021/acs.jctc.1c00690},  however this is not present in the experiments reported in this thesis where relaxation is studied.}. However, in solution state NMR at ambient temperatures, sample molecules undergo rapid diffusive tumbling. If we focus on a particular molecule, its orientation, and hence both $\mathcal{H}_{DD}$ and $\mathcal{H}_{CSA}$, as seen by it, becomes random functions of time.
\begin{gather}
\mathcal{H}_{\text{fluc}}(t) = \mathcal{H}_{DD}(t) + \mathcal{H}_{CSA}(t)
\end{gather}
The timescale of the rapid tumbling, and hence of the rapid fluctuations of $\mathcal{H}_{\text{fluc}}(t)$, are much faster than the dynamics caused by the spin Hamiltonian. Thus the spin system only sees an ensemble averaged (averaged over all possible orientation) $ \overline{\mathcal{H}_{\text{fluc}}}(t)$, which turns out to be zero for both the terms $\overline{\mathcal{H}_{DD}}(t) = \overline{\mathcal{H}_{CSA}}(t) = 0$. However, their second moments remain non-zero and thus they play a vital role in the relaxation of spin systems which will be described in details during the discussion on relaxation.  
\end{enumerate}

\subsection{Relaxation in Solution State NMR}
At the beginning of any the experiment, when the sample is inside the spectrometer in the presence of the Zeeman field $B_0\,\hat{z}$, the spin system rests in thermal equilibrium at ambient temperature $T$. The state is described by the thermal density operator $\rho^{th}$, which reads ($k_B$ representing the Boltzmann constant)
\begin{gather}
\rho^{th} = \frac{\expo{-\mathcal{H}/k_B T}}{\text{tr}(\expo{- \mathcal{H}/k_B T})} \approx \frac{\expo{- \mathcal{H}_A}/k_B T}{\text{tr}(\expo{- \mathcal{H}_A}/k_B T)} \approx \frac{\mathbbm{1}}{N} + \sum_{i=1}^{N}\underbrace{\left(\frac{\omega_{0}^{i} \hbar}{N k_{B}T} \right)}_{\epsilon_i} I_{iz},   \label{eq1:Rth}
\end{gather}
where the first approximation is justified because, as mentioned in Eq.~\Eqref{eq1:HA_decomp}, $\mathcal{H}_A$ dominates over all the other ($\mathcal{H}_B,\, \mathcal{H}_C(t)$ and $\mathcal{H}_{\text{fluc}}(t)$) terms of the spin Hamiltonian. Whereas at ambient temperatures of $T \approx 300$ K, the typical values of the purity factor read $\epsilon_i \approx 10^{-5}$, which justifies the second approximation. During an experiment, the spin state of the system is maneuvered as the experimenter wants, using the techniques of quantum control. However, the spin state always comes back to the thermal state $\rho^{th}$ in some characteristic time ($T_1$). This irreversible non-unitary dynamics of the spin system, by which it loses its memory of the initial state and always ends up in the thermal state $\rho^{th}$, is called the Relaxation dynamics, or thermalization.

I will describe a semi-classical approach to understand the pure relaxation dynamics for solution state NMR at ambient temperatures ($\approx 300$ K). By pure relaxation, I meant that no rf pulses are considered during the relaxation in this treatment, i.e $\mathcal{H}_C (t)$ is set to zero. We consider the relaxation of a two-spin homonuclear system to describe the theory. While the formalism presented below suffices the needs of this thesis, the reader may refer to \cite{kowalewski2017nuclear,abragam1961principles,10.1093/oso/9780198556527.001.0001} for a more detailed and general presentation of the theory of relaxation.

As we noticed in subsection \ref{subsec1:spinHam}, the spin Hamiltonian $\mathcal{H}(t)$ contains a part $\mathcal{H}_{\text{fluc}}(t)$ which depends on the molecular orientation with respect to the external static field. In solution state NMR experiments at ambient temperatures ($\approx 300$ K), the sample molecules undergo rapid tumbling, which causes the orientation of the molecule with respect to the static field to vary randomly, which causes $\mathcal{H}_{\text{fluc}}(t)$ to be a random function of time. Roughly speaking, this establishes a `coupling' between the spin system and the diffusive motion of the molecules caused by the thermal environment (called the `lattice' in NMR for historical reasons), which is the key to relaxation. 

To see how this works, We consider a sample whose molecule contains two spin $1/2$ nuclei of same species (two homo-nuclear qubits).The Hamiltonian, as seen by the spin system of a single molecule in the ensemble, reads
\begin{gather}
\mathcal{H}(t) = 
\mathcal{H}_\text{coh} + \mathcal{H}_{\text{fluc}}(t), \nonumber \\
\text{where}~\mathcal{H}_{\text{coh}}/\hbar = \underbrace{\omega_0\,\left(I_{1z} + I_{2z}\right)}_{\mathcal{H}_{\text{A}}/\hbar} - \underbrace{\Delta\,\left(I_{1z} - I_{2z}\right) + 2\pi J\, I_{1z}\,I_{2z}}_{\mathcal{H}_{\text{B}}/\hbar} 
\nonumber
\end{gather}
is the Hamiltonian mentioned in Eq.~\Eqref{eq1:HA_decomp}, and it does not depend on the molecular orientation. Hence, it remains invariant in time while the sample molecule undergoes rapid molecular tumbling due to diffusion in liquids at high temperature. Moreover, at a given instant of time $t$, $\mathcal{H}_{\text{coh}}$ is same for all the molecules of the ensemble, which are distributed uniformly (since the solution is isotropic) across random orientations. Whereas $\mathcal{H}_{\text{fluc}}(t)$ describes the interactions that depends on molecular orientation. It varies randomly with time since the molecular orientation undergoes rapid tumbling due to the diffusional motion of the sample molecule in the liquid solvent at finite temperatures. We fix a frame of reference, called the Lab frame $(x,y,z)$,  which stays fixed in time having $z$ axis coinciding with the Zeeman field. To take care of the random molecular tumbling, we consider the frame of reference formed by the principal axis system (PAS) of the chemical shift anisotropy tensor $(X,Y,Z)$ \footnote{assuming same PAS for both the nuclei} \cite{goldman2001quantum,kowalewski2017nuclear}, which will be called the molecular frame as it remains glued with the molecule and undergoes random rotation in time with respect to the lab frame $(x,y,z)$. For simplicity, we also consider that the vector $\hat{n}$ connecting both the nuclei, which also remains glued with the molecule and undergoes random rotation with respect to the lab frame, coincides with the $Z$ axis of the molecular frame \footnote{For a more general treatment, consider \cite{KUMAR2000191}}. The fluctuating Hamiltonian contains two interactions  : the intra-molecular dipole-dipole interaction $\mathcal{H}_{\text{DD}}(t)$ and the chemical shift anisotropy $\mathcal{H}_{\text{CSA}}(t)$ (assumed to be axially symmetric). Utilizing the freedom to choose the initial time, we consider that the molecular frame coincides with the Lab frame at $t=0$. At a later time $t$, the fluctuating Hamiltonian $\mathcal{H}_{\text{fluc}}(t)$ can be written in the Lab frame as:  \cite{goldman2001quantum,GOLDMAN2001160,kowalewski2017nuclear}
\begin{gather} 
\mathcal{H}_{\text{fluc}}(t) =   \mathcal{H}_{\text{DD}}(t) + \mathcal{H}_{\text{CSA}} (t),~~\text{where} \nonumber \\
\mathcal{H}_{\text{DD}}(t)/\hbar = -\underbrace{\frac{\mu_0}{4\pi}\frac{\gamma^2\,\hbar}{r^3}}_{b} \sum_{m=-2}^{2} A^{*}_{2m}\left(\theta(t),\phi(t) \right)\,T_{2m}(\vec{I_1},\vec{I_2}), \nonumber \\
\mathcal{H}_{\text{CSA}}(t)/\hbar = \underbrace{\frac{\gamma\,\delta}{3}}_{d} \sum_{m=-1}^{1} A^{*}_{2m}\left(\theta(t),\phi(t) \right)\,T_{2m}(\vec{B},\vec{I}_{1,2}), \nonumber  
\end{gather}
with $r$ being the distance between the two nuclei $(\approx 1-4 \, A^0)$, $\delta$ being the chemical shift anisotropy \cite{goldman2001quantum}, and both the spin vectors $\vec{I_1}$,$\vec{I_2}$ and the Zeeman field vector $\vec{B}$ is written in the Lab frame. The angles $\theta(t)$ and $\phi(t)$ are the co-latitude and longitude, respectively, of the molecular frame $\hat{Z}$ axis with respect to the Lab frame and are random functions of time. The exact forms of the spherical functions are
\begin{gather}
A_{0}(\theta,\phi) = \sqrt{\frac{3}{2}} \left(3\cos^2 \theta -1 \right) = A_{0}^*(\theta,\phi), \nonumber \\
A_{\pm1}(\theta,\phi)  =\mp \,3 \,\sin \theta \cos \theta \expo{\pm i\phi} = \mp A^{*}_{\mp 1}(\theta,\phi), \nonumber \\
A_{\pm 2}(\theta,\phi) = \frac{3}{2} \,\sin^2 \theta \expo{\pm 2i\phi}
= A^{*}_{\mp 2}. \label{eq:spher_har}
\end{gather}
Whereas the spherical tensor operators $T_{2m}(\vec{A},\vec{B})$ reads 
\begin{gather}
T_0(\vec{A},\vec{B}) = \frac{1}{\sqrt{6}} \left(3A_{z}\,B_{z} - \vec{A} \cdot \vec{B} \right) = T_{0}^{\dagger}(\vec{A},\vec{B}), \nonumber \\
T_{\pm1}(\vec{A},\vec{B}) = \mp \frac{1}{2}
\left(A_{\pm}B_{z} + A_{z}B_{\pm}  \right) = -T^{\dagger}_{\mp1}(\vec{A},\vec{B}) \nonumber \\
T_{\pm2}(\vec{A},\vec{B}) = \frac{1}{2} A_{\pm}\,B_{\pm} = T^{\dagger}_{\pm2}(\vec{A},\vec{B}) \label{eq:spher_tens}, 
\end{gather}
with $A_{\pm} := A_x \pm i A_{y}$ and same for $B_{\pm}$. Let $\varrho(t)$ be the density operator of the spin system of one particular molecule at time $t$, which, along with the Hamiltonian $\mathcal{H}(t)$ transforms as we go to the interaction frame defined by $U_0 = \exp(i\mathcal{H}_A t)$ as :
\begin{gather}
\mathcal{H}(t) \rightarrow \widetilde{\mathcal{H}}(t) = U_0(t)\mathcal{H}(t)\,U_{0}^{\dagger}(t) - i U_0(t)\dot{U_{0}^{\dagger}}(t)  = \mathcal{H}_{B} + \underbrace{U_0(t)\mathcal{H}_{\text{fluc}}(t)U^{\dagger}_0(t)}_{\widetilde{\mathcal{H}}_{\text{fluc}}(t)}, \nonumber \\
\varrho(t) \rightarrow \widetilde{\varrho}(t) = U_0(t)\varrho(t)\,U_{0}^{\dagger}(t). 
\end{gather}
The equation of motion in the interaction frame can be readily written as \cite{sakurai1986modern}
\begin{gather}
\frac{d\widetilde{\varrho}(t)}{dt} = -\frac{i}{\hbar}[\mathcal{H}_B+\widetilde{\mathcal{H}}_{\text{fluc}}(t),\widetilde{\varrho}(t)]. \label{eq:VN}
\end{gather}
The time by which $\widetilde{\varrho}(t)$ changes is typically determined by the norm of the generators of the motion, i.e $\dnorm{\mathcal{H}_\text{b}}$ and $\dnorm{\widetilde{\mathcal{H}}_{\text{f}}(t)}$, where $\mathcal{H}_\text{b}=\mathcal{H}_{\text{B}}/\hbar$ and $\widetilde{\mathcal{H}}_\text{f}=\widetilde{\mathcal{H}}_{\text{fluc}}/\hbar$.  We assume this time to be much larger than the typical time $(\tau_c)$ by which the rapidly fluctuating Hamiltonian $\widetilde{\mathcal{H}}_{\text{fluc}}(t)$ changes. Therefore, it is possible to consider a time $t$ \cite{GOLDMAN2001160} such that 
\begin{gather}
\tau_c \ll t \ll \frac{1}{\dnorm{\mathcal{H}_{\text{b}}}},\frac{1}{\dnorm{\widetilde{\mathcal{H}}_{\text{f}}(t)}}, \label{eq:COND}
\end{gather}
where $\widetilde{\varrho}(t) \approx \varrho(0)$. In such a time $t$, the formal solution of Eq.~\Eqref{eq:VN} can be written  as
\begin{gather}
\widetilde{\varrho}(t) = \varrho(0) - i\int_{0}^{t}dt_1 [\mathcal{H_\text{b}}+\widetilde{\mathcal{H}}_{\text{f}}(t_1),\widetilde{\varrho}(t_1)].
\end{gather}
Putting this back to Eq.~\Eqref{eq:VN} \cite{breuer2002theory,lidar2020lecturenotestheoryopen}, we get
\begin{gather}
\frac{d\widetilde{\varrho}(t)}{dt} = -i[\mathcal{H_{\text{b}}}+\widetilde{\mathcal{H}}_{\text{f}}(t),\varrho(0)]  - \int_{0}^{t} dt_1 [\mathcal{H_{\text{b}}}+\widetilde{\mathcal{H}}_{\text{f}}(t), [\mathcal{H_{\text{b}}}+\widetilde{\mathcal{H}}_{\text{f}}(t_1),\varrho(0)]] \nonumber \\
= -i[\mathcal{H_{\text{b}}}+\widetilde{\mathcal{H}}_{\text{f}}(t),\widetilde{\varrho}(t)]  - \int_{0}^{t} dt_1 [\mathcal{H_{\text{b}}}+\widetilde{\mathcal{H}}_{\text{f}}(t), [\mathcal{H_{\text{b}}}+\widetilde{\mathcal{H}}_{\text{f}}(t_1),\widetilde{\varrho}(t)]].
\label{eq:apr1}
\end{gather}
We note that in the experiment, we do not have any access to this density operator $\widetilde{\varrho}(t)$ belonging to one particular molecule. Therefore, we define the density operator $\widetilde{\rho}(t)=\overline{\widetilde{\varrho}(t)}$ describing the spin system's state of the ensemble of all the molecules in the tube, where $\overline{(\cdot)}$ dictates ensemble average. We assume that the density operator $\widetilde{\varrho}(t)$ and the $\widetilde{\mathcal{H}}_{\text{f}}(t)$ are statistically uncorrelated so that they can be averaged separately \cite{abragam1961principles,cavanagh1996protein}. After averaging, Eq.~\Eqref{eq:apr1} reads
\begin{gather}
\frac{d\widetilde{\rho}(t)}{dt} = \frac{d\overline{\widetilde{\varrho}(t)}}{dt} = -i[\mathcal{H_{\text{b}}},\widetilde{\rho}(t)] 
- \int_{0}^{t}dt_1 [\overline{\widetilde{\mathcal{H}}_{\text{f}}(t), [\widetilde{\mathcal{H}}_{\text{f}}(t_1)},\widetilde{\rho}(t)]],
\end{gather}
where we have kept only up to the first non-zero terms \cite{GOLDMAN2001160} of $\widetilde{\mathcal{H}}_{\text{b}}$ and $\widetilde{\mathcal{H}}_{\text{f}}(t)$, after noticing that $\overline{\widetilde{\mathcal{H}}_{\text{fluc}}(t)}=0$. This is because the spherical functions of Eq.~\Eqref{eq:spher_har} averages out to zero over an uniform distribution of $\theta(t)$ and $\phi(t)$. We assume that the fluctuating components of the Hamiltonian can be represented by a weakly stationary process such that after a change of variable $\tau = t-t_1$ :
\begin{gather}
 \int_{0}^{t}dt_1 [\overline{\widetilde{\mathcal{H}}_{\text{f}}(t), [\widetilde{\mathcal{H}}_{\text{f}}(t-\tau)},\widetilde{\rho}(t)]] 
= \int_{0}^{t}d\tau [\overline{\widetilde{\mathcal{H}}_{\text{f}}(0), [\widetilde{\mathcal{H}}_{\text{f}}(\tau)},\widetilde{\rho}(t)]]. \label{eq:stationary}
\end{gather}
We also assume the existence of a bath correlation time $\tau_c$, such that the ensemble average becomes negligible for $\tau \gg \tau_c$. This allows us the extend the integration limit from $\int_{0}^{t} d\tau$ to $\int_{0}^{\infty} d\tau$ :
\begin{gather}
- \int_{0}^{\infty}d\tau [\overline{\widetilde{\mathcal{H}}_{\text{f}}(t), [\widetilde{\mathcal{H}}_{\text{f}}(t-\tau)},\widetilde{\rho}(t)]] \nonumber \\
= -\frac{1}{2} \sum_{m=-2}^{2} J_{\text{DD}}(m\omega_0) [T_{2m}(\vec{I_1},\vec{I_2}),[T_{2m}^{\dagger}(\vec{I_1},\vec{I_2}),\widetilde{\rho}(t)]] \nonumber \\
-\frac{1}{2} \sum_{j,k=1}^{2}\sum_{m=-1}^{1} J_{\text{CSA}}(m\omega_0) [T_{2m}(\vec{B},\vec{I_j}),[T_{2m}^{\dagger}(\vec{B},\vec{I_k}),\widetilde{\rho}(t)]] \nonumber \\
- \frac{1}{2}\sum_{j=1}^{2}\sum_{m=-1}^{1} J_{\text{CC}}(m\omega_0) ([T_{2m}(\vec{I_1},\vec{I_2}),[T_{2m}^{\dagger}(\vec{B},\vec{I_j}),\widetilde{\rho}(t)]] \nonumber \\
+  ([T_{2m}(\vec{B},\vec{I_j}),[T_{2m}^{\dagger}(\vec{I_1},\vec{I_2}),\widetilde{\rho}(t)]]. \label{eq:redfield}
\end{gather}
Assuming an exponentially decaying correlation function \cite{cavanagh1996protein,GOLDMAN2001160}, the spectral density functions read
\begin{gather}
J_{\text{DD}}(m\omega_0) = b^2\int_{-\infty}^{\infty} \expo{-imt} \,\expo{-\tau/\tau_c} d\tau = \frac{6 b^2}{5} \frac{2\tau_c}{1+(m\omega_0 \tau_c)^2}, \nonumber \\
J_{\text{CSA}}(m\omega_0) =\frac{6 d^2}{5} \frac{2\tau_c}{1+(m\omega_0 \tau_c)^2},~\text{and} \nonumber \\
J_{\text{CC}}(m\omega_0) =-\frac{6 bd}{5} \frac{2\tau_c}{1+(m\omega_0 \tau_c)^2}. \label{eq:cl_spec}
\end{gather}
So far we have not taken into account the temperature of the lattice, and thus the equation in its current form does not predict the correct thermal steady state. We follow the process described in \cite{BENGS2020106645} and 
\begin{itemize}
\item add correction \cite{EGOROV1998469,PhysRevLett.4.239} to the spectral densities so that they obey the detailed balance \cite{PhysRevLett.120.180502,PhysRevE.98.052104} : $J(\omega)\rightarrow \mathcal{J}(\omega ) = \exp(-\beta \omega \hbar/2)J(\omega)$, where $\beta=1/k_B\,T$, and
\item replace all the double commutators of Eq.~\Eqref{eq:redfield} with respective Lindblad dissipator \cite{Annabestani2017,Karabanov18072014} as 
\begin{gather}
-\frac{1}{2}[A,[B,(\cdot)]] \rightarrow \mathcal{D}[A,B](\cdot) = A(\cdot)B - \frac{1}{2}(BA(\cdot) - (\cdot)BA). \nonumber
\end{gather}
\end{itemize}
After all these, the equation of motion for the evolution of the density matrix describing the full ensemble reads 
\begin{gather}
\frac{d\widetilde{\rho}(t)}{dt} = -\frac{i}{\hbar}[\mathcal{H}_{\text{B}},\widetilde{\rho}(t)] + (\Gamma_{\text{DD}}+\Gamma_{\text{CSA}}+\Gamma_{\text{CC}})\widetilde{\rho}(t); \label{eq:gskl} \\
\mbox{where},~\Gamma_{\text{DD}}(\cdot) = \sum_{m=-2}^{2} \mathcal{J}_{\text{DD}}(m\omega_0)\, \mathcal{D}[T_{2m}(\vec{I_1},\vec{I_2}),T_{2m}^{\dagger}(\vec{I_1},\vec{I_2})] (\cdot), \nonumber \\
\Gamma_{\text{CSA}}(\cdot) = \sum_{j,k=1}^{2}\sum_{m=-1}^{1} \mathcal{J}_{\text{CSA}}(m\omega_0)\,
 \mathcal{D}[T_{2m}(\vec{B},\vec{I_j}),T_{2m}^{\dagger}(\vec{B},\vec{I_k})] (\cdot), ~~\text{and} \nonumber \\
\Gamma_{\text{CC}}(\cdot) = \sum_{j=1}^2\sum_{m=-1}^{1} \mathcal{J}_{\text{CC}}(m\omega_0)\, (\mathcal{D}[T_{2m}(\vec{I_1},\vec{I_2}),T_{2m}^{\dagger}(\vec{B},\vec{I_j})] \nonumber \\ + 
\mathcal{D}[T_{2m}(\vec{B},\vec{I_j}),T_{2m}^{\dagger}(\vec{I_1},\vec{I_2})])(\cdot) \label{eq:LB}
\end{gather}
$\Gamma_{\text{DD}},\Gamma_{\text{CSA}}$ and $\Gamma_{\text{CC}}$ are called the dipolar, CSA and the cross correlation dissipator, respectively. Note that the equation is valid only for time $t$ satisfying Eq.~\ref{eq:COND} and the formal solution reads $\widetilde{\rho}(t) = \exp(\mathcal{L}t)\rho(0)$. As the generator $\mathcal{L}$, called the Liouvillian, is time independent, for a larger time the solution would read $\widetilde{\rho}(t) = \prod_{i=1}^{b}\exp(\mathcal{L}\Delta t)\rho(0)$, where $t=n\Delta t $ with each $\Delta t$ is such that Eq.~\ref{eq:COND} is satisfied. Since $\prod_{i=1}^{n}\exp(\mathcal{L}\Delta t) = \exp(\mathcal{L}\sum_{i=1}^{n}\Delta t) = \exp(\mathcal{L}t)$, the validity of Eq.~\Eqref{eq:gskl} extends well beyond condition \Eqref{eq:COND}, thanks to the assumptions that lead to the time independence of $\mathcal{L}$. 
 
\subsection{Measurement in NMR}
After the quantum circuit been run, the density operator of the spin system can be expanded in product-operator \cite{cavanagh1996protein} basis $\{I_{1x},I_{1y},I_{1z},...,2I_{1x}I_{2x},...\}$ as 
\begin{gather}
\rho = \sum_{n} \left(p_n\,I_{nz} + c_n\,I_{nx}  + d_n I_{ny}  \right) + ... \label{eq:prdop}
\end{gather}
The measurement process involves reading out all the coefficients $(p_n,c_n,d_n...)$ of the above expansion, which determines the density operator. The nuclear spin state of Eq.~\Eqref{eq:prdop} is let to evolve freely under the dominant Hamiltonian $\mathcal{H}_{\text{A}}=\sum_{n}\omega_0^n\,I_{nz}\hbar$. The transverse magnetization of the $n$'th nucleus, which is proportional to $I_{nx}$, rotates about $\hat{z}$ under $\mathcal{H}_{\text{A}}$ with angular frequency $\omega_0^n$, and induces a current in the receiver coil wrapped around the NMR tube with the same frequency. This high frequency signal is first split into two signals, and stored as real and imaginary parts \cite{cavanagh1996protein}: 
\begin{gather}
\cos{(\omega_0^nt + \phi)} \xrightarrow{\text{splitter}}\cos{(\omega_0^nt + \phi)} - i \cos{(\omega_0^nt + \phi)}, 
\end{gather}
where $\phi$ is a random phase error due to the arbitrariness in the switching time of the coil from transmitter mode to receiver mode at the start of signal acquisition. The real and imaginary signals are fed into mixers, where they get mixed with reference signals of frequency $\omega^n_{\text{ref}}$. The real and imaginary parts of the signal are mixed with reference signals that are $0^0$ and $90^0$ out of phase with them, respectively. After the mixing, they are passed through low-pass filters to obtain the real $\mbox{Re}(S_t)$ and the imaginary $\mbox{Im}(S_t)$ part of the acquired free induction decay (FID) : 
\begin{gather}
\xrightarrow{\text{mixer}} \cos(\omega_0^n t)\cos(\omega^n_{\text{ref}}\,t) - i\cos(\omega_0^n t)\sin(\omega^n_{\text{ref}}\,t) \nonumber \\
=\frac{1}{2}\cos[(\omega_0^n+\omega^n_{\text{ref}})t] + \frac{1}{2}\cos([\omega_0^n-\omega^n_{\text{ref}}]t) - \frac{i}{2}\sin[(\omega_0^n+\omega^n_{\text{ref}})t] \nonumber \\
+ \frac{i}{2}\sin[(\omega_0^n-\omega^n_{\text{ref}})t]
\nonumber \\
\downarrow {\text{low-pass filters}} \nonumber \\
\mbox{Re}(S_t) = \frac{1}{2}\cos[(\omega_0^n-\omega^n_{\text{ref}})t] \\
\mbox{Im}(S_t) = -\frac{1}{2}\sin[(\omega_0^n-\omega^n_{\text{ref}})t],
\end{gather}
with $\mbox{Re}(S_t)$ and $\mbox{Im}(S_t)$ being proportional to $c_n$ and $c_n^*$, respectively. They are Fourier transformed to obtain the NMR spectrum, and the area under their modulus-squared spectrum are measured to directly obtain the numbers $c_n$ and $c_n^*$. The art of NMR measurement lies in effectively converting other terms of the density operator into $I_x$ and $I_y$ as their coefficients can be directly measured in the above way and thereby to obtain the full density operator of the nuclear spin system.

\chapter{Enhanced non-macrorealism: Extreme violations of Leggett-Garg inequalities for a system evolving under superposition of unitaries}\markboth{Chapter 2:  LGI under Superposition of Unitaries}{}\label{LGI}

\begin{center}
\textit{`` Einstein was consistent, clear, down to earth and wrong. \\ Bohr was inconsistent, unclear, wilfully obscure and right.''} \\
~~~~~~~~~~~~~~~~~~-- John Bell ($1928-1990$)
\end{center}

\section{Introduction} \label{2sec:intro}
The twilight zone between classical and quantum worlds \cite{RevModPhys.75.715,zurek1993preferred,zurek2014quantum} has been under perpetual investigation since the dawn of quantum mechanics \cite{PhysRev.47.777, PhysicsPhysiqueFizika.1.195,feynman2000theory}. Quantum physics radically differs from its classical counterpart primarily due to the postulates of state description and measurement. For single systems, a possible construction of a classical worldview adheres to the notion of `macrorealism'  \cite{PhysRevLett.54.857,Leggett_2008,zhang2023comparing} which says that a system, capable of being in two or more distinct states, will always be in any one of them, and it is possible to determine this state with arbitrarily small perturbation on the subsequent dynamics. In \cite{PhysRevLett.54.857}, Leggett and Garg considered measuring two-point correlations $C_{ij}=\langle q(t_i)q(t_j) \rangle$ in a system, where $q(t_k)$ is the outcome of a dichotomous ($\pm 1$ valued) observable $\hat{Q}$ at time $t_k$, and $\langle \cdot \rangle$ means an average over multiple runs of the experiment. They found, for any choice of times $t_1,t_2$ and $t_3$,  the assumptions of macrorealism restrict the value of a particular linear combination of two-point correlators $K_3 = C_{12} + C_{23} - C_{13}$ as $-3 \leq K_3 \leq 1$, which is known as Leggett-Garg Inequality (LGI). However, quantum systems can defy macrorealism by surpassing the upper bound (classical) of LGI. Thus, violation of LGI acts as a certification of non-classicality  and it has been observed experimentally in a variety of quantum systems \cite{palacios2010experimental,PhysRevLett.107.130402,PhysRevA.87.052102,PhysRevLett.107.090401}.       

Early experimental \cite{palacios2010experimental,PhysRevLett.107.130402} and theoretical studies \cite{PhysRevLett.111.020403, fritz2010quantum} suggest that for a qubit, evolving under unitary dynamics, the maximum violation of LGI is constrained by the temporal Tsirelson bound (TTB), which reads $\left( K_3\right)_{\rm m} \leq 1.5$, where $(K_3)_{\rm m}$ denotes the maximum value of $K_3$. This shows that even quantum theory restricts $\left( K_3\right)_{\rm m}$ way below its algebraic maximum value $3$. However, recently, it was shown that it is possible to observe the violation of LGI beyond the TTB either by replacing the qubit with a multi-level system \cite{PhysRevLett.113.050401, wang2017enhanced, katiyar2017experimental} or by replacing unitary evolution with a non-Hermitian $\mathcal{PT}$ symmetric evolution on a qubit \cite{PhysRevA.103.032420, PhysRevA.108.032202, PhysRevA.109.042205,wu2023maximizing}.

The superposition of states allows a quantum system to exhibit non-macrorealistic behavior by violating LGI up to TTB. This provokes us to consider superposition of unitary time evolutions and ask what happens to the violation of LGI for a qubit undergoing such a transformation. We construct unitary transformations by doing superposition of time evolution unitaries. Subjecting a qubit under such superposition of unitaries, we find clear violation of LGI beyond the TTB, growing with the strength of superposition. We obtain a way of physically realizing these superposed unitaries using an ancillary system and experimentally demonstrate the enhanced non-macrorealism using nuclear magnetic resonance (NMR) techniques. As an application, we report robustness in the behavior of a qubit when it is subjected to such a transformation. At longer times, due to the presence of environmental noise, qubits get decohered \cite{hornberger2009introduction,joos2013decoherence,RevModPhys.75.715,zurek1993preferred} and the LGI violation disappears, indicating the emergence of classicality \cite{PhysRevLett.107.090401, PhysRevLett.107.130402}. However, our superposed unitaries provide robustness against such environmental noise by remarkably increasing the time to which LGI violation persists.           
 
\section{Superpositions of Unitary Operators}\label{2sec:supU}
\begin{figure}
\centering
\includegraphics[width=10cm, clip = true, trim={5.7cm 4.55cm 5.8cm 2cm}]{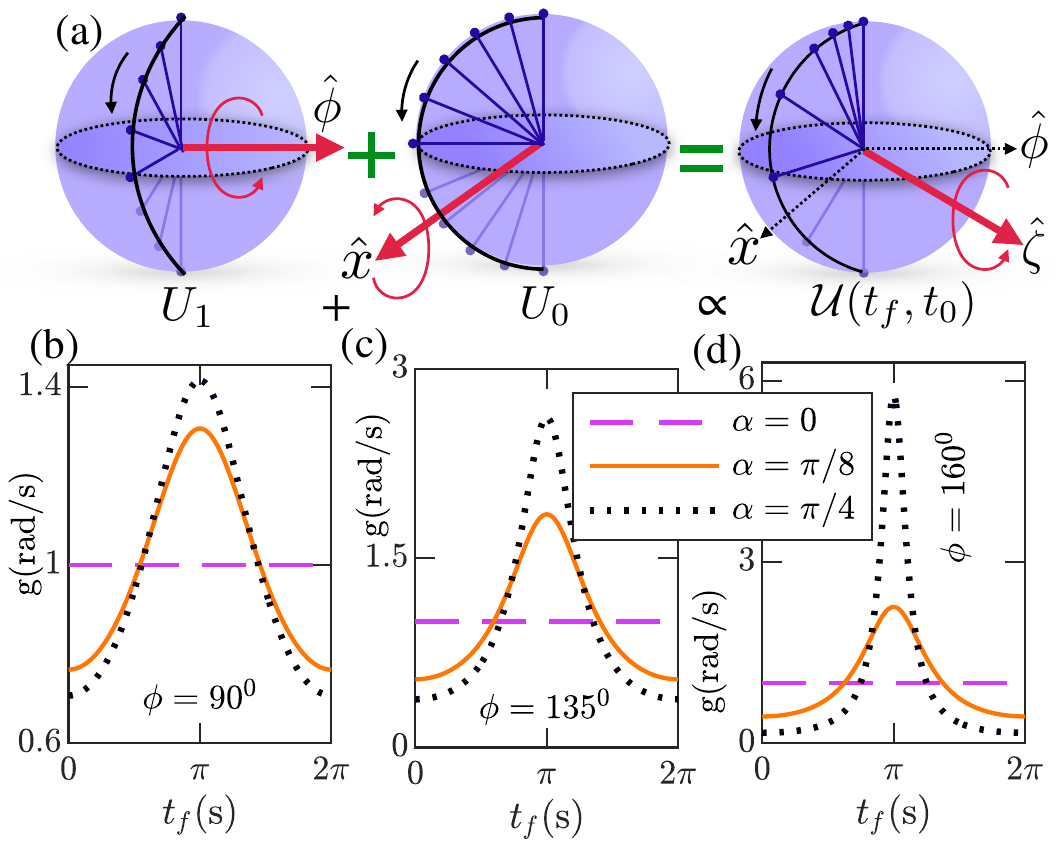}
\caption{(a) Superposition of two unitaries $U_0$ and $U_1$ (for $\phi=\pi/2$)
on the Bloch sphere shifts the rotation axis (thick arrow) and introduces nonlinearities in SOE (modeled by the gap in the spokes). (b-d) SOE for the initial state $\Ket{0}$ against final time $t_f$  for different values of $\alpha$ and $\phi$ at $\omega=1$.  }
\label{fig:graph_abst}
\end{figure}

Consider two unitary transformations $U_0=\exp(-\im \sigma_{x}\omega[t_f-t_0]/2)$ and $U_1=\exp(-\im \sigma_{\phi}\omega[t_f-t_0]/2)$, which transforms the state of a qubit from time $t_0$ to $t_f$. $\sigma_k$ denotes the Pauli matrix along $\hat{k}$, and $\hat{\phi}$ is an axis making an angle $\phi$ with $\hat{x}$ (see Fig.~\ref{fig:graph_abst}~(a)). $U_0$ and $U_1$ describe rotations of the Bloch vector about $\hat{x}$ and $\hat{\phi}$, respectively. We construct an effective unitary transformation by superposing them as
\begin{gather}
\mathcal{U}(t_f,t_0) \coloneq \left( \cos \alpha \, U_0 + \sin \alpha \, U_1 \right)/ \mathcal{N}(t_f,t_0),  
\label{eq:sup} \\
\text{and}~~ \mathcal{N}^{2}(t_f,t_0):=\text{tr}\left[\widetilde{\mathcal{U}}^{\dagger}_{+}(t_f,t_0) \, \widetilde{\mathcal{U}}_{+}(t_f,t_0) \right]/2, \label{eq:norm_def}
\end{gather}
with $\widetilde{\mathcal{U}}_{\pm}(t_f,t_0)=\cos \alpha \, U_0 \pm \sin \alpha \, U_1$, where $\alpha$ is the superposition parameter with no superposition for $\alpha = 0,\pi/2$ and maximum superposition at $\alpha=\pi/4$. On the other hand, $\phi$ is the angle between the rotation axis of $U_0$ and $U_1$, thereby indicating how `distinct' the unitaries that are being superposed.   The Superposed unitary is well defined as long as $\mathcal{N}^2(t_f,t_0)>0$, and we restrict $\alpha \in [0,\pi/2]$ and $\phi \in [0,\pi)$ to ensure the same for all $t_0$ and $t_f$. A general method for constructing such effective unitary transformation beyond qubits is presented in Sec.~\ref{2sec:supUgen}. 

To understand the action of $\mathcal{U}(t_f,t_0)$ on a random state $\Ket{\psi_0}$, we fix $t_0$ and continuously increase $t_f$ so that $\Ket{\psi_{t_f}}=\mathcal{U}(t_f,t_0)\Ket{\psi_0}$ traces a continuous trajectory $(\theta(t_f),\Phi(t_f))$ on the surface of the Bloch sphere. Here $\theta(t_f)$ and $\Phi(t_f)$ are the co-latitude and longitude of the Bloch vector at time $t_f$. We define the speed of evolution (SOE) of the Bloch vector as $\vec{g}(t_f) \coloneq \partial_{t_f}(\theta(t_f),\Phi(t_f))$. For simplicity, we choose $\Ket{\psi_0}=\Ket{0}$, the eigenstate of $\sigma_z$ with the eigenvalue $1$, which gets transformed as $\Ket{\psi_{t_f}} = \cos[f(t_f)/2]\Ket{0}-\im \sin[f(t_f)/2]\exp(\im\zeta)\,\Ket{1}$, where $f(t)$ and $\zeta$ depend on $\alpha$ and $\phi$. The SOE is now reduced to just one component $g(t_f)=\partial_{t_f} f(t_f)$, which is a nonlinear function of $t_f$. Therefore, the superposed unitary $\mathcal{U}(t_f,t_0)$ effectively describes a rotation of the Bloch vector with nonlinear SOE, with a shift in the axis of rotation $\hat{\zeta}$ (see Fig.~\ref{fig:graph_abst}~(a)). As shown in Fig.~\ref{fig:graph_abst}~(b-d)), the nonlinearity of SOE increases with increasing superposition $\alpha$. Also, at a given $\alpha$, the SOE gets more nonlinear as we make $U_0$ and $U_1$ increasingly distinct by raising $\phi$.  

To see this explicitly, We take two arbitrary directions $\hat{n} = n_x \hat{x} + n_y \hat{y} + n_z \hat{z}$ and $\hat{m} = m_x \hat{x} + m_y \hat{y} + m_z \hat{z}$ in the Bloch sphere and construct a superposition between $U_0 (t_f, t_0) = \exp(-\im \hat{\sigma}_n \omega \delta /2)$ and $U_1 (t_f, t_0) = \exp(-\im \hat{\sigma}_m \omega \delta /2)$ according to Eq.~\Eqref{eq:sup} to form $\mathcal{U}(t_f, t_0)$, where $\delta = t_f - t_0$. The un-normalized operator reads 
\begin{gather}
\widetilde{\mathcal{U}}_{+}(t_f,t_0) = \cos(\alpha) U_0(t_f,t_0) + \sin(\alpha) U_1(t_f,t_0) \nonumber \\
= \left( \cos(\alpha) + \sin(\alpha) \right) \cos \frac{\omega \delta}{2} \mathbbm{1} 
- \im \left( \cos(\alpha) \hat{\sigma}_n + \sin(\alpha) \hat{\sigma}_m \right) \sin \frac{\omega \delta}{2}.
\end{gather}
Using this, we find
\begin{gather}
\widetilde{\mathcal{U}}_{+}(t_f,t_0) \widetilde{\mathcal{U}}_{+}^{\dagger}(t_f,t_0) 
= 
\underbrace{ \left[ 1 + \sin(2 \alpha) \left( \cos^{2} \frac{\omega \delta}{2} + \hat{n} \cdot \hat{m} \, \sin^{2} \frac{\omega \delta}{2} \right)\right]}_\text{$\mathcal{N}^{2}(t_f,t_0)$}   \mathbbm{1}, 
\end{gather}
which ensures that the superposed operator $\mathcal{U}(t_f,t_0) =\widetilde{\mathcal{U}}(t_f,t_0) / \mathcal{N}(t_f,t_0) $ is always an unitary operator for arbitrary values of $t_f$ and $t_0$ and for arbitrary choices of directions $\hat{n},~\hat{m}$, as long as $-1 < \hat{n} \cdot \hat{m} \leq 1$, which is required to ensure the positivity of $\mathcal{N}^{2}(t_f,t_0)$ for all values of $t_0,~t_f$ and $\alpha \in [0,\pi/4]$ (see Sec.~\ref{2sec:supUgen}).

Without the loss of generality, we consider $\hat{n}:=\hat{x}$ and $\hat{m}:=\hat{\phi}$, where $\hat{\phi}$ is a unit vector making an angle $\phi \in [0,\pi)$ with $\hat{x}$. Taking $t_0=0$ and $t_f=t$, the superposed unitary can be expanded as 
\begin{gather}
\mathcal{U}(t,0) = \frac{\begin{bmatrix}
\left(\cos \alpha + \sin \alpha \right) \cos \frac{\omega t}{2} & -\im \left(\cos \alpha + \expo{-\im \phi} \sin \alpha \right) \sin \frac{\omega t}{2}  \\
-\im \left(\cos \alpha + \expo{\im \phi} \sin \alpha \right) \sin \frac{\omega t}{2} &
\left( \cos \alpha + \sin \alpha \right) \cos \frac{\omega t}{2}
\end{bmatrix}}{\sqrt{1+\sin(2\alpha)[\cos^{2}\frac{\omega t}{2} + \cos\phi \sin^{2}\frac{\omega t}{2}]}}   \nonumber \\
= \underbrace{\frac{\left(\cos \alpha + \sin \alpha \right) \cos \frac{\omega t}{2}}{\sqrt{1+\sin(2\alpha)[\cos^{2}\frac{\omega t}{2} + \cos\phi \sin^{2}\frac{\omega t}{2}]}}}_\text{$\cos [f( t)/2]$} \mathbbm{1} 
- \im  \underbrace{\frac{\sqrt{1 + \cos \phi \sin (2\alpha)}\sin \frac{\omega t}{2}}{\sqrt{1+\sin(2\alpha)[\cos^{2}\frac{\omega t}{2} + \cos\phi \sin^{2}\frac{\omega t}{2}]}}}_\text{$\sin[f( t)/2]$} \nonumber \\
\left( \underbrace{\frac{\cos \alpha + \cos \phi \sin \alpha}{\sqrt{1 + \cos \phi \sin (2\alpha)}}}_\text{$\cos \zeta$} \hat{\sigma}_x + \underbrace{\frac{\sin \alpha \sin \phi}{\sqrt{1 + \cos \phi \sin (2\alpha)}}}_\text{$\sin \zeta$} \hat{\sigma}_y \right) \nonumber \\
= \cos \left[\frac{f(t)}{2}\right] \mathbbm{1} - \im \sin \left[\frac{f(t)}{2}\right] \hat{\sigma}_{\zeta},
\label{eq:sup_detailed}
\end{gather}
where $\hat{\sigma}_{\zeta}:=\cos \zeta \hat{\sigma}_x + \sin \zeta \hat{\sigma}_y$. To define the speed of evolution (SOE) $g(t)$, we take the initial state $\Ket{\psi_0}:=\Ket{0}$ and let it transform under the superposed unitary as 
\begin{gather}
\Ket{\psi_t} = \mathcal{U}(t,0)\Ket{0} = \cos \left[\frac{f(t)}{2}\right] \Ket{0} -\im \sin \left[\frac{f(t)}{2}\right] \expo{\im \zeta} \Ket{1} \nonumber \\
= \cos \left[\frac{f(t)}{2}\right] \Ket{0} + \sin \left[\frac{f(t)}{2}\right] \expo{\im (\zeta - \pi/2)} \Ket{1}. \label{eq:rot_soe}
\end{gather}
Therefore, if we increase $t$ continuously from $0$, Eq.~\Eqref{eq:rot_soe} describes counter-clockwise rotation of the Bloch vector $\hat{z}$ about axis $\hat{\zeta}:=\cos \zeta \hat{x} + \sin \zeta \hat{y}$ in the Bloch Sphere. SOE $(g(t))$ can now be defined as the rate of change of the co-latitude (which is the only degree of freedom that is changing here) of the Bloch vector with $t$ as
\begin{align}
g(t) =  \partial_t f(t) =  \frac{-2\partial_t \cos [f(t)/2]}{\sin[f(t)/2]}.
\end{align}
Computing the derivatives,  we get
\begin{gather}
\partial_t \cos [f(t)]  = - \frac{\omega \left(1 + \sin (2\alpha) \cos \phi \right) \left( \cos \alpha + \sin \alpha \right) \sin \frac{\omega t}{2} }{2 \left[1+\sin(2\alpha)(\cos^{2} \frac{\omega t}{2} + \cos \phi \sin^{2} \frac{\omega t}{2}) \right]^{3/2}},
\end{gather}
which directly gives a closed form expression for the SOE as
\begin{gather}
g(t) = \frac{\omega \sqrt{1+\cos \phi \sin (2\alpha)} (\cos \alpha + \sin \alpha)}{1+\sin(2\alpha)(\cos^{2} \frac{\omega t}{2} + \cos \phi \sin^{2} \frac{\omega t}{2}) }. \label{apeq:soe}
\end{gather}
Of course, SOE can be defined more generally by taking a generic initial state as $\Ket{\psi_0}$ instead of $\Ket{0}$. In that case, both the co-latitude and longitude of the Bloch vector will change with $t$, and therefore $g(t)$ will read as the magnitude of the vector sum of these two rate of changes. However, the special case presented here suffices the requirements of this thesis.

Note that as $\mathcal{U}(t_f,t_0)$ depends non-linearly on $(t_f-t_0)$, the product law does not follow : $\mathcal{U}(t_3,t_1)\neq \mathcal{U}(t_3,t_2)\, \mathcal{U}(t_2,t_1)$. Thus, it does not describe closed system evolution. Therefore, to realize $\mathcal{U}(t_f,t_0)$, we consider the conditional evolution of the system qubit (S), depending on the state of an ancillary qubit (A) \cite{PhysRevA.85.022109,PhysRevLett.121.090503}. Under an interaction unitary $\mathcal{V}_{\text{AS}} = \proj{0} \otimes U_0 + \proj{1} \otimes U_1$,  S evolves under $U_0$ or $U_1$, conditioned on whether A is in state $\Ket{0}$ or $\Ket{1}$, respectively. Therefore, to realize a superposition between $U_0$ and $U_1$, we initialize A in the state $\Ket{\alpha}_{\text{A}}=\cos \alpha \Ket{0} + \sin \alpha \Ket{1}$, let it interact with S from $t_0$ to $t_f$ under $\mathcal{V}_{\rm{AS}}$, and finally post-select A in state $\Ket{+}=\left(\Ket{0}+ \Ket{1} \right)/\sqrt{2}$ (see Fig. \ref{fig:Th_Plots}~(a)), such that S gets transformed under the $\mathcal{U}(t_f,t_0)$.

As can be seen from the circuit presented in Fig. \ref{fig:Th_Plots} (a), the superposition between the unitaries an be realized using an ancillary qubit(A), initiated in an coherent state $\Ket{\alpha}_{\text{A}} = \cos \alpha \Ket{0} + \sin \alpha \Ket{1}$. For any initial state of the system $\rho_\text{S}$, the subsequent evolution under the quantum circuit shown traces as
\begin{gather}
\proj{\alpha}_{\text{A}}  \otimes  \rho_{\text{S}} \nonumber \\
\downarrow ~ \proj{0}_{\text{A}} \otimes \underbrace{\expo{-\im \omega \sigma_{x}t/2}}_{U_0} + \proj{1}_{\text{A}} \otimes \underbrace{\expo{-\im \omega \sigma_{\phi}t/2}}_{U_1} \nonumber \\
\cos^{2} \alpha \,\proj{0}_{\text{A}} \otimes U_0 \rho_{\text{S}} U^{\dagger}_0  + \sin^{2}  \alpha \, \proj{0}_{\text{A}} \otimes U_1 \rho_{\text{S}} U^{\dagger}_1 \nonumber \\
+ \cos \alpha \sin \alpha \left( \Ket{0}\Bra{1}_{\text{A}} \otimes U_0 \rho_{\text{S}} U^{\dagger}_1 + \Ket{1}\Bra{0}_{\text{A}} \otimes U_1 \rho_{\text{S}} U^{\dagger}_0  \right) \nonumber \\
\downarrow ~ \text{(post-selecting A in $\Ket{+}$ state)} \nonumber \\
\frac{\cos^{2} \alpha \, U_0 \rho_{\text{S}} U^{\dagger}_0 + \sin^{2} \alpha \, U_1 \rho_{\text{S}} U^{\dagger}_1 + \cos \alpha \, \sin \alpha \left(U_0 \rho_{\text{S}} U^{\dagger}_1 + U_1 \rho_{\text{S}} U^{\dagger}_0 \right)}{\text{tr}\, \left[ \cos^{2} \alpha \, U_0 \rho_{\text{S}} U^{\dagger}_0 + \sin^{2} \alpha \, U_1 \rho_{\text{S}} U^{\dagger}_1 + \cos \alpha \, \sin \alpha \left(U_0 \rho_{\text{S}} U^{\dagger}_1 + U_1 \rho_{\text{S}} U^{\dagger}_0 \right)\right]}
\label{sqe:halfway}
\end{gather}
Now, we define the operator $\widetilde{\mathcal{U}}_{+}:= \cos \alpha \, U_0 + \sin \alpha \, U_1$, and utilizing the property that $\widetilde{\mathcal{U}}^{\dagger}\, \widetilde{\mathcal{U}} \propto \mathbbm{1}$ (see Sec. \ref{2sec:supUgen} for a discussion on when this is possible in general) with proportionality constant $\mathcal{N}^{2}$, \Eqref{sqe:halfway} can be rewritten as 
\begin{gather}
\frac{\widetilde{\mathcal{U}}_{+}\,\rho_{\text{S}}\,\widetilde{\mathcal{U}_{+}^{\dagger}}}{\mathcal{N}^{2}} = \mathcal{U} \, \rho_{\text{S}} \, \mathcal{U}^{\dagger},    
\end{gather}
where $\mathcal{U}:=\widetilde{\mathcal{U}}_{+}/\mathcal{N}$ is the desired superposed unitary operator.

The circuit of Fig. \ref{fig:Th_Plots} (a) thus serves as a physical realization of the superposed unitary for any quantum simulator. Linear combination of unitaries like $\mathcal{U}(t_f,t_0)$ (generally non-unitary)  has also been explored in other contexts  \cite{LongGui-Lu_2006,LongGui-Lu_2008,LongGui-Lu_2009,Gudder2007,Zhang2010,LongGui-Lu_2008,LongGui-Lu_2009,Gudder2007}.

\section{LGI for a qubit under  superposed unitary $\mathcal{U}$}\label{2sec:lgi}
\begin{figure}
\centering
\includegraphics[width=10cm, clip = true, trim={6.2cm 9.7cm 6.2cm 0cm}]{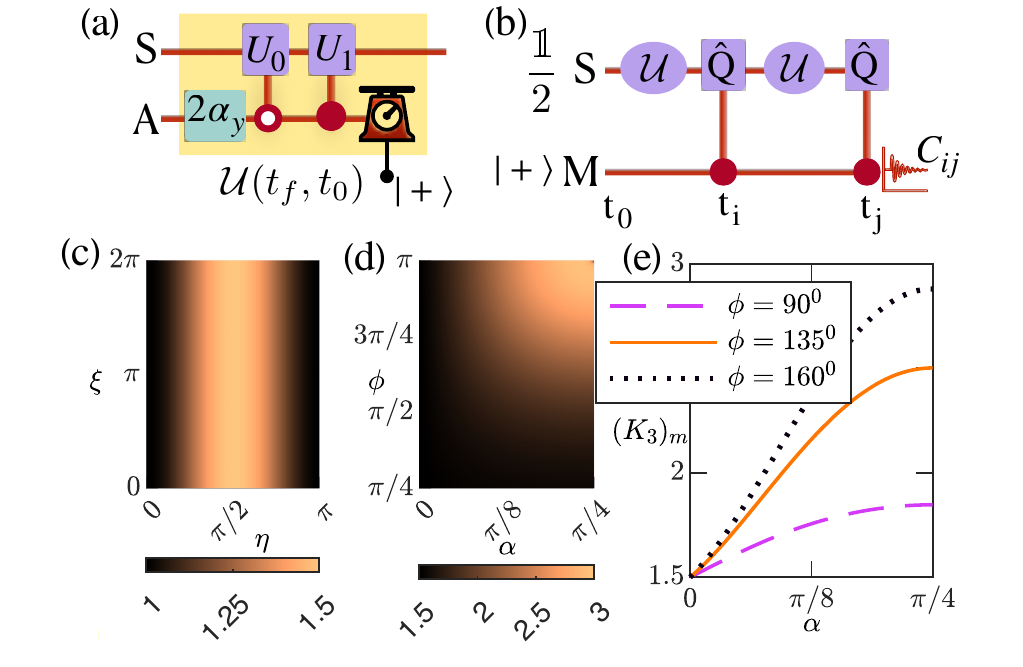}
\caption{(a) A quantum circuit for realizing the superposed unitary using an ancillary qubit (A), which, after interacting with S, gets post-selected in state $\Ket{+}$.  (b) The interferometric circuit to determine $C_{ij}$ corresponding to observable $\hat{Q}$ using an additional qubit M. (c) $(K_3)_{\rm m}$ against the longitude ($\xi$) and the co-latitude ($\eta$) of $\hat{k}$, confirming TTB under simple unitaries. (d) $(K_3)_{\rm m}$ vs $\alpha$ and $\phi$ showing violation of TTB under superposed unitaries.  (e) $(K_3)_{\text{m}}$ vs $\alpha$ for different  $\phi$ values showing violation of TTB explicitly. }
\label{fig:Th_Plots}
\end{figure}

Choosing the initial state  $\rho_{t_0}=\mathbbm{1}/2$, $C_{ij}$ for dichotomous observable $\hat{Q}$ can be measured using the interferometric circuit \cite{PhysRevLett.104.160501} shown in Fig.~\ref{fig:Th_Plots}~(b). Here the correlation between measurement outcomes of $\hat{Q}$ is stored in the phase of an additional qubit (M) initiated in state $\Ket{+}_{\rm{M}}$. The measurement of M's coherence $\mathcal{C}:=[ \langle \hat{\sigma}_{x} \rangle + \im \langle \hat{\sigma}_{y} \rangle ]/2$ directly gives $C_{ij}$ as
\begin{equation}
\mathrm{Re} \, [\mathcal{C}] = \frac{1}{2}\mathrm{tr} \left[ \hat{Q}\,\mathcal{U}(t_j,t_i) \hat{Q} \,\mathcal{U}^{\dagger}(t_j,t_i) \right] = C_{ij}.
\label{eq:circ_corr}
\end{equation} 

We set $t_0 = t_1=0,~t_2=t$ and $t_3=2t$ to determine $C_{12}(t)$, $C_{23}(t)$ and $C_{13}(t)$ for $\hat{Q}=\hat{\sigma}_z$.
First, we let S (with initial state $\mathbbm{1}/2$) to evolve under the simple unitary (no superposition) $U=\exp(-\im \omega \sigma_k t)$, and note the maximum of $K_3(t)$ over $t \in [0, 2\pi]/\omega$ as $(K_3)_{\text{m}}$, which is plotted for each possible $\hat{k}$ in Fig.~\ref{fig:Th_Plots}~(c) confirming the existence of TTB as $(K_3)_{\text{m}} \leq 1.5$. Next, we subject S (with the same initial state) under the superposed unitary
and again plot $(K_3)_{\text{m}}$ as a function of $\alpha$ and $\phi$ in Fig.~\ref{fig:Th_Plots}~(d), which shows increase of $(K_{3})_{\rm m}$ beyond TTB with increasing $\alpha$. This enhanced violation beyond TTB with rising $\alpha$ is plotted explicitly in Fig.~\ref{fig:Th_Plots}~(e) for $\phi = 90^o, 135^o$ and $160^o$. These results clearly show that for any given $\phi$, the superposition of unitary produces enhanced violation of LGI which increases with increasing superposition. Moreover, this violation increases with $\phi$ for a fixed $\alpha$ and $K_3$ approaches its algebraic maximum  $(K_3)_{\text{m}}=3$ at $\alpha = \pi/4$ as $\phi \rightarrow \pi$. 

\section{Experiments} \label{2sec:exp}
\begin{figure}
\centering
\includegraphics[width=10cm, clip=true, trim={7.48cm 2.25cm 9cm 2.2cm}]{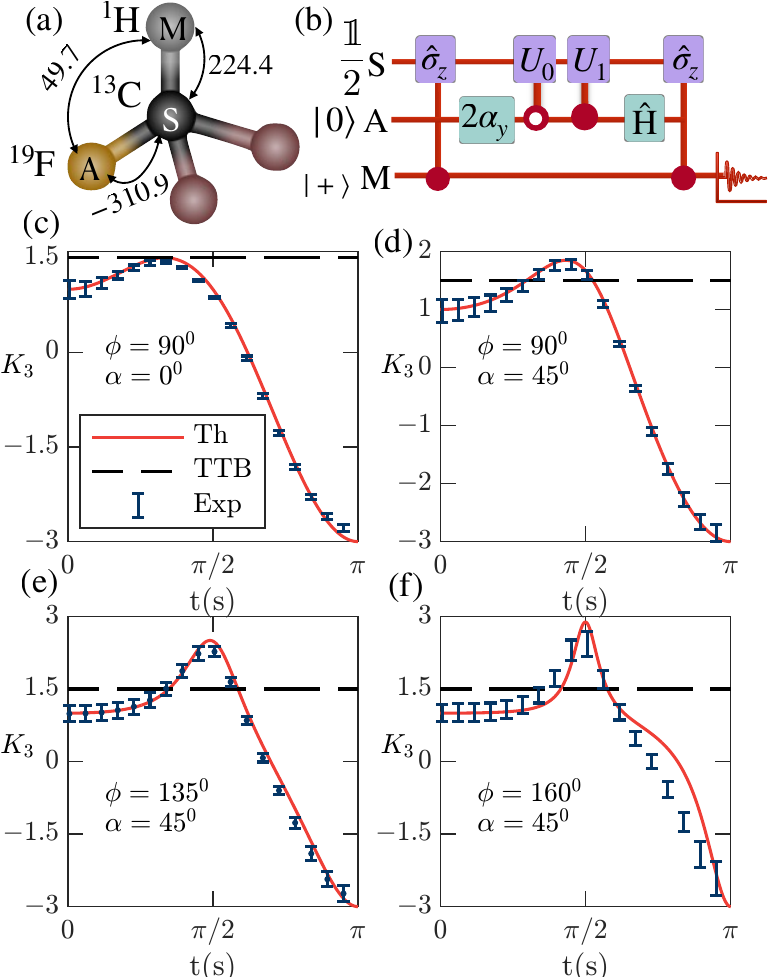}
\caption{(a) The molecular structure of DBFM with respective $J_{ij}$ values in Hz. (b) Quantum circuit for determining $C_{ij}$ ( $\hat{\rm{H}}$ is the Hadamard gate). (c-f) Experimentally measured values of $K_3$ vs theoretical curves (solid curves) for different values of $\alpha$ and $\phi$. Error bars account for systematic (rf inhomogeneity) and random errors (thermal noise). } 
\label{fig:circuits}
\end{figure}

We choose the three spin-$1/2$ nuclei of 13C-dibromofluoromethane (DBFM) molecule as our three qubit quantum register (see Fig.~\ref{fig:circuits}~(a)), identifying $^{13}$C  as S, $^{19}$F  as A, and $^{1}$H  as M. In a strong magnetic field of $11.7$ T (along $\hat{z}$), the liquid ensemble of DBFM, dissolved in Acetone-D6, rests in thermal equilibrium at an ambient temperature of $300$ K. Under high temperature-high field assumption  \cite{cavanagh1996protein}, the density matrix of the quantum register reads $\rho_{\mathrm{th}} = \mathbbm{1}/8 + \epsilon (\gamma_{\rm{C}} I_z^{\mathrm{S}} + \gamma_{\rm{F}} I_z^{\mathrm{A}} + \gamma_{\rm{H}} I_z^{\mathrm{M}})$, where $\gamma_i$ is proportional to the gyro-magnetic ratio of the $i$'th nucleus, $I_{z}^{v}:= \hat{\sigma}^{v}_z /2$, and the purity factor $\epsilon \approx 10^{-5}$.  In a triply-rotating frame, rotating at the resonant frequency of each nucleus, the Hamiltonian of the quantum register reads \cite{cavanagh1996protein,levitt2008spin}
$\mathcal{H}_{\mathrm{NMR}} = 2\pi \left( J_{\mathrm{SA}}I_z^{\mathrm{S}} I_z^{\mathrm{A}} +  J_{\mathrm{SM}} I_z^{\mathrm{S}} I_z^{\mathrm{M}} +  J_{\mathrm{AM}} I_z^{\mathrm{A}} I_z^{\mathrm{M}} \right)$.

We first initialize the quantum register into the pseudo-pure state  $\rho_{in} = (\mathbbm{1}/2)_{\mathrm{S}} \otimes \proj{0\,+}_{\mathrm{AM}}$ \cite{cory1997ensemble,gershenfeld1997bulk,cory1998nuclear}, after which, the quantum circuit of Fig.~\ref{fig:circuits}~(b) is used to perform the experiments, where we combine the circuit of Fig.~\ref{fig:Th_Plots}~(a) with the interferometric circuit of Fig.~\ref{fig:Th_Plots}~(b). Measuring real part of the coherence $\mathcal{C}_{\rm{M}}$ at the end of the circuit gives   
\begin{gather}
\rm{Re}\,[\mathcal{C}_{\rm{M}}] = \frac{1}{4} (T_{+}(t_j,t_i) + T_{-}(t_j,t_i)) \label{eqS}
\\
\mathrm{where,}~ T_{\pm}(t_j,t_i)  = \mathrm{tr} \left[ \hat{\sigma}_z \,\widetilde{\mathcal{U}}_{\pm}(t_j,t_i)\, \hat{\sigma}_z \frac{\mathbbm{1}}{2}\, \widetilde{\mathcal{U}}_{\pm}^{\dagger}(t_j,t_i)\right]. \nonumber
\end{gather}

To derive the Eq.~\Eqref{eqS}, consider the evolution of the quantum register under the quantum circuit shown in Fig \ref{fig:circuits}~(b):
\begin{gather}
\frac{1}{2} \left(\proj{0}+\proj{1}+\outpr{0}{1}+\outpr{1}{0}\right)_{\text{M}} \nonumber \\
\otimes  (\cos^{2}(\alpha)\proj{0}+\sin^{2}(\alpha)\proj{1}+\cos(\alpha)\sin(\alpha)(\outpr{0}{1}+ 
\outpr{1}{0}))_{\text{A}} \otimes \left(\frac{\mathbbm{1}}{2}\right)_{\text{S}} \nonumber \\ 
\bigg\downarrow \left(\proj{0}_{\text{M}} \otimes \mathbbm{1}_{\text{S}} + \proj{1}_{\text{M}} \otimes(\sigma_{z})_{\text{S}} \right) \nonumber \\
\frac{1}{2} (\proj{0}_{\text{M}} \otimes [\frac{\mathbbm{1}}{2}]_{\text{S}} + \proj{1}_{\text{M}} \otimes [\frac{\mathbbm{1}}{2}]_{\text{S}} 
(\Ket{0}\Bra{1}_{\text{M}} + \Ket{1}\Bra{0}_{\text{M}}) \otimes [\frac{\sigma_z}{2}]_{\text{S}}) \otimes (.)_{\text{A}} 
\nonumber \\   
\bigg\downarrow  \left(U_{0} \otimes \proj{0} +  U_{1} \otimes \proj{1}\right)_{\text{AS}} \nonumber \\
\frac{1}{2}( .... + \Ket{1}\Bra{0}_{\text{M}} \otimes (\cos^{2}(\alpha) \proj{0}_{\text{A}} \otimes [U_{0} \frac{\sigma_z}{2} U^{\dagger}_{0}]_{\text{S}} 
+ \sin^{2}(\alpha) \proj{1}_{\text{A}} \otimes [U_{1} \frac{\sigma_z}{2} U^{\dagger}_{1}]_{\text{S}} \nonumber \\
+ \cos(\alpha) \sin(\alpha) 
\outpr{0}{1}_{\text{A}} \otimes [U_{0} \frac{\sigma_z}{2} U^{\dagger}_{1}]_{\text{S}} + \outpr{1}{0}_{\text{A}} \otimes [U_{1} \frac{\sigma_z}{2} U^{\dagger}_{0}]_{\text{S}}) \nonumber \\
\bigg\downarrow \left(\proj{0}_{\text{M}} \otimes \mathbbm{1}_{\text{S}} + \proj{1}_{\text{M}} \otimes(\sigma_{z})_{\text{S}} \right) \nonumber \\
\frac{1}{2}( .... + \Ket{1}\Bra{0}_{\text{M}} \otimes (\cos^{2}(\alpha) \proj{0}_{\text{A}} \otimes [\sigma_{z}U_{0} \frac{\sigma_z}{2} U^{\dagger}_{0}]_{\text{S}} 
+ \sin^{2}(\alpha) \proj{1}_{\text{A}} \nonumber \\
\otimes [\sigma_{z}U_{1} \frac{\sigma_z}{2} U^{\dagger}_{1}]_{\text{S}}  + \cos(\alpha) \sin(\alpha) 
\outpr{0}{1}_{\text{A}} \otimes [\sigma_{z}U_{0} \frac{\sigma_z}{2} U^{\dagger}_{1}]_{\text{S}} + \outpr{1}{0}_{\text{A}} \otimes [\sigma_{z}U_{1} \frac{\sigma_z}{2} U^{\dagger}_{0}]_{\text{S}}) \nonumber \\
\bigg\downarrow \hat{\text{H}}_{\text{A}} \rightarrow \text{tr}_{\text{AS}}[.] \nonumber \\
\rho_{\text{M}} = \frac{1}{2}\proj{+} (T_{+} + T_{-}), \label{aeq:T}
\end{gather}
which results the (real) NMR signal $S = \mathrm{tr}[\sigma_{x}\rho_{M}] = \frac{1}{4}(T_{+} +T_{-})$ when M is measured. 

Finally, note that during NMR signal acquisition, the final three qubit state effectively evolves under $\mathcal{H}_{\rm{NMR}}$ of the main text. This causes the $T_{+}$ and $T_{-}$ of Eq.~\Eqref{aeq:T} to evolve at different frequencies. Therefore by taking a Fourier transform of the time domain signal we go to the frequency domain where we find $T_{+}$ and $T_{-}$ as separate spectral lines. This allows us to post-select the $T_{\pm}$ terms, which is equivalent of post selecting A in $\{\Ket{+},\Ket{-}\}$ basis. Just switching off two controlled $\hat{\sigma}_z$ gates, an identical run of the quantum circuit of fig.~\ref{fig:circuits}~(b) gives the measurement outcome $\rm{Re}\,[\mathcal{C}_{\mathrm{M}}] = (\mathcal{N}^{2}(t_j,t_i) + \widetilde{\mathcal{N}}^{2}(t_j,t_i))/4$, with  $\mathcal{N}^{2}(t_j,t_i)$ as defined in Eq.~\Eqref{eq:norm_def} and $\widetilde{\mathcal{N}}^{2}(t_j,t_i) = \mathrm{tr}[\mathcal{U}_{-}^{\dagger}(t_j,t_i)\mathcal{U}_{-}(t_j,t_i)]/2$. $\mathcal{N}^{2}(t_j,t_i)$ and $\widetilde{\mathcal{N}}^{2}(t_j,t_i)$ can again be measured separately as before. It can be easily seen from Eqs.~\Eqref{eq:sup} and \Eqref{eq:circ_corr} that $C_{ij}=T_{+}(t_j,t_i)/\mathcal{N}^{2}(t_j,t_i)$. This allows direct experimental measurement of $C_{ij}$ by setting appropriate $t_i$ and $t_j$ . Note that our method allows us to choose any value of the $\alpha$ and  $\phi$. 

As in the theory, we set $t_1 = 0,~t_2 = t,~t_3 = 2t$ and $\omega=1$ for our experiments. For no superposition, we set $\alpha=0$ and measure $K_3$ while increasing $t$ from $0$ to $\pi$. Results are shown in Fig.~\ref{fig:circuits}~(c) confirming the existence of TTB as expected. Next, we consider the maximum superposition $\alpha = \pi/4$  at $\phi = \pi/2$ and measure $K_3$ with increasing $t$ and observe a clear violation beyond TTB (see Fig. \ref{fig:circuits}~(d)). We increase $\phi$ to $135^0$ and $160^0$ at maximum superposition of $\alpha = \pi/4$ and measure $K_3$ with varying $t$. We observe enhanced violation beyond TTB with increasing $\phi$, as predicted theoretically. Experimentally measured value of  $(K_3)_{\rm m}$ reads $2.27 \pm 0.1$ for $\phi = 135^0$, showing enhanced violation by more than $7$ times the experimental uncertainty (Fig.~\ref{fig:circuits}~(e)), and $2.43 \pm 0.25$ for $\phi = 160^0$, which violates LGI beyond TTB by more than $3$ times the experimental uncertainty (Fig.~\ref{fig:circuits}~(f)). These results clearly and convincingly demonstrate the enhanced violation of LGI beyond TTB due to superposition of unitaries.

\section{Enhanced robustness} \label{2sec:robustness}
\begin{figure}
\centering
\includegraphics[clip=true,width=10cm,clip=true, trim={6.4cm, 12.4cm, 6.7cm, 12.45cm}]{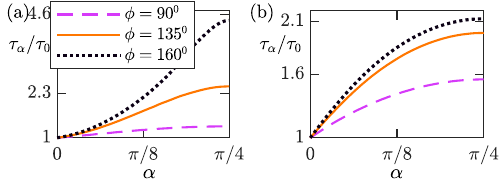} 
\caption{The gain in lifetime $\tau_{\alpha}/\tau_{0}$ ($\tau_0$ being lifetime at $\alpha=0$) vs $\alpha$ at different values of $\phi$ calculated from (a) Eq. \Eqref{eq:bloch} and (b) from applying noise on both S and A. In both cases, decay constant $\gamma=1/(4\pi)~\rm{s}^{-1}$.} 
\label{fig:dp}
\end{figure}

We now investigate the effect of dephasing on the dynamics under the superposed unitary. In Eq.~\Eqref{eq:circ_corr}, the observable $\hat{Q}(0)=\hat{\sigma}_z$ is subjected under $\mathcal{U}(t,0)$ to get transformed to $\hat{Q}(t)=\mathcal{U}^{\dagger}\,(t,0)\hat{\sigma}_{z}\,\mathcal{U}(t,0)$. Upon increasing $t$ continuously from $0$, this map describes rotation of the Bloch vector $\hat{z}$ (representing the observable) about $\hat{\zeta}$ with non-linear SOE $g(t)$. Writing the observable at time $t$ as $\hat{Q}(t)=\vec{S}(t) \cdot \vec{\sigma}$, with Bloch vector $\vec{S}(t):=(S_x(t),S_y(t),S_z(t))$ and Pauli vector $\vec{\sigma}:=(\hat{\sigma}_x,\hat{\sigma}_y,\hat{\sigma}_z)$, the rotation is described by the Bloch equation \cite{cavanagh1996protein,roberts1991bloch}, where dephasing can be phenomenologically introduced with decay rate $\gamma$ as 
\begin{equation}
\partial_{t} \vec{S}(t) = g(t)\hat{\zeta} \times \vec{S}(t) - \gamma \left( \hat{z} \times (\vec{S}(t) \times \hat{z}) \right). \label{eq:bloch}
\end{equation}
Eq.~\Eqref{eq:bloch} is solved with initial condition $\vec{S}(0)=(0,0,1)$ to get $\vec{S}(t)$ and $\vec{S}(2t)$, from which we directly compute $K_{3}(t) = \hat{z} \cdot (2\vec{S}(t)-\vec{S}(2t))$. 
The quantum lifetime of the system $\tau_{\alpha}$ is defined as the time up to which LGI violation persists under $\mathcal{U}$ at superposition $\alpha$. The gain in lifetime due to superposition is plotted against $\alpha$ for different values of $\phi$ in Fig.~\ref{fig:dp}~(a), which depicts the remarkable enhancement of robustness against dephasing with increasing superposition.   

However, in practice $\mathcal{U}(t,0)$ is realized via an ancillary system (A) as shown in Fig.~\ref{fig:Th_Plots}~(a), where S evolves jointly with A until post-selection.  for a practical realization of the superposed unitary $\mathcal{U}$, the system S and the ancilla A would both undergo noisy evolutions. To consider this aspect, we subject the joint state $\rho_{\rm{AS}}$ under a  GSKL master equation \cite{lidar2020lecturenotestheoryopen,breuer2002theory}
\begin{equation}
\partial_{t} \rho_{\mathrm{AS}}= -\im [\mathcal{H}_{\rm{AS}},\rho_{\mathrm{AS}}] + \frac{\gamma}{2} \sum_{j=\rm{S,A}}(\hat{\sigma}^{j}_{z}\, \rho_{\rm{AS}} \,\hat{\sigma}^{j}_{z} - \rho_{\rm{AS}}), 
\label{eq:lind_blad}
\end{equation}
where $\hat{\sigma}_{z}^{j}$ denotes the action of $\hat{\sigma}_{z}$ on $j$ and identity on the rest, and the interaction Hamiltonian reads $\mathcal{H}_{\rm{AS}}:=\proj{0} \otimes \omega \hat{\sigma}_{x}/2 + \proj{1} \otimes \omega \hat{\sigma}_{\phi}/2 $. In presence of noise $(\gamma > 0$), the initial joint state $\rho_{\rm{AS}}(0) =  \proj{\alpha_{\rm{A}}} \otimes \mathbbm{1}_{\rm{S}}/2 $ evolves under Eq.~\Eqref{eq:lind_blad} from $t=0$ to $t$. After which A is post selected in state $\Ket{+}_{\rm{A}}$, allowing S to transform under the map $\mathcal{W}(t,0)$, which describes the effective dynamics under superposed unitary and dephasing, and reads  $\mathcal{W}(t,0)\rho_{\text{S}} = \text{tr}_{\text{A}} [ \proj{+}_{\text{A}}\exp(\mathcal{L}t)\rho_{\text{AS}}(0)]/\text{tr}[ \proj{+}_{\text{A}}\exp(\mathcal{L}t)\rho_{\text{AS}}(0)]$, considering the  Liouvillian for Eq.~\Eqref{eq:lind_blad} as  $\mathcal{L}$. The correlators $C_{12},~C_{23}$ and $C_{13}$ are computed as before with $\mathcal{W}$ replacing $\mathcal{U}$ in Eq.~\Eqref{eq:circ_corr}. From this we compute $K_{3}(t)$ to find the gain in lifetime $\tau_{\alpha}/\tau_{0}$ with increasing $\alpha$. Results are plotted for different values of $\phi$ in Fig.~\ref{fig:dp}~(b), which shows that even the presence of additional noise in A does not destroy the robustness achieved by the superposition of unitaries against dephasing. I believe, the above theoretical analysis would pave the way for experimental investigations, such as those in \cite{PhysRevResearch.7.023045}, to further investigate and demonstrate the robustness achieved by superposing unitary operators.

\section{The General Construction of the Superposed Unitary} \label{2sec:supUgen}
Here we present the general method for constructing two the superposition between two unitary operators in a d-dimensional Hilbert space, and derive the necessary conditions required to ensure their validity.
Say, we have a $d$-dimensional quantum system with a Hilbert space $\mathcal{H}_d$. A set is formed by $n$ self-adjoint operators 
\begin{gather}
S_n \coloneqq \{ V_k : \mathcal{H}_d \rightarrow \mathcal{H}_d \, | \, V_k^{\dagger} = V_k, ~\text{and}~ \{V_k,V_l\}=2\delta_{kl}\mathbbm{1},\, \forall k,l \in \{1,2,...,n \} \}.
\end{gather}
The Hamiltonians $H_k$ of the system are chosen such that they are expandable in the elements of $S_n$ with real coefficients 
\begin{gather}
H_{k} = \sum_{j=1}^{n} r^{k}_{j} \, V_{j}, ~ r^{k}_{j} \in \mathbbm{R},~\text{and}~\norm{\vec{r}_k} = 1, 
\end{gather}
where $\vec{r}_k=(r^k_1, r^k_2, ... , r^k_n)$ is taken to be of unit norm to make the Hamiltonian idempotent, i.e $H_k^2 = \mathbbm{1}$. Given any of these Hamiltonians as the generator of time translation of the system, the evolution of the system from time $t_i$ to $t_j$ is described by the unitary operator $U_k(t_j,t_i)=\exp{(-\im H_k (t_j-t_i))}$ (setting $\hbar=1$). We call these as simple unitary operators. Note that because the Hamiltonians are idempotent, the exponential takes the closed form 
\begin{align}
U_{k}(t_j,t_i) = \cos (t_j - t_i) \mathbbm{1} - \im \sin(t_j - t_i) H_k. \label{seq:simp}
\end{align}
To form a superposition between any two of these simple unitaries,  we first construct the operator 
\begin{gather}
\widetilde{\mathcal{U}}_{+}(t_j,t_i) \coloneqq \sum_{k=1}^{2} c_k \, U_{k}(t_j,t_i),~\text{where}~c_1,c_2 > 0,~\text{and}~\sum_{k=1}^{2}c_{k}^{2}=1.    
\end{gather}
In order to form a superposition between $U_{1}(t_j,t_i)$ and $U_{2}(t_j,t_i)$ such that the resultant operator remains unitary, we need to ensure $\widetilde{\mathcal{U}}_{+}^{\dagger}(t_j,t_i) \,\widetilde{\mathcal{U}}_{+}(t_j,t_i) \propto \mathbbm{1}$, with a positive proportionality constant $\mathcal{N}^{2}(t_j,t_i)$. Denoting $t_j - t_i$ as $\theta_{ji}$, we write
\begin{gather}
\widetilde{\mathcal{U}}_{+}^{\dagger}(\theta_{ji})\,\widetilde{\mathcal{U}}_{+}(\theta_{ji}) = \left( \sum_{k=1}^{2} c_{k}U_{k}^{\dagger}(\theta_{ji}) \right) \left( \sum_{l=1}^{2} c_{l} U_{l}(\theta_{ji}) \right)  \nonumber \\
= \mathbbm{1} +  c_{1}c_{2} \,\left(U_{1}^{\dagger}(\theta_{ji})U_{2}(\theta_{ji}) + U_{2}^{\dagger}(\theta_{ji})U_{1}(\theta_{ji})\right) \nonumber \\
= \mathbbm{1} + c_{1}c_{2} \left(2 \mathbbm{1} \cos^{2} \theta_{ji} + \sin^{2} \theta_{ji} (H_1 H_2 + H_2  H_1) \right) \nonumber \\
= \mathbbm{1} +c_{1}c_{2} \left( 2 \mathbbm{1} \cos^{2} \theta_{ji} + \sin^{2} \theta_{ji} \left[ 2 \mathbbm{1} \vec{r}_1 \cdot \vec{r}_2 + \sum_{s \neq t} r_{s}^{1} r_{t}^{2} \{V_s,V_t \} \right]  \right) \nonumber \\
= \underbrace{ \left[1+2c_{1}c_{2} \left(\cos^{2} \theta_{ji} + \vec{r_1} \cdot \vec{r_2} \sin^{2} \theta_{ji}  \right)  \right]}_\text{$\mathcal{N}^{2}(\theta_{ji})$} \mathbbm{1}.
\end{gather}
As discussed earlier, we only need to ensure that the proportionality constant $\mathcal{N}^{2}(\theta_{ji}) > 0$. Notice, due to $\sum_{k}c_{k}^{2}=1$, the maximum value of the term $2c_1 c_2$ is $1$ (when $c_1=c_2=1/\sqrt{2}$). Therefore, $\mathcal{N}^{2}(\theta_{ji}) > 0$ as long as we ensure
\begin{gather}
\vec{r}_1 \neq - \vec{r}_2
\label{seq:cond}
\end{gather}
Choosing the $\vec{r}_1$ and $\vec{r}_2$ such that they obey Eq.~\Eqref{seq:cond}, we can now define the superposed unitary as 
\begin{gather}
\mathcal{U}(t_j,t_i) \coloneqq \frac{ \sum_{k=1}^{2} c_{k} U_{k} (t_j,t_i)}{\mathcal{N}(t_j,t_i)}.
\end{gather}
Incidentally, the superposed unitary $\mathcal{U}(t_j,t_i)$ could be decomposed as 
\begin{gather}
\mathcal{U}(t_j,t_i) = \frac{\sum_{k=1}^{2}c_{k} \left( \mathbbm{1} \cos \theta_{ji} - \im H_{k} \sin \theta_{ji} \right)}{\mathcal{N}(\theta_{ji})} \nonumber \\
= \frac{\left(\sum_{k} c_{k} \right)\cos \theta_{ji}}{\mathcal{N}(\theta_{ji})} \mathbbm{1} - \im \frac{\sqrt{\sum_{j}\widetilde{R}_{j}^2}\sin \theta_{ji}}{\mathcal{N}(\theta_{ji})} \left( \sum_{p} R_{p} V_{p} \right),
\label{seq:adra}
\end{gather}
where $\widetilde{R}_{j} = \sum_{k}r_{j}^{k}c_{k}$, and $R_j = \widetilde{R}_{j}/\sqrt{\sum_{i} \widetilde{R}_{i}^2}$. Of course, for the validity of this definition, we need to ensure that $\sum_{i} \widetilde{R}_{i}^2 >0$. Note that $\sum_{i} \widetilde{R}_{i}^2 = 1 +  (\vec{r}_1 \cdot \vec{r}_2) c_{1}c_{2} >0$ as long as $\vec{r}_1$ and $\vec{r}_2$ satisfies Eq.~\Eqref{seq:cond}. The resultant Hamiltonian $\widetilde{H}=\sum_{p} R_{p} V_{p} $ is again idempotent because $\sum_{p=1}^{2} R_{p} =1$ (by construction). It can be easily seen that the square of the coefficients of $\mathbbm{1}$ and $\widetilde{H}$ of Eq.~\Eqref{seq:adra} sums up to one, which allows us to identify them as :
\begin{gather}
\frac{\left(\sum_{k} c_{k} \right)\cos \theta_{ji}}{\mathcal{N}(\theta_{ji})} = \cos (\widetilde{\theta}_{ji}) ~ \text{and} ~ \frac{\sqrt{\sum_{j}\widetilde{R}_{j}^2}\sin \theta_{ji}}{\mathcal{N}(\theta_{ji})} = \sin(\widetilde{\theta}_{ji}).  
\end{gather}
Thus the superposed unitary can also be written in the form of Eq.~\Eqref{seq:simp} as 
\begin{gather}
\mathcal{U}(t_j,t_i) = \cos(\widetilde{\theta}_{ji}) \mathbbm{1} - \im \sin(\widetilde{\theta}_{ji}) \widetilde{H},
\end{gather}
with the only difference that $\widetilde{\theta}_{ji}$ is now  a non-linear function of the time interval $t_j-t_i$, which results non-linear speed of evolution (SOE) of the Bloch vector as discussed in the main text.

Note that we can now choose two such superposed unitary $\mathcal{U}_{i}(t_j,t_i)=\cos(\widetilde{\theta}_{ji}) \mathbbm{1} - \im \sin(\widetilde{\theta}_{ji}) \widetilde{H_i}$, for $i=1,2$. As long as the coefficient vectors $\vec{R}_1$ and $\vec{R}_2$ satisfies the condition of Eq.~\ref{seq:cond}, we can form a superposition between two superposed unitaries following the same process, and so on.

\section{Effect on Higher Order LGI $K_n$ :} \label{2sec:highLGI}
I primarily considered the effect of superposition of unitaries on three term LGI $(K_3)$. However, the above analysis can be readily generalized to the case of $n$ term LGI $K_n$ defined as 
\begin{gather}
K_n = C_{12} + C_{23} + C_{34} + ... + C_{(n-1)n} - C_{1n}. \label{eq:Kn_full}
\end{gather}
Due to the time translational symmetry of our superposed unitary $\mathcal{U}(t_f,t_0) = \mathcal{U}(t_f+t,t_0+t)$ (which follows directly from Eq. (1) of main text), the expression of $K_n$ in Eq.~\Eqref{eq:Kn_full} gets reduced to
\begin{gather}
K_n = (n-1) C_{12} - C_{1n}. \label{eq:Kn_short}
\end{gather}
\begin{figure*}
\centering
\includegraphics[width=12.5cm, clip=true, trim={0cm 0cm 0cm 0cm}]{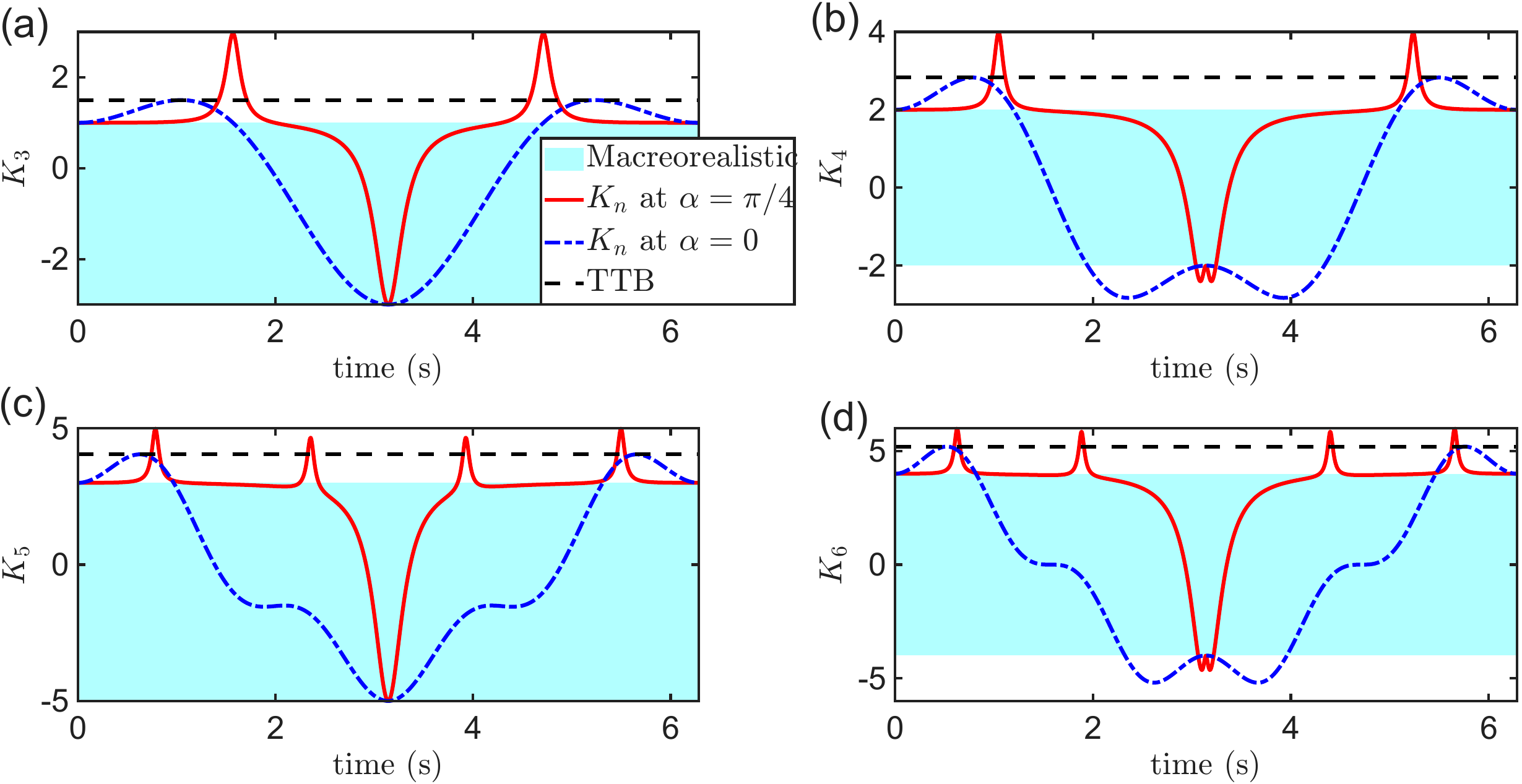}
\caption{Investigating higher order LGI's $K_n$ under superposition of unitary evolutions: Observing enhaced violation of LGI beyond the respective TTB. Here, the results are reported for $n=3,4,5$ and $6$, where $K_n$ stays below the TTB at zero superposition ($\alpha=0$), however goes beyond the TTB when the superposition is implemented ( $\alpha = \pi/4$ and $\phi=170^0$).} 
\label{fig:LGI_Extn}
\end{figure*}
Under the effect of superposition of unitary map $\mathcal{U}(t_f,t_0)$, the two time correlators $C_{12}$ and $C_{1n}$ can be once again directly computed using Eq.(3) of the main text by replacing $\mathcal{U}(t_f,t_0)$ with $\mathcal{U}(t,0)$ and $\mathcal{U}(t(n-1),0)$, respectively. From $C_{12}$ and $C_{1n}$, the n term LGI parameter $K_n$ can be directly computed from Eq.~\Eqref{eq:Kn_short}.

The computed results are displayed in Fig.~\ref{fig:LGI_Extn}. The classical or macrorealistic bound for $K_n$ reads $ -n \leq K_n \leq (n-2) $ for odd $n$, and $-(n-2) \leq K_n \leq (n-2)$ for even $n$. Whereas, the temporal Tsirelson's bound (TTB) for $K_n$ reads $K_n \leq (K_n)_{\text{TTB}} = n \cos(\pi/n)$ (see Ref. [14] in the main text). The computations are done at $\phi = 170^0$ for $n=3,4,5$ and $6$. The results confirm that the superposition of unitary evolutions lead to violation of TTB even in higher order LGI's, thereby demonstrating the generality of our investigated results (for $n = 3$).

\section{Conclusions} \label{2sec:concl}
We investigated temporal correlations in terms of LGI for qubit transforming under a superposition of unitaries. We found temporal correlations get stronger and violate LGI beyond TTB under such superposed unitaries. This enhanced non-macrorealism, as quantified by the violation of LGI beyond TTB, increases with increasing superposition between the unitaries. Also, at a given superposition, it increases as we make the angle between the axis of rotation larger between two unitaries that are being superposed. We present a method to physically realize the superposition of unitaries using an ancillary qubit and experimentally verify the enhanced non-macrorealism in NMR architecture. Moreover, we found that the dynamics under the superposed unitary have applications in providing more robustness against environmental noise by remarkably enhancing the time up to which the system violates LGI under decoherence. We believe that our work will pave the way for further investigations on applying these superposed unitaries in various quantum information processing tasks like \cite{PhysRevB.93.035441}, along with provoking foundational questions on the nature of temporal correlations in quantum systems in interaction with other ancillary system.

\chapter{Observing algebraic variety of Lee-Yang zeros in asymmetrical systems via a quantum probe}\markboth{Chapter 2: Determining Algebraic variety of Lee-Yang zeros}{}\label{chap_BrokenTS}

\begin{center}
\textit{``In physics, we are trying to uncover the rules of the game; \\ in mathematics, we create the rules of possible games. The surprise is how often they coincide.''} \\
~~~~~~~~~~~~~~~~~~~~~~~~~~~~~~~~~~-- Freeman Dyson 
\end{center}

\section{Introduction}
\label{sec:Introduction}
It is commonly understood that complex numbers merely play the role of a calculational tool in physics, while observable quantities are represented by real numbers.
In 1952, Lee and Yang published two landmark papers \cite{Yang:1952be,Lee:1952ig} showing the partition function of a system can become zero at certain points on the complex plane of its physical parameters. These zeros, now known as Lee-Yang (LY) zeros, provide a cohesive understanding of equilibrium phase transition \cite{ning2024experimental} as they correspond to the non-analyticity of free energies. However, it was widely believed that LY zeros can not be observed directly as they occur at complex values of physical parameters, and only when they approach the real axis, their presence gets disclosed as the system goes through a phase transition.

Nevertheless, this does not mean that complex LY zeros have nothing to say about the physical system. In fact, determination of LY zeros can play a key role in studying the thermodynamic behavior of complicated many-body systems as they fully characterize the partition function. Apart from that, recent studies \cite{	doi:10.1126/sciadv.abf2447, PhysRevResearch.1.023004} have paved the way for determining universal critical exponents of phase transitions from LY zeros, which is otherwise a computationally difficult problem due to critical slowdown. It was also observed that LY zeros can be employed to understand non-equilibrium phenomena like dynamical phase transitions \cite{PhysRevLett.118.180601},  along with other statistical studies \cite{PhysRevE.97.012115, PhysRevC.72.011901} like percolation   \cite{arndt2001directed} or complex networks \cite{krasnytska2015violation,krasnytska2016partition} and even protein folding \cite{PhysRevLett.110.248101,PhysRevE.88.022710}. Profound links between thermodynamics in the complex plane and dynamical properties of quantum systems have also been discovered \cite{PhysRevLett.110.135704,PhysRevLett.110.050601,PhysRevLett.110.230601,PhysRevLett.110.230602,wei2014phase}  in recent years. All these discoveries have made the determination of LY zeros for  general classical systems a crucial necessity in various disciplines of physics.

In 2012, by representing a classical Ising chain with a small number of spins, Wei and Liu showed \cite{PhysRevLett.109.185701} that the complex LY zeros for such a system can be mapped to the zeros of quantum coherence of an interacting probe.  Using this method, the experimental observation of LY zeros was directly achieved \cite{PhysRevLett.114.010601}. 

This opens up the possibility of using quantum simulation techniques to simulate classical systems and determine their LY zeros with a quantum probe, as demanded in many areas of physics. However, there are two major challenges to  overcome. First of all, the method proposed in \cite{PhysRevLett.109.185701} and successive experiments \cite{PhysRevLett.114.010601,doi:10.1126/sciadv.abf2447} reported observation of LY zeros lying on a single complex plane $\mathbb{C}$. This happens when the partition function is expandable as a univariate polynomial, known as Lee-Yang (LY) polynomial, in terms of the complexified physical parameter. This condition of LY polynomial being univariate is based on a symmetry assumption about the system. For example, in the case of the Ising chain, it assumes that all the spins will experience the same local (complexified) magnetic field. Unfortunately, a general system will not necessarily have this symmetry. Thus, the partition function, in general, is to be expanded as a multivariate LY polynomial in terms of complexified parameters. In the worst case, an $N$ spin Ising chain will have an $N$ variate LY polynomial if all the spins experience different local fields. In such a scenario, LY zeros of the system will form an \emph{algebraic variety} \cite{kollar2001simplest}  $V\subset(\mathbb{C}^*)^N$, where $\mathbb{C}^*=\mathbb{C}-\{0\}$.
Therefore, finding a method to experimentally determine the full \emph{algebraic variety} containing LY zeros of a general asymmetrical system is crucial. The second challenge to overcome is a quantum simulator of the classical system should have ways to simulate the system at a wide range of its physical parameters like temperature, internal couplings etc. This flexibility in initialization is crucial to uncover the full set of LY zeros corresponding to all physical situations.  However, the systems under study can get complicated and thus achieving full quantum control over system qubits can get challenging. Therefore, the desired method should not assume much experimental control over the system qubits so that it can be extended to complex systems in the future. 

In this work, we show the direct experimental determination of the \emph{algebraic variety} containing roots of a general multivariate LY polynomial for  asymmetrical Ising type systems. Mathematical developments in last few decades unveil that one can project the \emph{algebraic variety} to sets of coordinate-wise absolute values and arguments, called the \emph{amoeba} \cite{GKZ-94,Viro:2002} and \emph{coamoeba} \cite{Feng:2005gw,Nisse:2013a,Nisse:2013b,Nisse:2017}, respectively.
More precisely, we consider in this paper a polynomial of two complex variables, $f(z_1,z_2)$. The set of values $(z_1,z_2)$ which make $f(z_1,z_2)=0$ is what we call the \emph{algebraic variety}, which is a three (real)-dimensional object living in (four-real-dimensional) complex space $(\mathbb{C}^*)^2$. The \emph{amoeba} is obtained by taking logarithms of their absolute values $(\ln|z_1|,\ln|z_2|)$ of the variety, and the \emph{co-amoeba} is obtained by taking its arguments $(\arg z_1,\arg z_2)$. In this sense, the abstract variety living in a higher-dimensional complex space is studied by projecting into domains of lower dimensional real spaces.

We use qubits to simulate the classical asymmetrical Ising system which can be controlled through another qubit acting as a probe. We highlight, like an ideal quantum simulator, the system can be simulated at any arbitrary point on its \emph{amoeba} at any arbitrary temperature. Even the internal coupling of the Ising system can be set to a desired value ranging from ferromagnetic to anti-ferromagnetic regimes. This initialization is accomplished solely by manipulating the probe qubit, leaving the system qubits undisturbed. After effectively initiating the system at an arbitrarily chosen point on its \emph{amoeba}, corresponding points of \emph{coamoeba} can be directly sampled from the time evolution of the probe's coherence. By iterating the process, one samples the \emph{coamoeba} across the \emph{amoeba} to obtain the full \emph{algebraic variety}. Thus both preparation and detection are achieved through the probe alone. We demonstrate the method experimentally via a three-qubit NMR register by taking two of them as system and the third one as probe.  Sampling of \emph{coamoeba} at different points on \emph{amoeba} is performed directly from time domain NMR signal without any need for extensive post-processing of experimental data.  Apart from extending the range of quantum simulations to other areas of physics where determining LY zeros of general classical system is pivotal, our work also brings pure mathematical structures like \emph{amoeba} and \emph{coamoeba} into the realm of quantum simulations by physically sampling them for a given LY polynomial.

The rest of this chapter is organized as follows. In Sec.~\ref{sec:Ising}, we introduce the asymmetric Ising system and show how to use qubits as system and probe such that the \emph{amoeba} and the \emph{coamoeba} of the \emph{algebraic variety} corresponding to the LY polynomial of the system relate to probe qubit's coherence. After describing the methodology of sampling the \emph{algebraic variety} in Sec.~\ref{sec:Method}, we discuss how to initialize the system qubits at any desired point on the \emph{amoeba} at any value of physical parameters like temperature and coupling constant by operating on only the probe in Sec.~\ref{sec:Initialization}. We present the experimental results in Sec .~\ref {sec:Results}, and the derivation of the connection between the zeros of the probe spin's coherence with the LY zeros is presented in Sec .~\ref {appenA}. The methodology of sampling the co-amoeba is presented in Sec.\ref{AppenB}, while the same for amoeba is given in Sec\ref{AppendC} and \ref{AppendD}, before concluding in Sec.~\ref{sec:Conclusion}.  

\begin{figure}
	\centering 
	\includegraphics[scale=0.57]{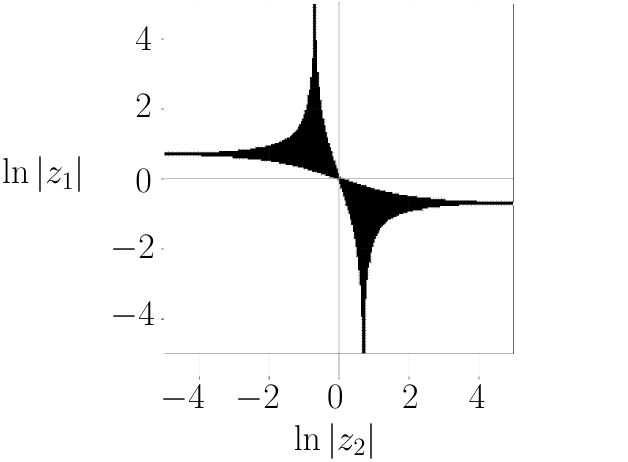}
	\includegraphics[scale=0.67]{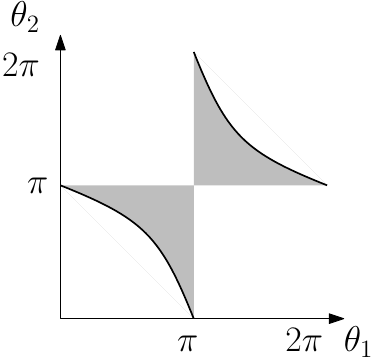} \\
	\caption{The \emph{amoeba} (top) and \emph{coamoeba} (bottom) of the LY polynomial $f(z_1,z_2)=1+\half z_1+\half z_2+z_1z_2$~ \cite{sadyT}.}
	\label{fig:example}
\end{figure}	

\section{Two-spin Ising model and (co)amoebas} \label{sec:Ising}
To explain the method, we consider our system to be a classical Ising chain consisting of only two sites, $A$ and $B$, coupled to each other with strength $J$. Without assuming any symmetry, let the magnetic fields at sites $A$ and $B$ be $h_A$ and $h_B$, respectively. This classical system can be mimicked by using two spin $1/2$ systems as qubits with Hamiltonian $H_{AB}=-J\sigma_z^A\sigma_z^B-h_A\sigma^A_z-h_B\sigma^B_z$, where $\sigma_i^X$ is the Pauli $i$-operator of spin $X$.

In a thermal bath of inverse temperature $\beta$, the partition function of $AB$ is 
\begin{align}
Z(\beta,h_A,h_B)=&\expo{\beta J+\beta h_A+\beta h_B}\big(1+\expo{-2\beta J-2\beta h_A}+
\nonumber \\
&~~~~~\expo{-2\beta J-2\beta h_B}+\expo{-2\beta h_A-2\beta h_B}\big).
\end{align}
We extend $Z$ into the complex domain by letting 
\begin{align}
	\expo{-2\beta h_A}=z_1, \expo{-2\beta h_B}=z_2; ~ \mathrm{where} ~~  (z_1,z_2) \in (\mathbb{C}^*)^2. \label{eq:complex_extension}
\end{align}
Further, by identifying $\Gamma =\expo{-2\beta J}\in\mathbb{R}_+$, the partition function becomes $Z=(\Gamma z_1z_2)^{-1/2}f(z_1,z_2)$, where 
\begin{align}
	f(z_1,z_2)=1+\Gamma z_1+\Gamma z_2+z_1z_2 \label{eq:LYpolynomial}
\end{align}
is the two-spin bivariate \emph{LY polynomial}. Zeros of the partition function now correspond to the vanishing of this polynomial, which defines the \emph{algebraic variety} $V_f=\{(z_1,z_2)\in\{\mathbb{C}^*\}^2\;\big|\; f(z_1,z_2)=0 \}$.

For any complex number $(z_1,z_2)$ lying on $V_f$, information about its absolute value is studied using a log map $\mathrm{Log}:(\mathbb{C}^*)^2\rightarrow\mathbb{R}^2$ by taking $(z_1,z_2)\mapsto(\ln|z_1|,\ln|z_2|)$. The image $\mathcal{A}_f=\mathrm{Log}(V_f)$ is called \emph{amoeba}. On the other hand, information about the phase is studied using the argument map $\mathrm{Arg}:(\mathbb{C}^*)^2\rightarrow S^1\times S^1$ where $(z_1,z_2)\mapsto(\arg {z_1},{\arg z_2})$. The image $\mathrm{co}\mathcal{A}_f=\mathrm{Arg}(V_f)$ is called the \emph{coamoeba}. For example, the \emph{amoeba} and \emph{coamoeba} for $\Gamma=\half$ in Eq.\Eqref{eq:LYpolynomial} are shown in Fig.~\ref{fig:example}.

\begin{figure}
	\centering
	\includegraphics[height = 3cm, trim = {0cm, 10.5cm, 2.21cm, 2cm},clip]{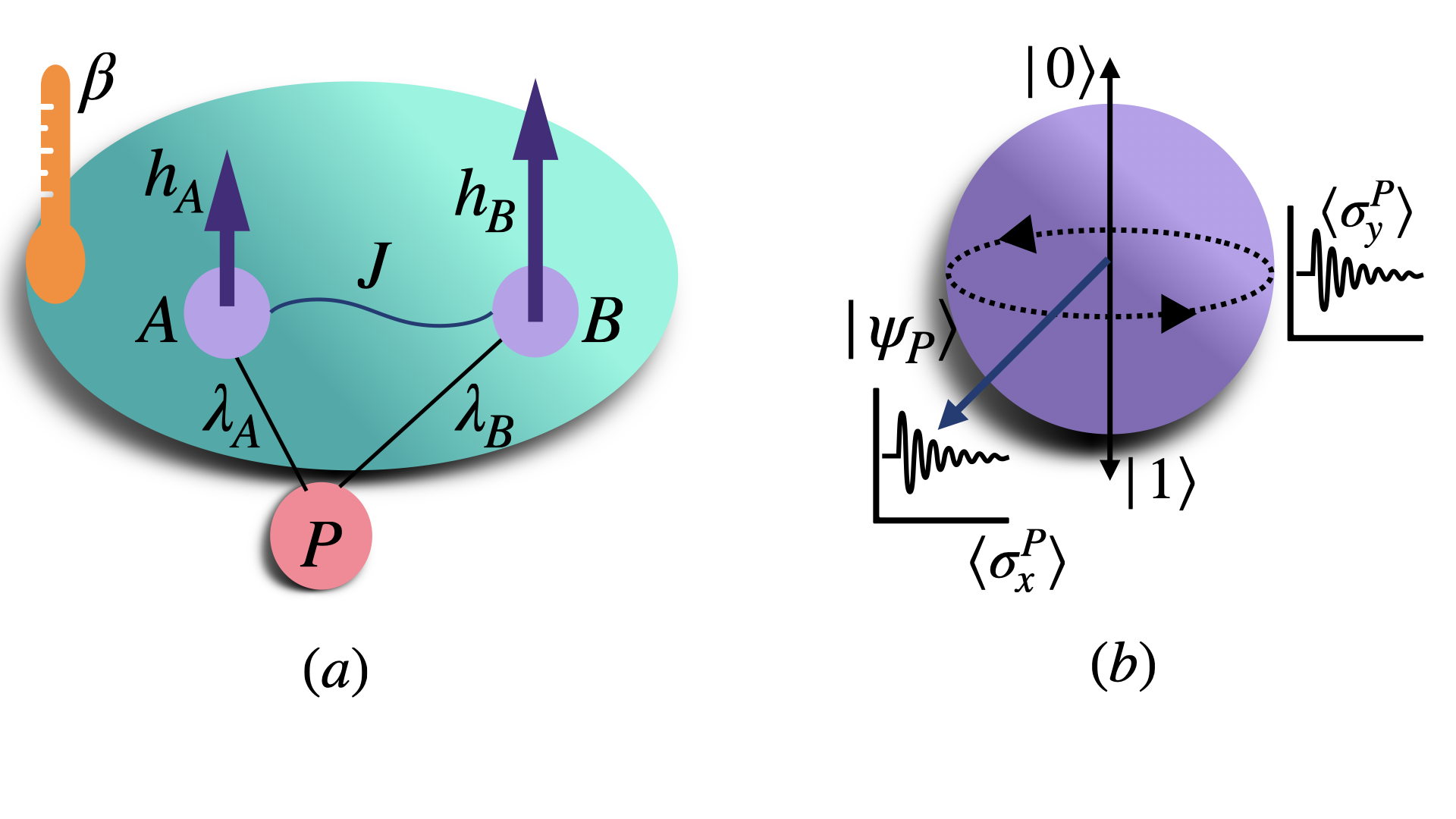} \\
	(a)  \hspace{4cm}  (b)
	\caption{(a) A two-spin Ising system in a thermal bath at inverse temperature $1/\beta$ coupled to a probe spin.  (b) As the probe evolves under the interaction with the system, it's coherence is recorded by measuring $\langle \sigma_x^P \rangle $ and $\langle \sigma_y^P \rangle$ with time.}
	\label{fig_SpinBath}
\end{figure}

To observe these complex zeros, we adopt the procedure of Wei and Liu \cite{PhysRevLett.109.185701} by introducing another spin-$1/2$ particle, which acts as a quantum probe. Its states $\{\ket{0},\ket{1}\}$ (here $\ket{0}$ and $\ket{1}$ are eigenvectors of $\sigma_z$ with eigenvalues $\pm 1$, respectively) span a two-dimensional Hilbert space $\mathcal{H}_{P}$. The total space is now $\mathcal{H}_{\mathrm{tot}}=\mathcal{H}_{P}\otimes\mathcal{H}_{AB}$. However, here we allow the probe to couple asymmetrically with $A$ and $B$. This is because, in our case, $h_A$ and $h_B$ are not necessarily equal, and hence $z_1$ and $z_2$ must be distinct variables, necessitating the description of Lee-Yang zeros in $(\mathbb{C}^*)^2$ rather than in $\mathbb{C}$.

Suppose that the probe spin $\sigma^P_z$ couples to $\sigma_z^A$ and $\sigma_z^B$ with coupling strengths $\lambda_A$ and $\lambda_B$, respectively. This situation is depicted in Fig.~\ref{fig_SpinBath}~(a). The interaction Hamiltonian is {}
\begin{align}
	H_{\mathrm{int}}=\lambda_A\sigma^P_z\sigma_z^A + \lambda_B\sigma^P_z\sigma_z^B. \label{eq:Hint}
\end{align}
Here we used the notation $\sigma^P_z\sigma^A_z=\sigma^P_z\otimes\sigma_z^A\otimes\mathbbm{1}$ and $\sigma^P_z\sigma^B_z=\sigma^P_z\otimes\mathbbm{1}\otimes\sigma_z^B$. The full Hamiltonian now becomes $H=\mathbbm{1}\otimes H_{AB}+H_{\mathrm{int}}$, or 
\begin{equation}
	H=-J\sigma^A_z\sigma^B_z-(h_A-\lambda_A\sigma^P_z)\sigma^A_z-(h_B-\lambda_B\sigma^P_z)\sigma_z^B. \label{eq:Hth}
\end{equation}
The full initial state at time $t=0$ is taken to be
\begin{align}
\rho(0)=\ket{\psi_P}\bra{\psi_P}\otimes\frac{\exp(-\beta H_{AB})}{Z(\beta,h_A,h_B)},\label{eq:initial-state}
\end{align}
where $\ket{\psi_P}$ is a pure coherent quantum state in the probe's $\sigma_z$ basis. Here, by a coherent state we mean that $\ket{\psi_P}$ has superposition in $\{\ket{0},\ket{1}\}$ basis. It subsequently evolves under the Hamiltonian in Eq.~\Eqref{eq:Hth} as $\rho(t) = \mathcal{U}(t)\rho(0)\mathcal{U}^{\dagger}(t)$, where $\mathcal{U}(t) = \expo{-\im t H/\hbar}$. As we only observe the evolution of the probe, we trace out $AB$ to get $\rho_{P}(t)=\mathrm{tr}_{AB}\rho(t)$.
Under this evolution, the coherence $L$ of probe at time $t$ can be shown to take the form 
\begin{align}
	L(t)&= \frac{\hbar^2}{4}\left|\langle\sigma_x^P(t)\rangle+\im\langle\sigma_y^P(t)\rangle \right|^2  = \mathcal{C}\left|f(z_1,z_2)\right|^2. \label{eq:Lt}
\end{align}
derivation can be found in Sec.~\ref{appenA}. Here  $\mathcal{C}$ is a constant over time and $f(z_1,z_2)$ is the two-spin  bivariate LY polynomial of Eq.~\Eqref{eq:LYpolynomial}, after identifying
\begin{align}
z_1&=\exp\{-2\beta h_A+4\im\lambda_At/\hbar\},
\nonumber \\
z_2&=\exp\{-2\beta h_B+4\im\lambda_Bt/\hbar\}, ~\mbox{and} \nonumber \\
\Gamma&=\exp\{-2\beta J\}.
\end{align}
Note that the variables $z_1$ and $z_2$ become complex as the time evolution introduces a complex phase. Furthermore, we note 
\begin{subequations}
	\begin{align}
		\ln|z_1|&=-2\beta h_A,\hspace{1.05cm} \ln|z_2|=-2\beta h_B, ~~\mathrm{and} \label{eq:am}\\
		\theta_1(t)&=\arg z_1=\frac{4\lambda_A t}{\hbar},\quad \theta_2(t)=\arg z_2=\frac{4\lambda_B t}{\hbar}. \label{eq:coam}
	\end{align}
\end{subequations}
Corresponding to the set of points $\{(z_1,z_2)\}$ for which $|f(z_1,z_2)|^2=0$, the set $\{(\ln|z_1|,\ln|z_2|)\}$  forms the \emph{amoeba} while the set $\{(\arg z_1,\arg z_2)\}$ forms the \emph{coamoeba}.

\section{Method for Observing The Algebraic Variety}  \label{sec:Method}
Our method for determining the \emph{algebraic variety} $V_f$ works as follows: we first initiate the system in the state given by  Eq.~\Eqref{eq:initial-state} at arbitrary values of $\beta h_A$ and $\beta  h_B$, which according to Eq.~\Eqref{eq:am}, fix a point on the \emph{amoeba} space. As we shall see later, the preparation is achieved through the probe, assuming no control over the system qubits. Next, we let the probe interact with the system and observe its coherence. If the coherence is non-zero at all finite times, we conclude that there are no LY zeros at that value of $(\beta h_A,\beta h_B)$; hence, the chosen point does not belong to the \emph{amoeba}. On the other hand, if we find the coherence vanishing at time instants $\{t_i\}$, the point $(\beta h_A,\beta h_B) \equiv (\ln|z_1|,\ln|z_2|)$ belongs to the \emph{amoeba}. Moreover, by Eq.~\Eqref{eq:coam},  $\{t_i\}$ can be mapped to $\{(\theta_1(t_i),\theta_2(t_i))\}$, which lie on a $S^1\times S^1$ torus as the points of \emph{coamoeba}, corresponding to the point  $(\ln|z_1|,\ln|z_2|)$ on \emph{amoeba}. Upon multiple iterations of this procedure, one can sample the \emph{coamoeba} across the \emph{amoeba} to extract the full $V_f$ for the multivariate LY polynomial. For more details regarding the sampling of \emph{coamoeba}, see Sec~\ref{AppenB}.

Another way of seeing the fact that complex LY zeros leave footprint in the real time dynamics of the probe would be in terms of correlation. Mutual information \cite{nielsen2010quantum} between probe and the system qubits captures the total correlation between them and is defined as 
\begin{align}
I_{P:AB}(t) = S_{P}(t) + S_{AB}(t) - S_{PAB}(t),   \label{eq:MI}
\end{align}
where $S_{X}(t) = -\mathrm{tr}\left[\rho_X(t) \ln \rho_X(t)\right]$ is the Von-Neumann entropy \cite{nielsen2010quantum} of  $X$ at time $t$. By evolving the initial state of Eq.~\Eqref{eq:initial-state} under the Hamiltonian of Eq.~\Eqref{eq:Hth}, it can be seen that $S_{AB}$ and $S_{PAB}$ do not evolve over time. Therefore, according to Eq.~\Eqref{eq:MI}, $I_{P:AB}$ becomes maximum at times only when the probe is maximally mixed, i.e $\rho_{P} = \mathbbm{1}/2$. The reduced density matrix of the probe can be expanded in Pauli basis as
\begin{align}
\rho_P(t) = \frac{1}{2} \left(\mathbbm{1} + \sum_{i \in \{x,y,z\}} c_i(t) \sigma_{i}^P \right), \label{eq:MIstate}
\end{align}
where $c_i (t)= \langle \sigma^P_i (t) \rangle$.  It can be easily seen that $\partial_{t}c_z=0$ in the evoltion under the Hamiltonian of Eq.~\Eqref{eq:Hth} . If the intial state of the probe is such that $c_z(0)=0$, then its reduced density matrix becomes maximally mixed at time points where $\langle \sigma_x^P \rangle$ and $\langle \sigma_y^P \rangle$ simultaniously vanish, i.e points corresponding to LY zeros. Therefore, only at the time points corresponding to LY zeros, the correlation between the probe and the system becomes maximum.

\section{System Initialization via a Quantum Probe} \label{sec:Initialization}
\begin{figure}
	\centering
	\includegraphics[ width=8cm, trim={0cm 4cm 0cm 0cm},  clip=true]{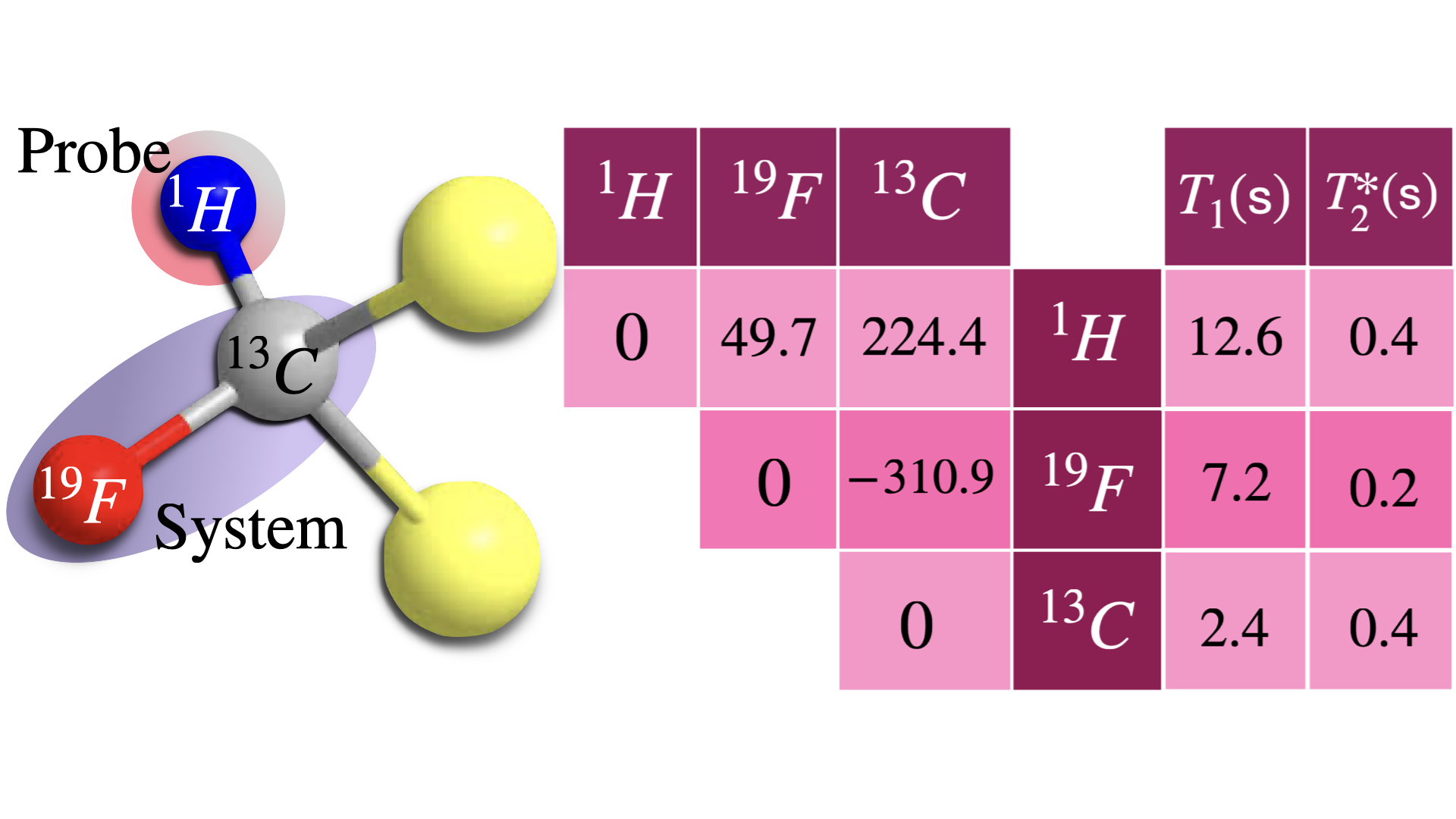} \\
	(a) \hspace{3.3cm} (b) 
	\caption{(a) The molecular structure of DBFM with labeled spins. (b) Values of NMR Hamiltonian parameters (resonance offsets in diagonal and coupling constants $J_{ij}$ in off-diagonal elements, in Hz) along with relaxation time constants.} 
	\label{fig:molecule}
\end{figure}

\subsection{NMR register}
For a concrete demonstration via NMR,  we use a liquid ensemble of three spin-1/2 (three-qubit) nuclear register of $^{13}C$-Dibromofluoromethane (DBFM) (see Fig.~\ref{fig:molecule}~(a)), dissolved in Acetone-D6, by identifying $^{1}H$ as probe $P$ and $^{19}F$, $^{13}C$ as system spins $A, B,~$respectively. In a high static magnetic field of $B_0 = 11.7$~T, their Larmor frequencies have magnitudes $\omega_i = \gamma_{i}B_0$, where $\gamma_i$ are the gyromagnetic ratios \cite{cavanagh1996protein}.  The three spins also interact mutually via scalar coupling with strengths $J_{ij}$ as tabled in Fig.~\ref{fig:molecule}~(b). In terms of the spin operators $I_\alpha^i = \sigma_\alpha^i/2$, the lab-frame NMR Hamiltonian is of the form $H_{\mathrm{NMR}} = H_I + H_{RF}$, with internal part $H_{I} = -\sum_{i} \hbar \omega_{i} I_z^i + 2\pi \hbar \sum_{i \neq j} J_{ij} I_z^i I_z^j$  
and the probe control part $H_{RF}=-\hbar \Omega_{P}(t) I_{x}^{P} $. 
Here $\Omega_{P}(t) = \gamma_i B_P(t)$  represents the control amplitude achieved through the magnetic component $B_P(t)$ of the applied circularly polarized RF field resonant with the probe's Larmor frequency. Notice, the control Hamiltonian contains only the probe's component $\Omega_{P}(t)$, while we set $\Omega_{A}(t)=\Omega_{B}(t)=0$ to ensure that the method works without assuming any experimental control over the system.

We allow the liquid ensemble register to equilibrate at an ambient temperature of $T = 300$ K inside a $500$ MHz Bruker NMR spectrometer.
As $\hbar \omega_i \ll k_{B}T$, the density operator in thermal state becomes  $\rho_{\text{th}}  = \exp(-\beta H_I)/ \mathrm{tr} [\exp(-\beta H_I)] \approx \mathbbm{1}/8 + \epsilon\rho_{\text{th}}^\Delta$, where $\rho_{\text{th}}^\Delta =  I_{z}^{H} + (\gamma_{F}/\gamma_{H}) I_{z}^{F} + (\gamma_{C}/\gamma_{H}) I_{z}^{C}$ is the deviation thermal state and 
the purity factor $\epsilon \approx 10^{-5}$.  From the thermal state,  preparation of the initial state in Eq.~\Eqref{eq:initial-state}  at any chosen values of $\left\{\beta h_A, \beta h_B\right\}$ is to be achieved, subject to constraint that the system ($^{19}F,~^{13}C$) remains inaccessible. From now onward the probe's state of Eq.~\Eqref{eq:initial-state} is taken to be $\ket{\psi_P} = (\ket{0} + \ket{1})/\sqrt{2}$  

\begin{figure*}
\centering
\includegraphics[ width=15cm, trim={12cm 0cm 14cm 0cm},  clip=true]{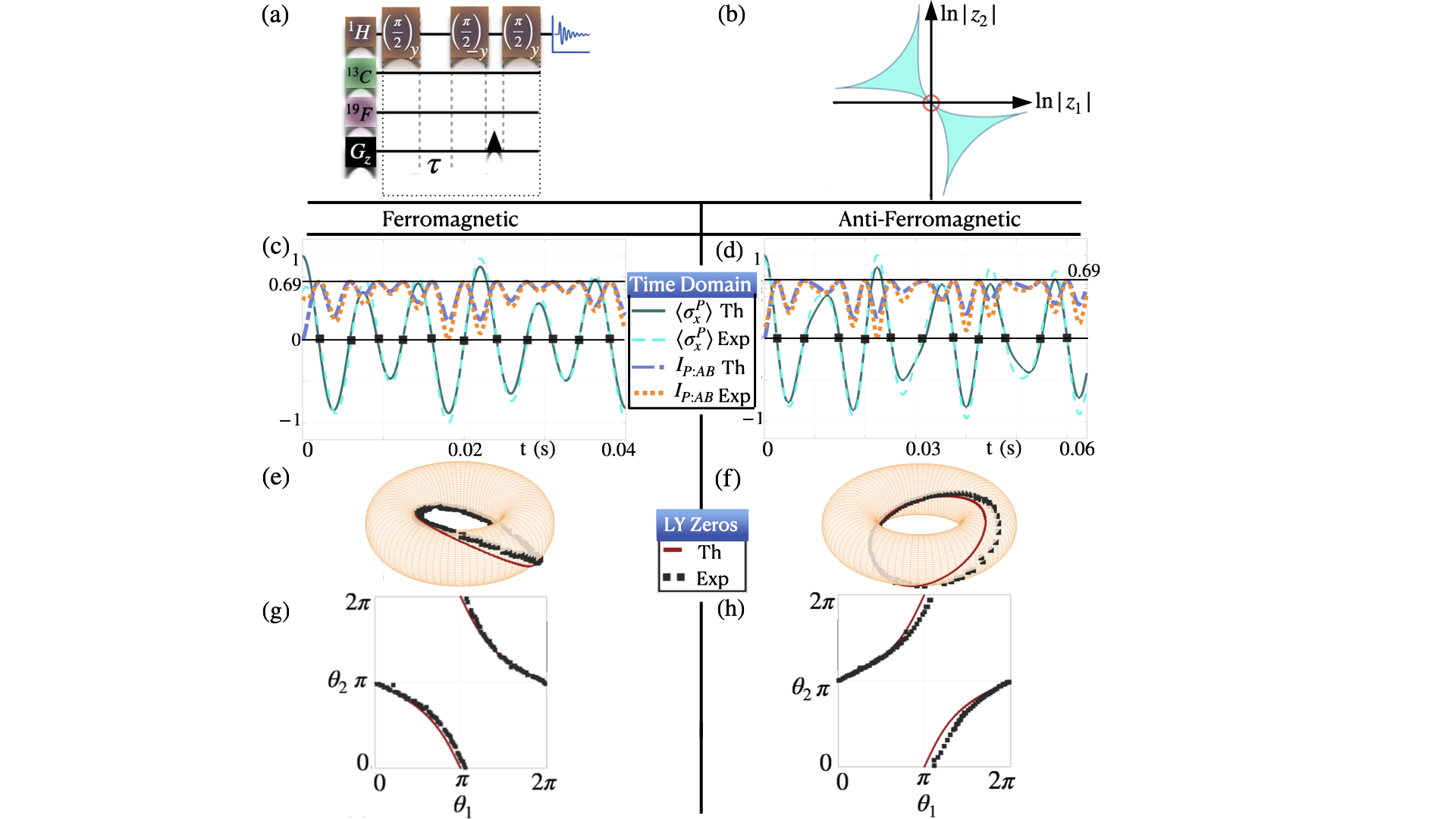} \\
\caption{Pulse sequence (a) for preparing the system in state of Eq.~\Eqref{eq:initial-state} for $h_A = h_B = 0$. We set $\tau = 8.76$ ms or $\tau =9.013$ ms to prepare $\beta J= 0.5$ (ferromagnetic case) or $\beta J=-0.5$ (anti-ferromagnetic case) respectively.  The corresponding \emph{amoeba} point is at the origin of the complex plane as illustrated in (b).  (c-d) The real parts of FIDs (dashed lines) corresponding to $\langle \sigma_x^P(t) \rangle$  are plotted along with the simulated curves (solid lines) for (c) ferromagnetic and (d) anti-ferromagnetic cases respectively. As $\langle \sigma_y^P (t) \rangle$ is identically zero in these cases, the FID null points (solid squares) correspond to LY zeros. Theoretical and experimental value of  the mutual information $I_{P:AB}$ between the probe and the system are also plotted, which reach their maxima (horizontal line at $0.69$) only at the LY zeros. (e-f) Theoretically calculated (solid line) and experimentally observed (filled squares) \emph{coamoeba} on a 2-torus for (e) ferromagnetic and (f) anti-ferromagnetic cases respectively.  The corresponding planar visualizations are in (g-h). }
\label{fig:ImExp}
\end{figure*}

\subsection{Initializing at the Origin of Amoeba Plane}

First consider the preparation for $h_A = h_B = 0$, which corresponds to the origin of \emph{amoeba} in accordance to Eq.~\Eqref{eq:am}. We note that the factor $\exp(-\beta H_{AB})$ of Eq.~\Eqref{eq:initial-state} can be expanded as 
\begin{align}
\expo{-\beta H_{AB}}  = &\expo{\beta J \sigma_z^A \sigma_z^B} \expo{\beta h_A \sigma_z^A} \expo{\beta h_B \sigma_z^B} \nonumber \\
 = C_a \mathbbm{1}_4 + & C_b \sigma_z^A + C_c \sigma_z^B + C_d \sigma_z^A  \sigma_z^B, \nonumber \\
\mathrm{where,} ~~ C_a &= [\cosh(\beta J) \cosh(\beta h_A) \cosh(\beta h_B) \nonumber \\
&+ \sinh(\beta J) \sinh(\beta h_A) \sinh(\beta h_B)], \nonumber \\
C_b &=  [\cosh(\beta J) \sinh(\beta h_A) \cosh(\beta h_B) \nonumber\\
&+ \sinh(\beta J) \cosh(\beta h_A) \sinh(\beta h_B)],  \nonumber \\
C_c &=  [\cosh(\beta J) \cosh(\beta h_A) \sinh(\beta h_B) \nonumber\\
&+ \sinh(\beta J) \sinh(\beta h_A) \cosh(\beta h_B)],  \nonumber \\
C_d &=  [\cosh(\beta J) \sinh(\beta h_A) \sinh(\beta h_B) \nonumber\\
&+ \sinh(\beta J) \cosh(\beta h_A) \cosh(\beta h_B)]  \label{eq:series}. 
\end{align}
Substituting the above to Eq.~\Eqref{eq:initial-state} and setting $h_A=h_B=0$, the initial state in this case becomes 
\begin{gather}
\rho(0) = \mathbbm{1}/8 + (1/2) \tanh (\beta J) I_z^A I_z^B +  \left[\cosh(\beta J) I_x^P + \sinh (\beta J) 4 I_x^P I_z^A I_z^B\right]/Z(\beta,0,0).
\end{gather}
Here the first two terms can be suppressed as they remain invariant under time evolution governed by $H_I$, and also do not contribute to probe's coherence in Eq.~\Eqref{eq:Lt}. Therefore, the target state for initialization reads
\begin{equation}
\rho^{\Delta}(0) =  \left[\cosh(\beta J) I_x^P + \sinh (\beta J) 4 I_x^P I_z^A I_z^B\right]/ Z(\beta,0,0),	\label{eq:targIm}
\end{equation}
where $\Delta$ is used as superscript to indicate that redundant terms are suppressed as mentioned earlier.  For the same reason we ignore the identity term of $\rho_{\text{th}}$ and consider ony the deviation thermal state  $\rho^{\Delta}_{\mathrm{th}}$. We have to realize the target initial state $\rho^{\Delta}(0)$ of Eq.~\Eqref{eq:targIm}  from the deviation thermal state $\rho_\mathrm{th}^\Delta$. In other words, for initialization, the target state of Eq.~\Eqref{eq:targIm} is to be achieved starting from $\rho_\mathrm{th}^\Delta(0)$ with the constrain that the system remains inaccessible. For this purpose, we employ the pulse sequence shown in Fig.~\ref{fig:ImExp}~(a) which operates on probe spin of $^{1}H$ only. Detailed derivation of the pulse sequence is given in Sec.~\ref{AppendC1}. Here, the delay time $\tau$ of the pulse sequence is a free parameter that depends directly on the value of $\beta J \in \mathbb{R}$  of the target state as $\exp(\beta J)=\cos(\pi(J_{PA}+J_{PB})\tau)/\cos(\pi(J_{PA}-J_{PB})\tau)$. This direct dependency  allows us to initiate the system at any value of $\beta J$ in both ferromagnetic ($\beta J \in \mathbb{R}_{+}$) and anti-ferromagnetic ($\beta J \in \mathbb{R}_{-}$) regime just by varying the delay $\tau$ suitably, as discussed in detail in Sec.~\ref{AppendC1}.

\subsection{Initialization on non-origin points of Amoeba Plane}

\begin{figure*}
	\centering
	\includegraphics[width=15cm, trim={12.12cm 0cm 13.9cm 0cm},  clip=true]{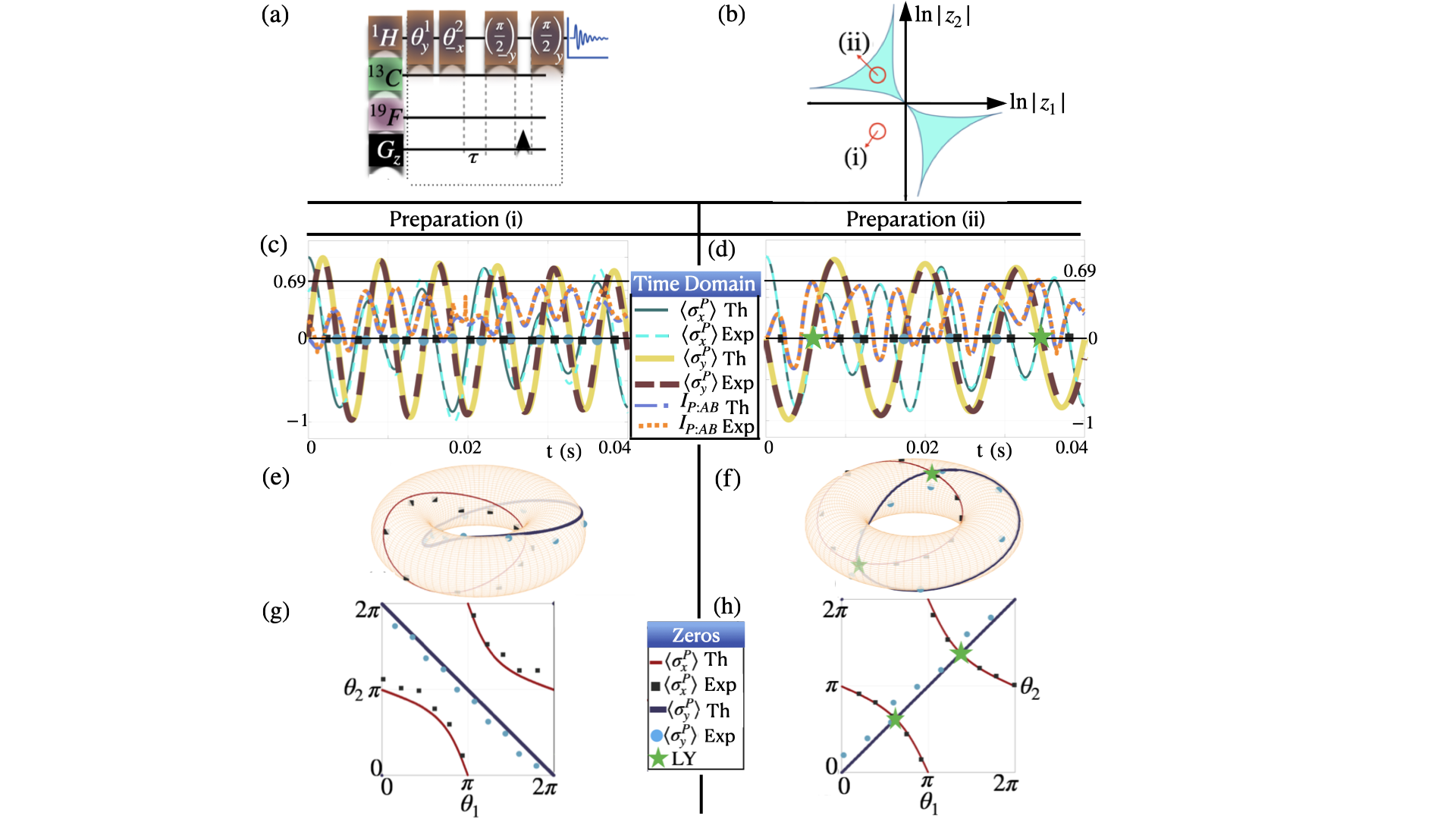} \\
	\caption{
		Pulse sequence (a) for preparing the system in state of Eq.~\Eqref{eq:initial-state} for non zero values of $h_A$ and $h_B$. In particular, the initialization is performed for two distinct points on \emph{amoeba} plane (b) :(i) $h_A=h_B=0.1$ and (ii) $h_A=-h_B=0.1$, considering $\beta J = 0.5$ in both cases.
		(c-d) The real parts (dashed thin lines) and imaginary parts (dashed thick lines) of FID corresponding to $\langle \sigma_x^P(t) \rangle$  and  $\langle \sigma_y^P(t) \rangle$, respectively,  are plotted along with the simulated curves (solid lines, thin and thick correspondingly) for prepared initial state of (i) in (c) and (ii) in (d). Simultaneous null points of real (solid square) and imaginary part (solid circle) of FID correspond to LY zeros, which exists for (ii) (stars) while absent for (i). Theoretical and experimental curves of the mutual information $I_{P:AB}$ between the probe and the system are also plotted, which reach their maxima (horizontal line at $0.69$) only at the LY zeros. (e-f) Theoretically calculated (solid line) and experimentally observed null points of real (solid square) and imaginary (solid circle) part of FID are plotted on the 2-torus for (i)(e) and (ii)(f). The desired \emph{coamoeba} points lie in the intersetion, which can be seen in (f). The corresponding planar visualizations are in (g-h). }
	\label{fig:CompExp}
\end{figure*}

To sample \emph{coamoeba} corresponding to non-origin points on \emph{amoeba}, we need $h_A$ and $h_B$ to take non-zero values. The state of Eq.~\eqref{eq:initial-state} can be directly computed as $\rho(0) =  (\mathbbm{1}/2 + I_x^P) \otimes (\expo{-\beta H_{AB}}/Z(\beta J,h_A,h_B))$, which, upon substituting the value of $\expo{-\beta H_{AB}}$ from Eq.~\Eqref{eq:series} yields 
\begin{equation}
\rho^\Delta (0) = C_a I_x^P + C_b 2I_x^P I_z^A + C_c 2I_x^P I_z^B + C_d 4I_x^P I_z^A I_z^B.  \label{eq:target}
\end{equation}	
Here, we have used `tilde' as before to indicate that only those terms that contribute to the probe's coherence are considered whereas $\mathbbm{1}$ and $I_z^A I_z^B$ are suppressed.
$C_i$'s of Eq.~\Eqref{eq:target} are hyperbolic functions of $\beta h_A$, $\beta h_B$ and $\beta J$ as defined in Eq.~\Eqref{eq:series}. Again, starting from the thermal deviation $\rho_{\text{th}}^\Delta$, target state $\rho^\Delta (0)$ of Eq.~\Eqref{eq:target} can be prepared with only probe control using the pulse sequence shown in Fig.~\ref{fig:CompExp}~(a). The derivation of pulse sequence is given in Sec.~\ref{AppendD1} which explains how the pulse sequence achieves the desired initialization. As there are three variables $\left( \beta J,~\beta h_A,~\beta h_B\right)$  to fix the target state of Eq.~\Eqref{eq:target}, the pulse sequence of Fig.~\ref{fig:CompExp}~(a) is also having three free parameters $\theta^1,~\theta^2$ and $\tau$, which can be set suitably to prepare any desired state. Given a specific target state, how to obtain the correct values of three control parameters $\theta^1,~\theta^2$ and $\tau$ is discussed in Sec.~\ref{AppendD1}.

\section{Experimental Results} \label{sec:Results}

As explained above, LY zeros are extracted from the time points where probe coherence vanishes, which, according to Eq.~\Eqref{eq:Lt}, leads to simultaneous vanishing of $\langle \sigma_{x}^P\rangle$ and $\langle \sigma_{y}^P\rangle$.  As the prepared state $\rho^\Delta (0)$ evolves freely under the internal Hamiltonian $H_I$, we measure the probe expectation values $\langle \sigma_{x}^P(t)\rangle$ and $\langle \sigma_{y}^P(t)\rangle $ in a rotating frame synchronous with the probe's Larmor precession \cite{cavanagh1996protein,levitt2008spin}. In this frame, the effective Hamiltonian becomes
\begin{align}
	H_{\mathrm{eff}} = 2\pi \hbar \left( J_{PA} I_z^{P} I_{z}^{A} + J_{PB} I_z^{P} I_{z}^{B}  \right),  \label{eq:Heff}	
\end{align}
where $A$-$B$ interaction is suppressed since $\rho^\Delta(0)$ of Eq.~\Eqref{eq:targIm} and \Eqref{eq:target} do not evolve under it.  By comparing this Hamiltonian with the interaction Hamiltonian $H_{\mathrm{int}}$ in Eq.~\Eqref{eq:Hint}, we identify $\lambda_A = \pi J_{PA} \hbar /2$ and $\lambda_B = \pi J_{PB} \hbar /2$.  Feeding this to Eq.~\Eqref{eq:coam}, experimentally measured time points corresponding to vanishing probe coherence are directly mapped to the \emph{coamoeba} torus, and thus, physical sampling of \emph{coamoeba} is achieved. 

First, we sample the \emph{coamoeba} corresponding to the origin (circled in Fig.~\ref{fig:ImExp}~(b)) of the \emph{amoeba}'s plane. Following Eq.~\Eqref{eq:am}, we set $h_A = h_B = 0$ and accordingly prepare $\rho^\Delta(0)$ of Eq.~\Eqref{eq:targIm}, which subsequently evolves under $H_\mathrm{eff}$.  As $\langle \sigma^P_y(t) \rangle$ is identically zero in this case (see Sec.~\ref{AppendC1} for calculation),  LY zeros are determined just by measuring $\langle \sigma_x^P (t) \rangle$ and noting the time points where it vanishes. 
An essential advantage of the NMR quantum testbed is, being an ensemble architecture, it directly gives $\langle \sigma^P_{x(y)}  (t)\rangle$ as the real (imaginary) part of the NMR signal known as the free induction decay (FID) \cite{fukushima2018experimental}. 
Hence, after performing the initialization via a probe in ferromagnetic and anti-ferromagnetic regimes, we record their FIDs, whose real parts are shown in Fig.~\ref{fig:ImExp}~(c) and (d), respectively. 
In the figure, time points corresponding to LY zeros are marked with solid squares, which are readily identified as FID null points. No additional data processing is required for this purpose. 
We sampled the corresponding \emph{coamoeba(s)} from collected FID up to $300$ ms (see Sec.~\ref{AppendC2}). However, for clarity, the time domain signal for ferromagnetic and anti-ferromagnetic cases in Fig.~\ref{fig:ImExp}~(c) and (d) are shown only up to $40$ and $60$ ms, respectively. The mutual information $I_{P:AB}$ between the probe and the system is calculated directly from experimental FID and are plotted against their simulated values in Fig.~\ref{fig:ImExp} (c,d). It is shown that the maximum of the mutual information occurs only at the LY zeros as predicted by Eq.~\Eqref{eq:MI}, thus confirming the footprint of complex LY zeros in real-time correlation dynamics between the system and the probe.
The FID null points are mapped to the \emph{coamoeba} by Eq.~\Eqref{eq:coam}. The sampled \emph{coamoeba} for ferromagnetic ($\beta J= 0.5$) and anti-ferromagetic ($\beta J = -0.5$) regimes are shown on the $2$-torus in Fig.~\ref{fig:ImExp}~(e) and (f)  respectively.  Their  equivalent modulo $2\pi$ planar visualizations are shown in Fig. ~\ref{fig:ImExp}~(g) and (h), respectively. We observe a fairly good agreement with the theoretically computed \emph{coamoeba} within the experimental limitations.

It is worth noting that no state tomography is needed for these experiments, and the LY zeros emerge directly from the NMR FID without further data processing. Thus, each experiment taking less than a second yields a dense set of LY zeros.

We now demonstrate the sampling of \emph{coamoeba} for two non-origin points, (i) $\beta h_A = \beta h_B = 0.1$, which is outside \emph{amoeba}, and (ii) $\beta h_A = -\beta h_B = 0.1$, which is inside \emph{amoeba} as shown in Fig.~\ref{fig:CompExp}~(b).  Again we note that $\langle \sigma_x^P (t) \rangle$ and $\langle \sigma_y^P (t) \rangle$ are just real and imaginary components of the probe ($^1H$) FID. Therefore in each case, we initialize the system considering $\beta J=0.5$, let it evolve under $H_\mathrm{eff}$ while recording the FID as shown in Fig.~\ref{fig:CompExp}~(c-d).We extract the time points at which $\langle \sigma_x^P(t) \rangle$ and $\langle \sigma_y^P(t) \rangle$ vanish, and map them to a 2-torus via Eq.~\Eqref{eq:coam}. In case (i), as shown in Fig.~\ref{fig:CompExp}~(c,e,g), the zeros of $\langle \sigma_x^P(t) \rangle$ and $\langle \sigma_y^P(t) \rangle$ do not intersect indicating the absence of L-Y zeros, thereby confirming that the point (i)  does not belong to the \emph{amoeba}.  
However in case (ii), the null points of $\langle \sigma_x^P(t) \rangle$ (real FID) and $\langle \sigma_y^P(t) \rangle$ (imaginary FID) intersect twice as marked by stars in Fig.~\ref{fig:CompExp} (d,f,h), confirming the existence of two distinct \emph{coamoeba} points. The mutual information calculated from the NMR FID is plotted along with its simulated values for case (i) [(ii)] in Fig.~\ref{fig:CompExp} (c) [(d)]. It reaches its maximum twice for case (ii) at times corresponding to simultaneous null points of real and imaginary components of the FID. However, as there are no simultaneous null points of real and imaginary FID in case (i), the mutual information never becomes maximum in this case. These experimental observations confirm the prediction of Eq.~\Eqref{eq:MI} that the correlation between the probe and the system reaches its maximum only at points corresponding to LY zeros. It is worth highlighting that the existence and non-existence of LY zeros for case (i) and (ii) , respectively, can be directly observed just by looking at quadrature NMR FID shown in Fig.~\ref{fig:CompExp}~(c-d) without any data processing.
Again, in both cases, we see a reasonably good agreement between the theoretical predictions and experimental values. This method of high-throughput extraction of LY zeros can be used for efficient sampling of \emph{coamoeba} for a large set of \emph{amoeba} points, thereby determining the \emph{algebraic variety} $V_f$ at any desired precision. 

\section{Relation Between Spin Coherence and Zeros of Bivariate LY Polynomials} \label{appenA}
The initial state of the probe and system is given in Eq.~\Eqref{eq:initial-state} in main text. We evolve this state to get $\rho(t) = \mathcal{U}(t)\rho(0)\mathcal{U}^{\dagger}(t)$, where $\mathcal{U}(t) = \exp{(-\im tH/\hbar)}$ for the Hamiltonian $H$ given in Eq.~\Eqref{eq:Hint}:
\begin{figure*}
	\centering
	\includegraphics[width = 4.1cm]{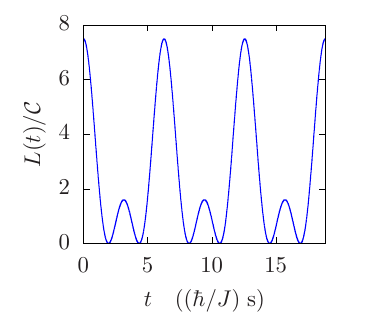} 
	\includegraphics[width = 3.2cm]{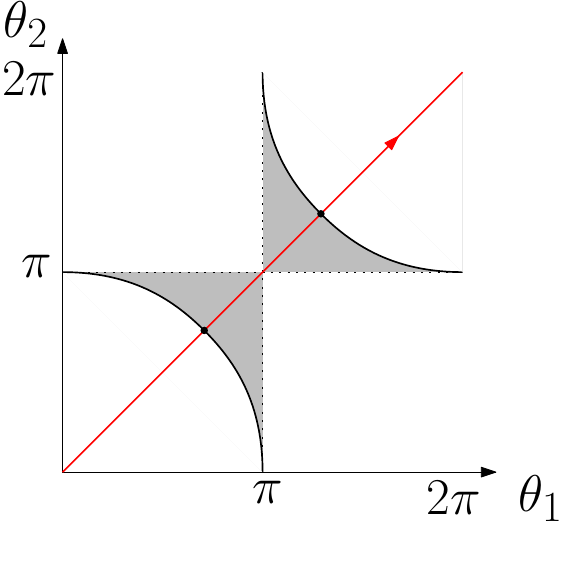}
	\includegraphics[width=4.1cm]{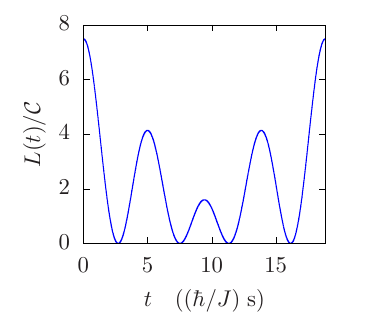} 
	\includegraphics[width=3.2cm]{Images/fig6b.pdf}
	(a) \hspace{3cm} (b) \hspace{3cm} (c) \hspace {3cm} (d) \\  
	$m = 1$  \hspace{6cm} $m = \frac{1}{3}$
	\caption{The sampling of a \emph{coamoeba} for $\lambda_A=J/4$, $\lambda_B=mJ/4$, and $\beta=0.5$, for (a-b) $m=1$, and (c-d) $m=\frac{1}{3}$, plotted for $0\leq t\leq6\pi\frac{\hbar}{J}$. (a) and (c) show the plots of $L(t)$ vs $t$, while (b) and (d) depict the parameterized lines Eq.~\Eqref{eq:coam} in the main text, overlaid with a sketch of the \emph{coamoeba}. The points when $L(t)=0$ correspond to the intersection of the lines with the boundary of the \emph{coamoeba}.}
	\label{fig:rational}
\end{figure*}

\begin{align}
	\rho_P(t) & = \mathrm{tr}_{AB} \left( \mathcal{U}(t) ~ \rho(0) ~ \mathcal{U}^{\dagger}(t) \right)
	\nonumber \\
	& =  \frac{ \langle 0 | \psi_P\rangle \langle \psi_P | 0 \rangle ~ + ~ \langle 1 | \psi_P \rangle \langle \psi_P | 1 \rangle  }{Z(\beta,h_A,h_B)} \nonumber \\
	& + \frac{ \langle 0 | \psi_P \rangle \langle \psi_P | 1 \rangle }{Z(\beta,h_A,h_B)} [ \mathrm{e}^{\beta(h_A+h_B+J)} \mathrm{e}^{-2\im(\lambda_A + \lambda_B)t/\hbar} \nonumber\\
	&+  \mathrm{e}^{\beta(h_A-h_B-J)} \mathrm{e}^{-2\im(\lambda_A - \lambda_B)t/\hbar} \nonumber \\
	&  +\mathrm{e}^{\beta(-h_A+h_B-J)} \mathrm{e}^{-2\im(-\lambda_A + \lambda_B)t/\hbar} \nonumber \\
	&+  e^{\beta(-h_A-h_B+J)} \mathrm{e}^{-2\im(-\lambda_A - \lambda_B)t/\hbar}] + \mathrm{h.c}  \nonumber
\end{align}
From this we get (using the complexified variables $z_1$,~$z_2$ and $\Gamma$ defined in the main text) : 
\begin{align}
	\langle \sigma_x^P (t) \rangle = \frac{ \langle 0 | \psi_P \rangle \langle \psi_P | 1 \rangle }{Z(\beta,h_A,h_B) (\Gamma z_1 z_2)^{\frac{1}{2}}} \nonumber \\
	\left( 1 + \Gamma z_1 + \Gamma z_2 + z_1 z_2 \right) + \mathrm{h.c} \label{eq:sigmax} \\
	\mathrm{and} ~~  \langle \sigma_y^P (t) \rangle = \im\frac{ \langle 0 | \psi_P \rangle \langle \psi_P | 1 \rangle }{Z(\beta,h_A,h_B) (\Gamma z_1 z_2)^{\frac{1}{2}}} \nonumber \\
	\left( 1 + \Gamma z_1 + \Gamma z_2 + z_1 z_2 \right) + \mathrm{h.c} \label{eq:sigmay}. 	
\end{align}	
Therefore, we get the probe coherence $L$ as a function of time as :
\begin{align}
	L(t) = & \frac{\hbar^2}{4} |\langle \sigma_x^P (t) \rangle + \im\langle \sigma_y^P (t) \rangle |^2 \nonumber \\
	& = \frac{\hbar^2 |\langle 1 | \psi_P \rangle \langle \psi_P | 0 \rangle|^2 }{Z^{2}(\beta,h_A,h_B) (\Gamma z_1^{*} z_2^{*})} |1 + \Gamma z_1^* + \Gamma z_2^* + z_1^* z_2^*|^2 \nonumber \\
	& = \mathcal{C} |f(z_1,z_2)|^2,
\end{align}
where, $\mathcal{C} = \frac{\hbar^2 |\langle 1 | \psi_P \rangle \langle \psi_P | 0 \rangle|^2 }{Z^{2}(\beta,h_A,h_B) (\Gamma z_1^{*} z_2^{*})} $, and $f(z_1,z_2) = 1 + \Gamma z_1 + \Gamma z_2 + z_1 z_2$ is a bivariate LY polynomial. Hence the derivation of Eq.~\Eqref{eq:Lt} is complete.
\section{Sampling of Coamoeba: Methodology} \label{AppenB}
 Since, at points where the LY polynomial $|f(z_1,z_2)|^2=0$, the arguments of the variables represent points of the \emph{coamoeba}, Eq.~\Eqref{eq:coam}  are straight lines in the coamoeba plane, with slope $\lambda_B/\lambda_A$, parametrised by $t$. As the time $t$ elapses, a point traverses the coamoeba plane along a straight line at velocity $\{4\lambda_A/\hbar,4\lambda_B/\hbar\}$. The spin coherence $L(t)$ of Eq.~\Eqref{eq:Lt} vanishes whenever $(\theta_1(t),\theta_2(t))$ coincides with the corresponding point(s) on the coamoeba.

\begin{figure}
		\centering
		\includegraphics[width=4.2cm]{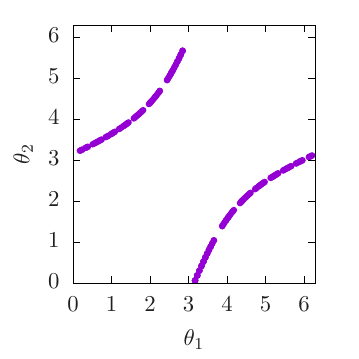}
		\includegraphics[width=4.2cm]{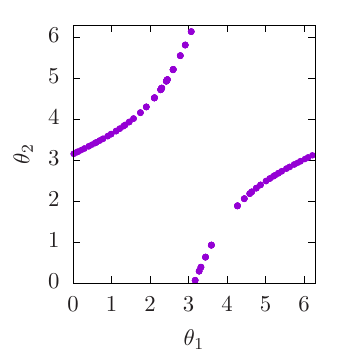} \\
		(a) \hspace{4cm} (b)
		\label{fig_Jnet_msqrt2}
	\caption{Physical sampling of a coamoeba in the anti-ferromagnetic case, $J<0$, where $\lambda_A=-J/4$, $\lambda_B=-mJ/4$, and $\beta=0.5$ for (a) $m = \pi$ and (b) $m = \sqrt{2}$. The points are obtained by recording the times where $L(t)$ vanishes in the interval $0\leq t\leq 500\hbar/|J|$.}
	\label{fig_Jneg}
\end{figure}

As an example, we plot in Fig.~\ref{fig:rational} $L(t)$ vs $t$, along with the parametric lines $\{\theta_1(t),\theta_2(t)\}$ for the case $\lambda_A=J/4$, $\lambda_B=mJ/4$, $\beta=0.5$, and $h_A=h_B=0$;  for the cases $m=1$ and $m=\frac{1}{3}$. We see that $L(t)$ vanishes whenever the line $\{\theta_1(t),\theta_2(t)\}$ intersects the  \emph{coamoeba} boundary. For the case $m=1$ (Fig.~\ref{fig:rational}~(b)), the line only intersects the  \emph{coamoeba} boundary twice before it repeats after a period of $2\pi\frac{\hbar}{J}$. For $m=\frac{1}{3}$ (Fig.~\ref{fig:rational}~(d)), the line starts from the origin as segment \textsf{A} where it intersects the  \emph{coamoeba} once, then continues as segment \textsf{B}, intersecting the  \emph{coamoeba} twice, and finally as segment \textsf{C}, where it intersects the  \emph{coamoeba} one more time before it repeats the segments \textsf{A}, \textsf{B}, \textsf{C} after a period of $6\pi\frac{\hbar}{J}$.
\begin{figure*}
	\centering
	\includegraphics[width=3.5cm]{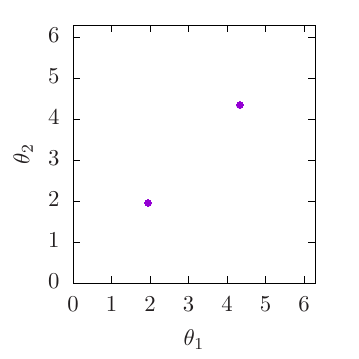}
	\includegraphics[width=3.5cm]{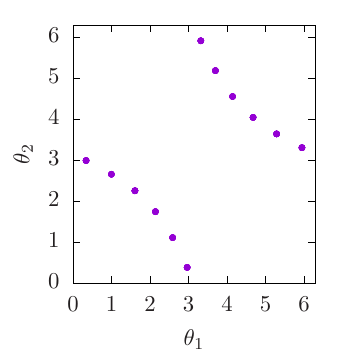}
	\includegraphics[width=3.5cm]{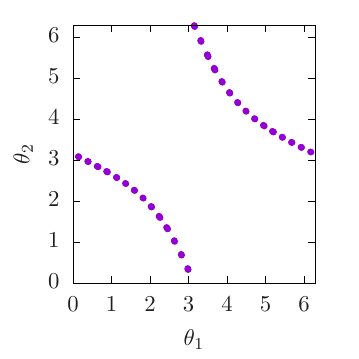}
	\includegraphics[width=3.5cm]{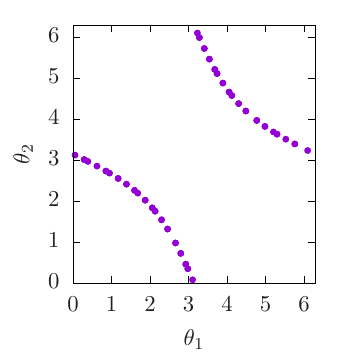}
	(a) \hspace{3cm} (b) \hspace{3cm} (c) \hspace {3cm} (d) \\
	$m=1$ \hspace{2.3cm} $m=\frac{7}{5}$ \hspace{2.3cm} $m=\pi$ \hspace {2.3cm} $m=\sqrt{2}$
	\label{fig_Jpos_msqrt2}
	\caption{Physical sampling of a coamoeba in the ferromagnetic case, $J>0$, with $\lambda_A=J/4$, $\lambda_B=mJ/4$, and $\beta=0.5$ for various $m$. The points are obtained by recording the times where $L(t)$ vanishes in the interval $0\leq t\leq 100\hbar/J$.}
	\label{fig_Jpos}
\end{figure*}
Thus, an experimental procedure to sample a  \emph{coamoeba} can be done as follows. A two-spin system is prepared in a heat bath of temperature $1/\beta$, and coherence $L(t)$ of a quantum probe coupled to the system is measured. We then record the times when $L(t)$ vanishes. This gives a sequence, say, $t_1,t_2,t_3,\ldots$. By Eq.~\Eqref{eq:coam}, this will form a collection of points on the torus. As explained above, these mark the intersection points of the line with the coamoeba. As long as $m=\frac{\lambda_B}{\lambda_A}$ is not rational, (which, in an experimental situation, is most likely the case) we can eventually collect enough points to form the shape of the coamoeba. Figure \ref{fig_Jpos} demonstrates this for the ferromagnetic case $J>0$ with $t$ running in the interval $0\leq t\leq 100\hbar/J$. The anti-ferromagnetic case is shown in Fig.~\ref{fig_Jneg}, where $t$ is $0\leq t\leq 500\hbar/|J|$.
In our case, $m$ is irrational and equals to $J_{PB}/J_{PA}$, whose values are listed in Fig.~\ref{fig:molecule}~(b).

\section{Experiments for \texorpdfstring{$h_A = h_B = 0$}{hA=hB=0} } \label{AppendC}
\subsection{Preparation via Probe :} \label{AppendC1}
By setting $h_A=h_B=0$ in Eq.~\eqref{eq:target}, we get the target state 
\begin{equation}
	\rho^\Delta (0) = \cosh(\beta J) I_x^P + \sinh(\beta J) 4I_x^P I_z^A I_z^B. \label{eq:ImTar}
\end{equation}  
Here we have dropped the proportionality factor of $1/Z(\beta,0,0)$. To prepare this, we start with the NMR thermal deviation state $\rho_\mathrm{th}^{\Delta}(0)$: 
\begin{align}
I_z^P \xrightarrow{(\pi/2)^P_y} I_x^P \xrightarrow{\tau} \cos(\pi J_{PA} \tau) \cos(\pi J_{PB} \tau) I_x^P + \nonumber\\
\cos(\pi J_{PA} \tau) \sin(\pi J_{PB} \tau) 2 I_y^P I_z^B  \nonumber \\
+\sin(\pi J_{PA} \tau)  \cos(\pi J_{PB} \tau) 2I_y^P I_z^A \nonumber\\
- \sin(\pi J_{PA} \tau)\sin(\pi J_{PB} \tau) 4I_x^P I_z^A I_z^B \nonumber \\ 
\downarrow{(\pi/2)^P_y - \mathrm{Grad.} - (\pi/2)^P_{-y}} \nonumber \\
\cos(\pi J_{PA} \tau)\cos(\pi J_{PB} \tau) I_x^P \nonumber\\
 - \sin(\pi J_{PA} \tau)\sin(\pi J_{PB} \tau) 4I_x^P I_z^A I_z^B \label{eq:Imprep}
\end{align} 
Thus the target state is achieved by applying pulses on the probe alone considering the system to be inaccessible. Here, by $\tau$ we meant free evolution under $H_{\mathrm{eff}}$ (given in the main text) and Grad. represents a pulsed filed gradient pulse along the $z$-direction. In particular, the prepared state of Eq.~\Eqref{eq:Imprep} equals the target state of Eq.~\Eqref{eq:ImTar} (apart from trivial proportionality factors) when 
\begin{align}
	\frac{\cos(\pi(J_{PA}+J_{PB})\tau)}{\cos(\pi(J_{PA}-J_{PB})\tau)} = \expo{2\beta J} \label{rq:ImCond}.
\end{align}
Therefore for any value of $\beta J$, we can find the corresponding $\tau$ satisfying the above equation. Setting that delay time $\tau$ in the pulse sequence, the initial state can be readily prepared. For example, we prepare the state for $\beta J = 1/2$ by setting $\tau = 8.7646484$ ms. In this value of $\tau$, the fidelity between Eq.~\Eqref{eq:ImTar} and Eq.~\Eqref{eq:Imprep} becomes $0.99$. On the other hand, to initiate the state for $\beta J = -1/2$, we set $\tau=9.0136719$ ms in which case the respective fidelity is $\approx 1$.
\subsection{Analysis :} \label{AppendC2}
To sample the \emph{coamoeba} we initiate the system as mentioned above both for ferromagnetic ($\beta J = 1/2$) and anti-ferromagnetic ($\beta J = -1/2$) cases. After the initiation, as mentioned in the main text,  we just need to let it evolve under its NMR internal effective Hamiltonian $H_{\mathrm{eff}}$ (form is given in the main text). By direct computation, we get
\begin{align}
\langle \sigma_x^P (t) \rangle  = \mathrm{tr} [\sigma_x^P \expo{-\im H_{\mathrm{eff}}t/\hbar} (	\cosh(\beta J) I_x^P \nonumber\\
+ \sinh(\beta J) 4I_x^P I_z^A I_z^B) \expo{\im H_{\mathrm{eff}}t/\hbar}] \nonumber \\
\propto \left\{ \expo{\beta J} \cos[\pi(J_{PA} + J_{PB})t] + \expo{-\beta J} \cos[\pi(J_{PA} - J_{PB})t]\right\}, \label{eq:Imreal} \\
\mathrm{and}, ~~  \langle \sigma_y^P (t) \rangle  = \mathrm{tr} [\sigma_y^P \expo{-\im H_{\mathrm{eff}}t/\hbar} (	\cosh(\beta J) I_x^P \nonumber\\
+ \sinh(\beta J) 4I_x^P I_z^A I_z^B) \expo{\im H_{\mathrm{eff}}t/\hbar}] = 0. 
\end{align}

\begin{figure}
	\centering
	\includegraphics[width = 10cm, trim={0cm 0cm 23cm 0cm},  clip=true]{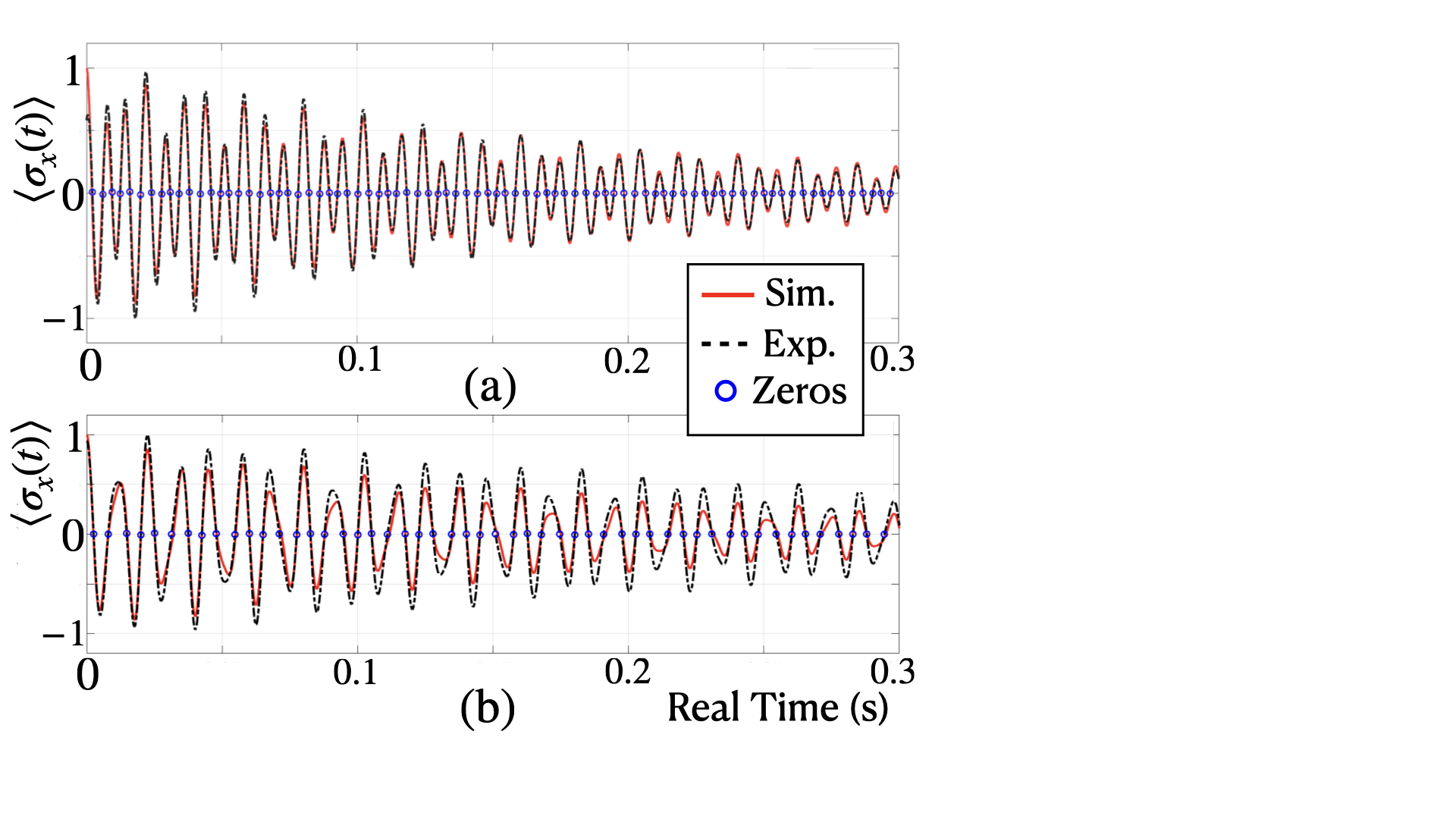} 
	\caption{(a) Simulated and Experimental FID for $\beta J = 1/2$. (b) Same for $\beta J = -1/2$. Zeros of FID are marked with circles} \label{ImFID}
\end{figure}

As $\langle \sigma_y^P (t) \rangle$ is zero throughout,  we just need to find the the time points where $\langle \sigma_x^P (t) \rangle$ vanishes. In Fig.~\ref{ImFID}, we plot the real part of direct NMR FID on top of the predicted FID by analytical expression of Eq~\Eqref{eq:Imreal} and observe they match really well. (To correct the initial phase error due to electronic switching time etc, we have performed a zeroth order phase correction on the data). Experimentally observed time points, where the coherence vanishes, are noted and mapped to the coamoeba torus via Eq.~\Eqref{eq:coam}.

\section{Experiments for \texorpdfstring{$h_A,h_B~\neq~0$}{hA,hB=/=0}} \label{AppendD}
\subsection{Preparation via Probe :} \label{AppendD1}
To initialize the system at any arbitrary non-zero value of $h_A$ and $h_B$, full state mentioned in Eq.~\Eqref{eq:target} becomes our target. Following the sequences of pulses given in Fig~\ref{fig:CompExp}~(a), we achieve this preparation. Here we prove how the target is achieved by the pulse sequence of Fig.~\ref{fig:CompExp}~(a), starting from thermal equilibrium :
\begin{align}
	I_z^P \xrightarrow{\theta^1_y} \cos(\theta^1)I_z^P + \sin(\theta^1) I_x^P 
	\xrightarrow{\theta^2_{-x}} \cos(\theta^1)\cos(\theta^2) I_z^P \nonumber\\
	+  \cos(\theta^1)\sin(\theta^2) I_y^P +  \sin(\theta^1) I_x^P \nonumber \\
	\downarrow{\tau} \nonumber \\
	\cos(\theta^1)\cos(\theta^2) I_z^P + 	\cos(\theta^1)\sin(\theta^2) 
	 [\cos(\pi J_{PA} \tau)\nonumber\\
	 \cos(\pi J_{PB} \tau) I_y^P  - \cos(\pi J_{PA} \tau)\sin(\pi J_{PB} \tau) 2I_x^PI_z^B \nonumber \\
	 - \sin(\pi J_{PA} \tau)\cos(\pi J_{PB} \tau) 2I_x^PI_z^A - \sin(\pi J_{PA} \tau) \nonumber\\ 
	 \sin(\pi J_{PB} \tau) 4I_y^PI_z^AI_z^B] +
	  \sin(\theta^1)[\cos(\pi J_{PA} \tau) \nonumber \\
	  \cos(\pi J_{PB} \tau) I_x^P 
	  + \cos(\pi J_{PA} \tau)\sin(\pi J_{PB} \tau) 2I_y^PI_z^B \nonumber \\
	  + \sin(\pi J_{PA} \tau)\cos(\pi J_{PB} \tau) 2I_y^PI_z^A - \sin(\pi J_{PA} \tau)  \nonumber\\ 
	  \sin(\pi J_{PB} \tau) 4I_x^PI_z^AI_z^B] \nonumber \\
	 \downarrow{(\pi/2)^P_y - \mathrm{Grad.} - (\pi/2)^P_{-y}} 
	 \nonumber \\
	 [\sin(\theta^1)\cos(\pi J_{PA} \tau)\cos(\pi J_{PB} \tau)] I_x^P \nonumber \\
	  +  [-\cos(\theta^1)\sin(\theta^2)\sin(\pi J_{PA} \tau)\cos(\pi J_{PB} \tau)] 2I_x^PI_z^A \nonumber \\
	 [-\cos(\theta^1)\sin(\theta^2)\cos(\pi J_{PA} \tau)\sin(\pi J_{PB} \tau)] 2I_x^PI_z^B \nonumber \\
	  +[-\sin(\theta^1)  \sin(\pi J_{PA} \tau)\sin(\pi J_{PB} \tau)] 4I_x^PI_z^AI_z^B   \label{eq:compPrep}.
\end{align}
Comparing this prepared state with the target state of Eq.~\Eqref{eq:target}, we note that to prepare for a given value of $\{\beta h_A ,\beta h_B ,\beta J\}$, we need to find corresponding values of free parameters $\{\theta^1,\theta^2,\tau\}$ from the bellow four equations :
\begin{subequations}
	\begin{align}
	C_a =   [\sin(\theta^1)\cos(\pi J_{PA} \tau)\cos(\pi J_{PB} \tau)]  \label{opa} \\
	C_b= [-\cos(\theta^1)\sin(\theta^2)\sin(\pi J_{PA} \tau)\cos(\pi J_{PB} \tau)]  \label{opb} \\ 
	C_c=  [-\cos(\theta^1)\sin(\theta^2)\cos(\pi J_{PA} \tau)\sin(\pi J_{PB} \tau)] \label{opc} \\ 
	C_d= [-\sin(\theta^1)  \sin(\pi J_{PA} \tau)\sin(\pi J_{PB} \tau)]  \label{opd}
	\end{align}
\end{subequations}
Therefore it boils down to a three parameter estimation problem, given a set of four equations. In particular, if we denote right hand sides of Eq.~\Eqref{opa},\Eqref{opb},\Eqref{opc}, and \Eqref{opd} as $\{c_a,c_b,c_c,c_d\}$, respectively, then we can define the optimization problem as :
\begin{align}
	\mathrm{Optimize~for} ~ \{\theta^1,\theta^2,\tau\} ~~\mathrm{such~that,} \nonumber \\
	f = |\vec{C}-\vec{c}| = \sum_{i=\{a,b,c,d\}} |C_i - c_i| ~\mathrm{is~minimized.}  
\end{align}        
Any available optimization algorithm can be employed to minimize $f$. For example, we have used genetic algorithm. Of-course, as the system gets larger the optimization problem will become difficult unless the system posses some symmetry. However, our method allows us to reduce this optimization problem into two optimization problems with lesser complexity. To explain how that can be done we compute $\langle \sigma_x^P (t) \rangle$ and  $\langle \sigma_y^P (t) \rangle$ as the state of Eq.~\Eqref{eq:target} evolves under $\mathcal{U}(t) = \exp(-\im H_{\mathrm{eff}}t/\hbar)$:
\begin{subequations}
	\begin{align}
		&\langle \sigma_x^P (t) \rangle = \mathrm{tr}[\mathcal{U}\rho_\Delta(0)\mathcal{U}^{\dagger} \sigma_x] \nonumber \\
		&= C_a \cos(\pi J_{PA}t)\cos(\pi J_{PB}t) - C_d \sin(\pi J_{PA}t)\sin(\pi J_{PB}t) \label{lopf}\\
		&\langle \sigma_y^P (t) \rangle = \mathrm{tr}[\mathcal{U}\rho_\Delta(0)\mathcal{U}^{\dagger} \sigma_y] \nonumber \\
		&= C_b \sin(\pi J_{PA}t)\cos(\pi J_{PB}t) + C_c \cos(\pi J_{PA}t)\sin(\pi J_{PB}t). \label{lops}
	\end{align}
\end{subequations}
From Eq.~\Eqref{lopf} and Eq.~\Eqref{lops}, we note that only $c_a$ and $c_d$ contributes in the measurement of $\langle \sigma_x^P (t) \rangle$ while $c_b$ and $c_c$ term contributes in $\langle \sigma_y^P (t) \rangle$ measurement. This mutual exclusivity can be exploited to prepare Eq.~\Eqref{eq:target} via optimizing for first and last term alone without caring for the other two terms and measuring $\langle \sigma_x^P (t) \rangle$. Then, by the same logic, we optimize only for the second and third term of  Eq.~\Eqref{eq:target} and measure  $\langle \sigma_y^P (t) \rangle$.

As mentioned in the main text, we perform our experiments at two non-zero values of $h_A$ and $h_B$, namely  $\beta J = 1/2, \beta h_A = \beta h_B = 0.1$ and $\beta J = 1/2, \beta h_A = - \beta h_B = 0.1$.  For each of these cases,  we perform two sets of experiments by extracting $\langle \sigma_x^P(t) \rangle$ and $\langle \sigma_y^P(t) \rangle$ separately to demonstrate the before-mentioned optimization-splitting.  
Using the optimized values, pulse sequence of Fig.~\ref{fig:CompExp}~(b) is employed for state initialization. After initialization is achieved, $\langle \sigma_x^P(t) \rangle$ and $\langle \sigma_y^P (t) \rangle$ are recorded as the NMR register evolves freely under $H_{\mathrm{eff}}$ (form is given in the main text). After zeroth order phase correction on the experimental data, interpolated signals are plotted in Fig.~\ref{fig:CompExp}~(g-h). Their zero points (marked by circles) are then mapped to the 2-torus by Eq.~\Eqref{eq:coam}. 

\section{Conclusions}  \label{sec:Conclusion}
For the continued advancement of quantum technologies in the coming years, it is imperative that their applications extend to a broader spectrum of scientific domains by addressing challenges beyond the confines of problems related to quantum physics alone. Following the spirit, we showed a method of using qubits to simulate asymmetrical classical Ising systems at any arbitrary value of its temperature and coupling constant for determining its LY zeros in a wide range of physical situations. Most importantly, in our method, both initialization and determination of LY zeros are achieved through a quantum probe interacting with the system qubits while system qubits themselves left untouched. We believe, this feature of our protocol makes it easier to generalise for more complex systems where controlling system qubits become intractable. A lot of recent works presented a wide range of applications of LY zeros in solving problems across areas like statistical studies  \cite{PhysRevE.97.012115, PhysRevC.72.011901} of equilibrium (phase transition, critical phenomena etc \cite{doi:10.1126/sciadv.abf2447, PhysRevResearch.1.023004}) and non-equilibrium (dynamical phase transitions etc \cite{PhysRevLett.118.180601}) statistical physics, percolation   \cite{arndt2001directed}, complex networks \cite{krasnytska2015violation,krasnytska2016partition} and even protein folding \cite{PhysRevLett.110.248101,PhysRevE.88.022710}. Therefore a method for extracting full \emph{algebraic variety} containing LY zeros of a general asymmetrical classical system has become essential to implement these studies in real situations. We believe, our method of using quantum simulation technique with control over a single qubit alone to do the task is a pioneering step in bringing all those different areas of physics to the sphere of quantum simulation. Experimental validation using a three-qubit NMR register demonstrates the feasibility of our protocol. It is worth noting that this protocol samples the \emph{amoeba} and \emph{coamoeba} for a given LY polynomial and thereby can provoke applications of quantum simulations in the domain of pure mathematics. Apart from applications, this work also uncovers the rich aesthetic structure of the LY zeros by physically sampling the \emph{algebraic variety} that contains them.

\chapter{Direct Experimental Observation of Quantum Mpemba Effect without Bath Engineering}\markboth{Chapter 3:  Experimental Obsewrvation of Mpemba Effect}{}\label{chap_Mpemba}

\begin{center}
``\textit{Somehow, after all, as the universe ebbs toward its final equilibrium in the featureless heat bath of maximum entropy, \\ it manages to create interesting structures}" \\
~~~~~~~~~~~~~~~~~~~~~~~~~~~~~~~~~~~~~~~~~~~ -- James Gleick
\end{center}

\section{Introduction} \label{Mpsec:intro}
\begin{figure}
\centering
\includegraphics[width=12cm,  clip=true, trim={0cm 11.7cm 8.8cm 2cm}]{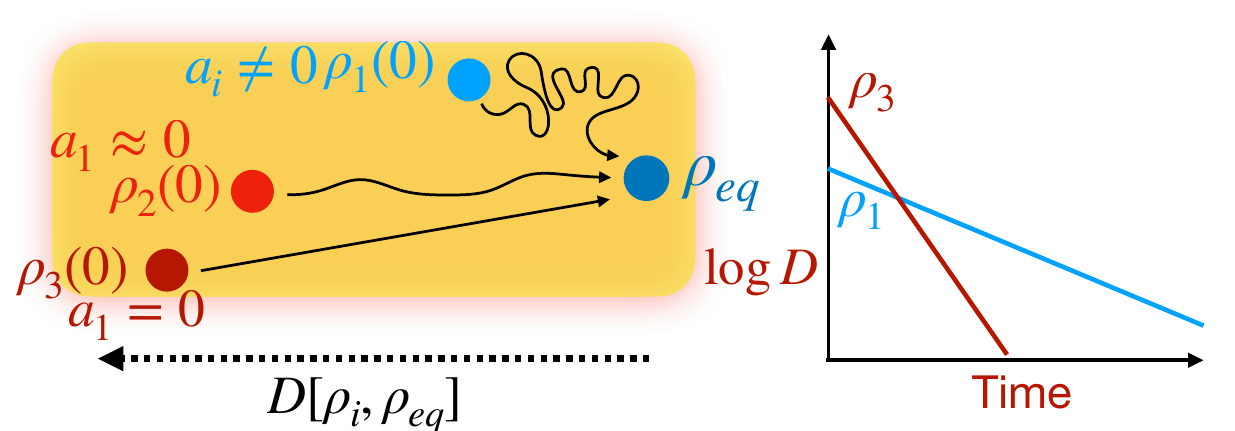}
\caption{(left) The Mpemba effect occurs when a generic state $\rho_1$, overlapping ($a_i$ is the overlap with $i$'th decay mode with $i=1$ being the slowest) with all decay modes, relaxes much slowly than a state that is far away from the steady state $\rho_{eq}$ w.r.t a metric $D$, but have near zero ($\rho_2$) or zero ($\rho_3$) overlap with the slowest decay mode. (right) A typical plot of the observation of Mpemba effect with respect to metric $D$, where the crossing indicates that the far state has overtaken the nearer state and reached the equilibrium faster.} 
\label{fig:grpabs}
\end{figure}

Mpemba effect refers to a situation where an initially hotter system, when brought into contact with a cold bath, reaches the thermal equilibrium more rapidly than an initially cooler system. Even though this rather surprising phenomenon finds its mentions in various texts throughout history  \cite{alma991000934559708966,999697569602121,Bacon1962,bacon1902novum,groves2009now}, the first systematic studies by Mpemba and Osborne \cite{E_B_Mpemba_1969}, as well as Kell \cite{10.1119/1.1975687} in 1969 sparked huge interest in the community, followed by a series of  investigations on a variety of systems \cite{Greaney2011,Ahn2016,PhysRevLett.119.148001,Keller_2018,Hu2018,doi:10.1073/pnas.1819803116,chaddah2010overtakingapproachingequilibrium}. Despite being studied \cite{PhysRevLett.129.138002,Holtzman2022,PhysRevLett.131.017101,PhysRevLett.132.117102,PhysRevLett.124.060602,walker2023mpembaeffecttermsmean,walker2023optimaltransportanomalousthermal,bera2023effectdynamicsanomalousthermal,PhysRevE.109.044149,PhysRevLett.134.107101,teza2025speedupsnonequilibriumthermalrelaxation} over many decades, the very conditions for the occurrence of this effect continues to be questioned \cite{Vynnycky2010,Burridge2016,burridge2020observing,Bechhoefer2021} and lacks uniform agreement. First details-independent uniform understanding of Mpemba effect came  by trying to find its presence when an out-of-equilibrium system approaches equilibrium via Markovian evolution \cite{doi:10.1073/pnas.1701264114}. Considering a distance function $D$, which measures how far the system is from the equilibrium, Mpemba effect can be formally defined as a scenario when the system that is far from equilibrium relaxes quickly than a system starting relatively nearer to equilibrium. At long times, the dynamics of such a system is determined by its slowest and second slowest decay modes. Mpemba effect can show up if the far state has much smaller overlap with the slowest decay mode in comparison to the nearer state. The far state can then effectively relax quickly through the second slowest decay mode, while the near state undergoes a slow relaxation through the slowest decay mode (see Fig.~\ref{fig:grpabs}). A stronger version of it, known as strong Mpemba effect, \cite{PhysRevX.9.021060}  happens when the overlap between the far state and the slowest decay mode becomes exactly zero. This transparent explanation of Mpemba effect in Markovian setting resolved many debates regarding its origin, and also got experimental confirmation \cite{Kumar2020,doi:10.1073/pnas.2118484119}, even though certain observations of Mpemba effect still lies outside the Markovian setting and call for system specific explanations \cite{VYNNYCKY2015243,10.1119/1.18059,ESPOSITO2008757,10.1119/1.2996187,C4CP03669G,Jin2015,https://doi.org/10.1002/crat.2170230702}.    

The understanding of Mpemba effect in Markovian systems prompted investigations regarding its existence in the quantum realm. When a quantum system relaxes to equilibrium by interacting with a sufficiently large and weakly coupled reservoir, its evolution is governed by a GSKL master equation \cite{breuer2002theory,lidar2020lecturenotestheoryopen}, whose solution can be written down as a sum of the decay modes much like the classical case \cite{doi:10.1073/pnas.1701264114}. A generic initial quantum state will have overlap with all the decay modes and hence undergoes a slow relaxation  with the timescale dictated by the slowest decay mode. An exponentially accelerated relaxation \cite{PhysRevLett.127.060401,bao2025acceleratingquantumrelaxationtemporary,PhysRevA.106.012207,PhysRevE.108.014130} can be obtained by applying a suitable unitary operator to the state such that its overlap with the slowest decay mode gets minimized, or become zero (see Fig.~\ref{fig:grpabs}).  In this process, if the action of the unitary also pushes the system much farther from the unique steady state but still it relaxes faster than a nearer state, then it is referred to as the quantum Mpemba effect
\cite{PhysRevLett.127.060401,bao2025acceleratingquantumrelaxationtemporary,PhysRevA.106.012207,PhysRevE.108.014130,PhysRevResearch.3.043108,PhysRevResearch.6.033330,PhysRevLett.133.136302,PhysRevLett.131.080402,PhysRevA.110.022213,PhysRevLett.133.140404,Ares2025,PhysRevA.110.042218,Longhi2025mpembaeffectsuper,PhysRevA.111.022215,zhao2025noiseinducedquantummpembaeffect,chang2024imaginarytimempembaeffectquantum,10.1063/5.0234457,kheirandish2024mpembaeffectquantumoscillating,PhysRevB.111.125404,PhysRevLett.134.220402,FURTADO2025170135}. 
Even though there have been reports of Mpemba like effects beyond Markovian
settings  \cite{PhysRevLett.133.010402,5d6p-8d1b,PhysRevLett.134.220403,PhysRevLett.133.140405,PhysRevLett.133.010401,Caceffo_2024,PhysRevB.111.104312,Chalas_2024}, in this article we will primarily focus on the  Markovian evolution.         

Experimental demonstration of quantum Mpemba effect in markovian setting have recently been achieved using trapped ions \cite{PhysRevLett.133.010403,Zhang2025}, where the relaxation is realized by coupling the system states with one or more metastable states. Since the system relaxes through engineered decay channels in the respective experiments, it does not thermalize in the end. Therefore, corresponding to the classical case, the key question of whether quantum Mpemba effect can occur or not in natural thermalization of quantum systems remains open. In the work reported in this chapter, we provide an affirmative answer to this question by observing quantum Mpemba effect in thermalization of nuclear spins. Considering dipolar relaxation as the dominant decoherence process in nuclear spins, we derive the conditions which can lead to the occurrence of quantum Mpemba effect. In the experiments, we prepare nuclear spin states satisfying those conditions and then allow them to relax without applying any pulses to avoid any external influence to the natural thermalization. The fact that they end up in the thermal state is also experimentally confirmed through tomography. We also experimentally demonstrate that even `genuine quantum Mpemba effect' \cite{PhysRevLett.133.140404}, where a system with higher free energy attains equilibrium faster than the one with lower free energy, can happen naturally during thermalization. Our results establishes that both quantum Mpemba effect, and `genuine quantum Mpemba effect' are  natural phenomena that can occur as quantum system thermalises without the requirement of any bath engineering or specially induced relaxation.       

\section{Thermalization of Nuclear spins} \label{Mpsec:therm}
We consider an ensemble of homonuclear two spin $1/2$ systems in the presence of a strong magnetic field $B_0\,\hat{z}$. The spin Hamiltonian of such systems reads \cite{cavanagh1996protein,levitt2008spin,keeler2011understanding,Wong2014}
\begin{equation}
\frac{1}{\hbar}\mathcal{H} = \left(\omega_0 - \frac{\Delta}{2}\right) I_{1z} +\left(\omega_0 + \frac{\Delta}{2} \right) I_{2z}  + 2\pi J I_{1z} \, I_{2z}, \label{eq:nmr_Ham}
\end{equation}
where  $I_{kl}$ is the spin angular momentum operator of the $k$'th spin along $\hat{l}$ direction. 
In the presence of a  strong field, the Larmor frequency $\omega_0/(2\pi)$ remains the dominant parameter (few hundreds of MHz) while resonance offset ($\Delta/(2\pi)$) and scalar coupling $(J)$ can be treated as perturbations ($\approx 0.1 -  1$ KHz). The sample molecules, with nuclear spin Hamiltonian of Eq.~\Eqref{eq:nmr_Ham}, are dissolved in some  solvent to prepare a dilute isotropic solution, which is kept at temperature $T$. Initially, the nuclear spins rest in thermal equilibrium state 
\begin{gather}
\rho^{th} = \expo{-\mathcal{H}/k_B T}/Z(T) \approx \left[ \mathbbm{1} + 2\epsilon \left( I_{1z} + I_{2z} \right)\right]/4, \label{eq:thstate}
\end{gather}
where $Z(T)=\text{tr}[\exp(-\mathcal{H}/k_B T)]$ is the partition function with Boltzmann constant $k_B$. Since the temperature $T$ is usually very high ($k_BT \gg \omega_0 \hbar$), we only consider up to the first order in $\epsilon = -\omega_0\hbar/(2k_BT)$. The thermal state populations thus differ from maximally mixed state $\mathbbm{1}/4$ by first order in $\epsilon$ : using trace distance \cite{barnett2009quantum,10268381,Nielsen_Chuang_2010} $D(\rho_A,\rho_B) = |\rho_A-\rho_B|/2$, with $|X| = \text{tr}\sqrt{X^{\dagger}X}$, to measure the distance between two states $\rho_A$ and $\rho_B$, we notice $D(\rho^{\text{th}}, \mathbbm{1}/4) \propto \epsilon \approx 10^{-5}$. However, modern day NMR spectrometers are sensitive enough to amplify the signal coming from this small polarization $(\epsilon)$ with sufficient noise suppression
\cite{doi:10.1073,PhysRevA.62.052314}.

When the nuclear spin state is perturbed from equilibrium, it always comes back to the thermal state $\rho^{th}$ of Eq.~\Eqref{eq:thstate} in some characteristic timescale ($T_1$). The irreversible dynamics of this thermalization process can be described by a GSKL master equation \cite{BENGS2020106645,breuer2002theory}. In the interaction frame of $U_0(t)=\exp(i\omega_0(I_{1z}+I_{2z})t)$, the density operator and the Hamiltonian transform as $\rho \rightarrow \widetilde{\rho}(t)=U_0(t) \rho(t)U_{0}^{\dagger}(t)$, and $\mathcal{H}\rightarrow \widetilde{\mathcal{H}} = U_0(t) \mathcal{H}\,U_{0}^{\dagger}(t) - i\,U_0(t)\dot{U_{0}^{\dagger}}(t)$ (the upper dot $\dot{(\,)}$ means partial derivative w.r.t. $t$), respectively. The master equation in this frame takes the following form 
\begin{gather}
\dot{\widetilde{\rho}}(t) = -i\left[\widetilde{\mathcal{H}}, \widetilde{\rho}(t) \right] +  \sum_{m=-2}^{2} \mathcal{K}(m\omega_0)\,\Gamma\left[T_{2m},T^{\dagger}_{2m}  \right] \widetilde{\rho}\,(t), \nonumber \\
\text{where,}~~\Gamma[A,B](\cdot)=A(\cdot)B-\frac{1}{2}\{BA,(\cdot)\},
\label{eq:GSKL}
\end{gather}  
and $\{T_{2m}\}$ for $m\in\{-2,-1,0,1,2\}$ are the rank-$2$ irreducible spherical tensor operators \cite{Kimmich1997,sakurai1986modern} :
\begin{gather}
T_0 = \frac{1}{\sqrt{6}} \left(3I_{1z}\,I_{2z} - \vec{I_1} \cdot \vec{I_2} \right) = T_{0}^{\dagger}, \nonumber \\
T_{\pm1} = \mp \frac{1}{2}
\left(I^{1}_{\pm}I_{2z} + I_{1z}I^{2}_{\pm}  \right) = -T^{\dagger}_{\mp1} \nonumber \\
T_{\pm2}= \frac{1}{2} I^{1}_{\pm}\,I^{2}_{\pm} = T^{\dagger}_{\pm2} \label{eq:spher_tens},    
\end{gather}
where $\vec{I_{n}} = (I_{nx},I_{ny},I_{nz})$ ,  and $I_{\pm}^{n}=I_{nx} \pm I_{ny}$ for the $n$'th spin. We have considered the dipolar relaxation \cite{abragam1961principles,10.1063/1.469808,ernst1990principles,GOLDMAN2001160} here which is usually the most dominating relaxation mechanism in two spin systems. Assuming an exponential decay of two-time bath correlations beyond a characteristic time $\tau_c$, the spectral density function can be written as      
\begin{equation}
\mathcal{K}(x) = K(x)\,\expo{x\epsilon/\omega_0}, ~\text{where}~K(x)=\frac{12\,b^2 \tau_c}{5(1+x^2\tau_{c}^2)}
\end{equation}
is the spectral density computed from semiclassical theory \cite{mr-3-27-2022,abragam1961principles} (with $b$ being the dipolar coupling constant), which is multiplied by the temperature correction \cite{PhysRevLett.4.239,doi:10.1021/jp9919314} to ensure quantum detailed balance \cite{BENGS2020106645,PhysRevE.98.052104,PhysRevE.98.052104}. The formal solution of Eq.~\Eqref{eq:GSKL} reads $\widetilde{\rho}(0)=\exp(\mathcal{L}t)\rho(0)$, where the generator $\mathcal{L}$ is called the Liouvillian \cite{Nakano2011,PhysRevLett.130.230404} and the spectrum of $\mathcal{L}$ ultimately determines the possible occurrence of Mpemba effect. 

For small  molecules dissolved in liquid solvents at ambient temperatures ($T \approx 290-340$ K), we observe extremely fast random diffusion motion causing the bath correlation time to be very small \cite{cavanagh1996protein,levitt2008spin} in comparison to the dynamics under $\mathcal{H}_0$, which is called `extreme narrowing' $\omega_0\,\tau_c \ll 1$. Under this condition, we get $K(m\omega_0)\approx K_0 = 12b^2 \tau_c/5$. This, and the fact that experiments are done at high temperatures ($\mathcal{O}(\epsilon^2)\approx 0$), allow us to write the spectral density function as $\mathcal{K}(m\omega_0)\approx K_0 (1+m\,\epsilon)$.  

\section{Eliminating the effect of dephasing} \label{Mpsec:dephel}
The Zeeman field $B_0$ is usually slightly inhomogeneous along the length of the solution ($\approx 4-6$ cm). As the sample molecules undergo random diffusion motion across that length, their spin states experience a randomly fluctuating magnetic field along $\hat{z}$ and gets quickly decohered. This process, known as inhomogeneous dephasing, generally adds to the dipolar relaxation mentioned in Eq.~\Eqref{eq:GSKL}. As we are interested in investigating quantum Mpemba effect in the natural thermalization of nuclear spins caused by their interaction with the thermal bath, we would like to eliminate the effect of this  inhomogeneous dephasing process.

To do so, we first note that the two-spin density operator can be expanded in a basis $\{\xi_i\}_{i=1}^{16}$, which breaks up into three blocks $\{\xi_i\} = \{\xi_i^{(0)}\}_{i=1}^{6}\,\cup\,\{\xi_i^{(1)}\}_{i=1}^{8}\,\cup \{\xi_i^{(2)}\}_{i=1}^{2}$ such that the elements of $\{\xi_i^{(m)}\}$ form the $m$-quantum block satisfying the commutation relation $[\xi_{i}^{(m)},I_{1z}+I_{2z}]=m \, \xi_{i}^{(m)}$. The elements of density operators in the 1- and 2-quantum blocks suffer maximum inhomogeneous dephasing as they pick up larger random phases under a fluctuating magnetic filed  along $\hat{z}$ in a given interval of time. However, the zero-quantum block (ZQB) mostly contains population terms ($\{\proj{00},\proj{01},\proj{10},\proj{11}\}$) which do not evolve under $\hat{z}$ field since $\Ket{0(1)}$ is the eigenstate of $\sigma_z$ with eigenvalue $1(-1)$. There are only two coherence terms ($\outpr{01}{10}, \outpr{10}{01}$) present in the ZQB which gets affected by the inhomogeneous dephasing. As the Liouvillian of Eq.~\Eqref{eq:GSKL} also admits this block diagonal structure $\mathcal{L}= 
\mathcal{L}_{0}\,\oplus \mathcal{L}_{1} \oplus \mathcal{L}_2$, where $\mathcal{L}_{i}$ is $\mathcal{L}$ projected in the $i$'th quantum block for $i=0,1$ or $2$, desnity matrix elements from different blocks do not get mixed up in the thermalization process. This remarkable symmetry, known as the Redfield kite structure \cite{abragam1961principles,cavanagh1996protein}, allows us to restrict the dynamics only in the ZQB so that the subsequent relaxation happens solely due to the natural thermalization of the nuclear spins, provided we ensure the coherence terms of ZQB to remain zero throughout.

\begin{figure}
\centering
\includegraphics[width=9.1cm, clip=true, trim={0cm 2.1cm 9.5cm 2.2cm}]{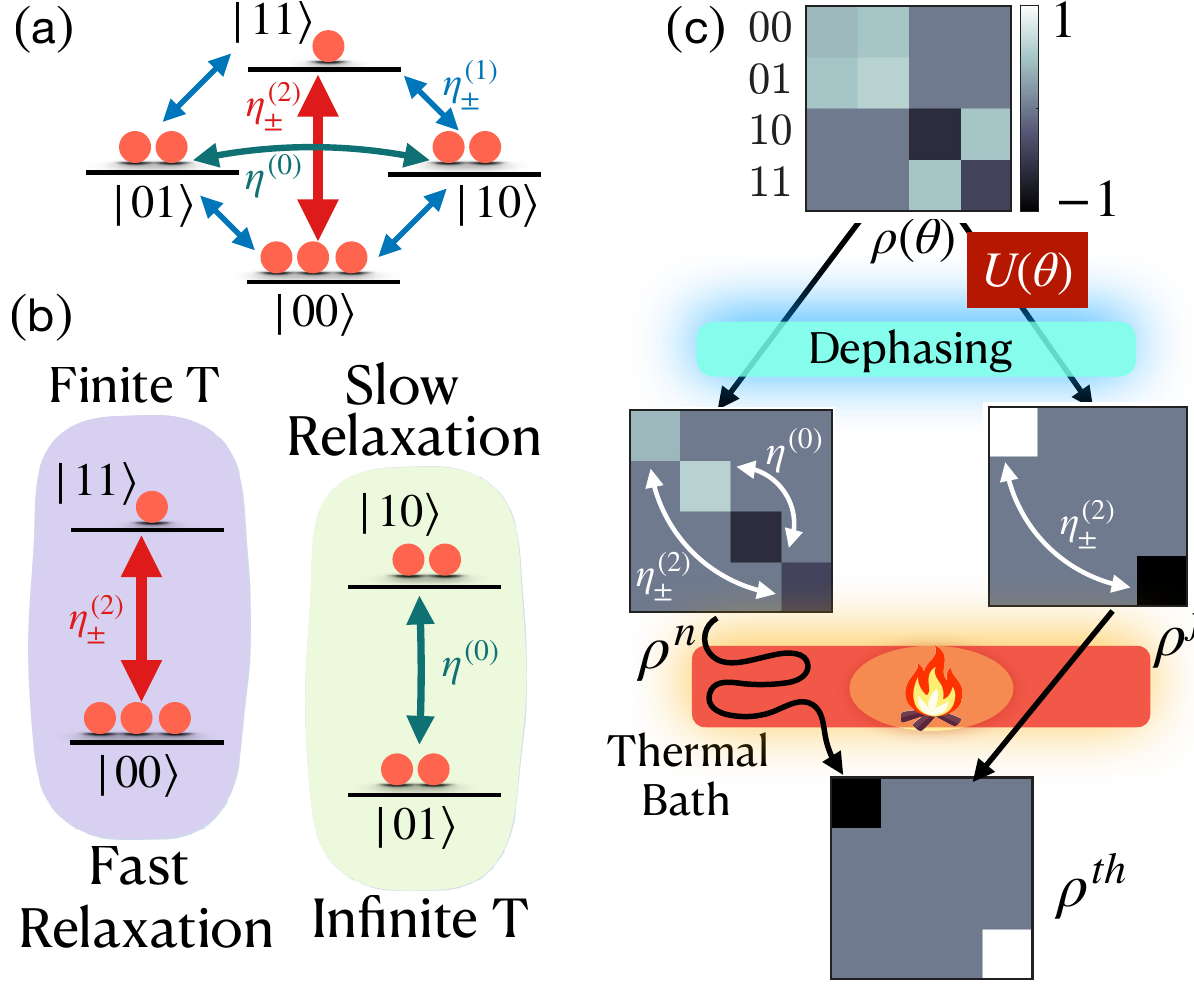} 
\caption{(a) The energy eigenstates of the spin Hamiltonian $\mathcal{H}(t)$ (considering $\omega_0$ to be negative), with various transitions induced by $\mathcal{L}_p$. (b) For particular choices of initial states, the dynamics under $\mathcal{L}_p$ can be interpreted as thermalization of two independent qubits. (c) A schematic view of the experiments that observe quantum Mpemba effect.} 
\label{fig:thseq}
\end{figure}

At infinite temperature ($\epsilon=0$), it can be shown (see Sec.~\ref{Mpsec:append} for details) that if the initial state $\rho(0)$ contains only population terms satisfying 
\begin{equation}
p_{00}+p_{11} = p_{01} + p_{10},
\label{eq:incond}
\end{equation}
then the coherence terms remain zero in the subsequent dynamics, where $p_{ab}$ is the coefficient of the $\proj{ab}$ term in $\rho(0)$. However, since we are working at high but finite temperature,  we have to consider up to first order of $\epsilon$, as in Eq.~\Eqref{eq:thstate}. Interestingly, if we start with only population terms respecting Eq.~\Eqref{eq:incond}, the coherence terms generated in the evolution will be of the order of $(K_0/\Delta)\epsilon$ (see Sec.~\ref{Mpsec:append} for details), which is a very small number and can thus be neglected as long as $K_0 \ll \Delta$. For such initial states, which can be conveniently represented as a vector $\vec{p}(0):=(p_{00},p_{01},p_{10},p_{11})$, the time evolution can be described as (see Sec. \ref{Mpsec:append} for details)
\begin{equation}
\partial_t \vec{p}(t)=\mathcal{L}_p\, \vec{p}(t) \implies \vec{p}(t) = \vec{p}_{th} +  \sum_{n=1}^{3}a_n\, \expo{\lambda_nt}\, \vec{v}_n, \label{eq:restrict}
\end{equation}
where the overlaps $a_n=\vec{w}_n\cdot \vec{p}(0)$, the population at thermal state is $\vec{p}_{th}=(1+2\epsilon,1,1,1-2\epsilon)/4$, and the decay rates are $\{\lambda_1,\lambda_2,\lambda_3\}=-(K_0/24)\{5,6,15\}$, with $\vec{v}_n(\vec{w}_n)$ being the right (left) eigenvector of $\mathcal{L}_p$ w.r.t decay rate $\lambda_n$. The transitions between the population terms $\proj{ab}\rightarrow\proj{cd}$ for $a,b,c,d \in \{0,1\}$ induced by $\mathcal{L}_p$ can be divided into three categories depending on $m=(c+d)-(a+b)$ (see Fig.~\ref{fig:thseq}~(a)). For $m=\pm1$, we get single quantum transitions with decay rates $\eta_{\pm}^{(1)}=K_0(1\mp \epsilon)/16$, for $m=\pm2$ we get double quantum transitions with decay rates $\eta_{\pm}^{(2)}=K_0(1\mp 2\epsilon)/4$, and for $m=0$ we get zero quantum transitions with decay rate $\eta^{(0)}=K_0/24$. Note that the transition probabilities obey quantum detailed balance, where temperature corrections are considered up to the first order in $\epsilon$. 

Interestingly, if we choose initial states such that the dynamics happens only through $\eta_{\pm}^{(2)}$ and $\eta^{(0)}$, then the thermalization process of two spins can be described as independent thermalization of two qubits (see Fig.~\ref{fig:thseq}~(b)), with the first one (composed of states $\Ket{00}$ and $\Ket{11}$) is in contact with a bath at temperature $T$ while the second one (composed of states $\Ket{01}$ and $\Ket{10}$) remains in contact with an infinite temperature bath. Since $\eta_{\pm}^{(2)} > \eta^{(0)}$, a state which can go to thermal equilibrium only through the double quantum transitions will relax much faster than a state which requires single or zero quantum transitions to thermalize. This observation will help us to find suitable nuclear spin states which can show quantum Mpemba effect.       

\section{Observing Quantum Mpemba Effect} 
To experimentally investigate quantum Mpemba effect, we consider the two spin-1/2 $^{1}$H nuclei of  $2$-Chloroacrylonitrile (CAN), dissolved in Dimethyl Sulfoxide (DMSO) ($20~\mu$L of CAN in $600~\mu$L of DMSO), as our two spin system, which sits in a Zeeman field of $B_0 = 11.7$ T at an ambient temperature of $T=295$ K inside a $500$ MHz Bruker NMR spectrometer. The spin Hamiltonian and the equilibrium thermal state of the system is described by Eqs.~\ref{eq:nmr_Ham} and \ref{eq:thstate}, respectively, with $\Delta/2\pi= 89$ Hz, $J=3.24$ Hz, and $\omega_0/2\pi = 500.02$ MHz. 
We consider the state 
\begin{gather}
\rho(\theta)= \mathbbm{1}/4 + 2\epsilon[\cos^2 \theta (I_{1z}+I_{2z})+ \nonumber \\
\sin^2{\theta}(I_{1z}-I_{2z})+  \sin(2\theta)I_{2y}]/4,~\text{for}~\theta \in (0,\pi/2), \label{eq:intern_state}
\end{gather}
to study the relaxation dynamics. The experimental procedure is sketched in Fig.~\ref{fig:thseq}~(c)(see Appendix B for details). After preparation, the state of Eq.~\Eqref{eq:intern_state} is subjected to a strong inhomogeneous magnetic field using a pulsed field gradient (PFG), which eliminates the terms of the density matrix that are susceptible to inhomogeneous dephasing and ensures the subsequent evolution to happen solely under natural thermalization as described by Eq.~\Eqref{eq:GSKL}. The state of Eq.~\Eqref{eq:intern_state} gets dephased by the PFG and becomes   $\rho^{n}(\theta) = \mathbbm{1}/4 + 2\epsilon (I_{1z} + I_{2z} \cos{2\theta})/4$, which is then allowed to thermalize naturally. Note that $\rho^{n}(\theta)$ satisfies Eq.~\Eqref{eq:incond}, and hence it evolves under $\mathcal{L}_p$ according to Eq.~\Eqref{eq:restrict}.  As shown in Fig.~\ref{fig:thseq}~(c), it requires the fast ($\eta_{\pm}^{(2)}$) as well as the slow ($\eta^{(0)}$) transition for $\rho^{n}(\theta)$ to reach the thermal equilibrium. The trace distance from $\rho^{n}(\theta)$ to the equilibrium thermal state $\rho^{th}$ reads $D(\rho^{n}(\theta),\rho^{th}) = \epsilon |1-\cos{2\,\theta}|$. However, by applying an unitary operator $U(\theta)$ on $\rho(\theta)$, we  push it further away from equilibrium, which passes through the dephasor and becomes  $\rho^{f} = \mathbbm{1}/4 - \epsilon(I_{1z}+I_{2z})/2$  such that $D(\rho^{f},\rho^{th}) = \epsilon > D(\rho^{n}(\theta),\rho^{th})$. The transformed state $\rho^f$ could thermalize only via the fast $\eta_{\pm}^{(2)}$ transition. In other words, $\rho^{f}$ possesses zero overlap with the slowest decay mode $\lambda_1$ of $\mathcal{L}_p$, and thus, despite of being much further away, should reach equilibrium in shorter time than $\rho^{n}$.       

\begin{figure}
\centering
\includegraphics[width=11cm, clip=true, trim={0.7cm 2.1cm 13.7cm 2.4cm}]{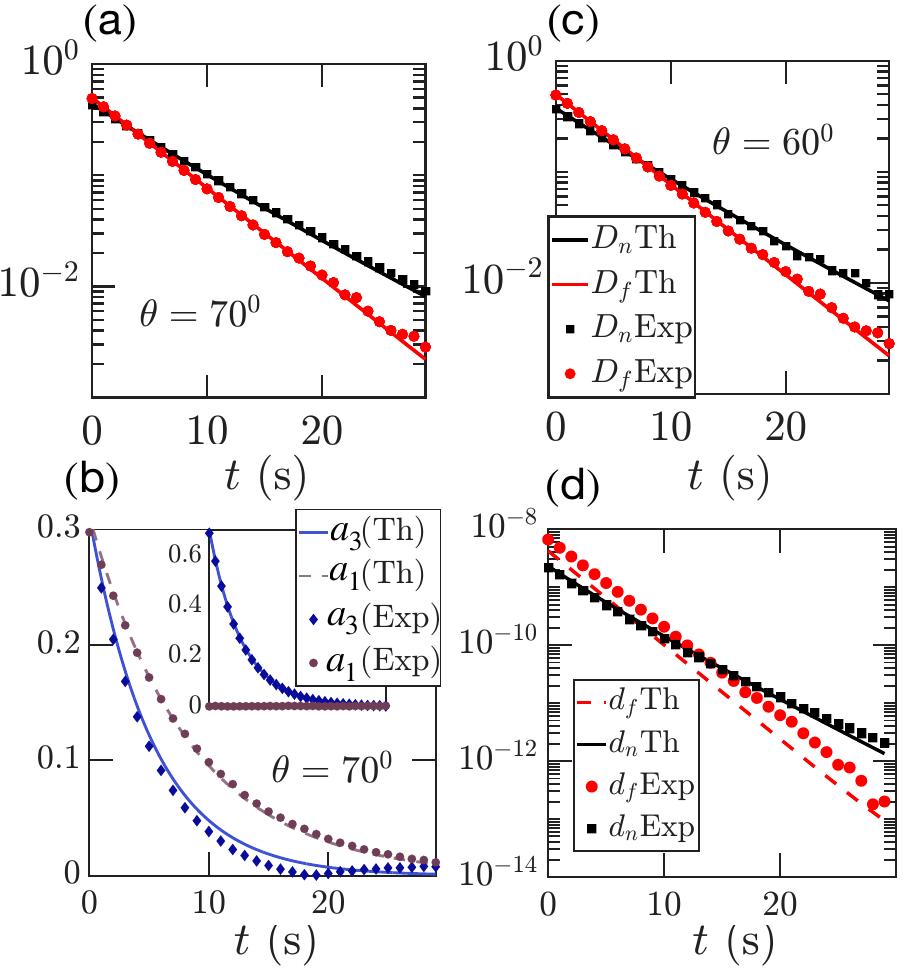} 
\caption{(a,c) Occurrence of Quantum Mpemba Effect where the far state $\rho^f$ relaxes much faster than the nearer state $\rho^n$ for different values of $\theta$. As time elapses, the distance of $\rho^f$ and $\rho^n$ from thermal state $\rho^{th}$ is measured by the trace distance $D_n$ and $D_f$, respectively. (b) $\rho^n$ overlaps with both decay modes $\lambda_1$ and $\lambda_3$ while (inset) $\rho^f$ overlaps only with the fast decay mode $\lambda_3$, the data(s) are for experiment at $\theta=70^0$. The y axis is scaled by multiplying with $1/(2\epsilon)$ in (a-c). (d) Experimental demonstration of `genuine quantum Mpemba effect' where an initial state $\rho^f$ with higher relative entropy (w.r.t $\rho^{th}$), as measured by $d_f$ attains equilibrium faster than a state $\rho^n$ with lower relative entropy, as measured by $d_n$. The experimental errors lie well within the marker size used. The theoretical predictions are made at $\tau_c=2.1$ ps and $b=5.903$ KHz.} 
\label{fig:exp}
\end{figure}

The experimental results are shown in Fig~\ref{fig:exp} (a) and (c), which confirms the theoretical prediction and demonstrates the quantum Mpemba effect in natural thermalization of nuclear spins. We also experimentally measure the overlap of each state $\rho^f$ and $\rho^n$ with the respective decay modes. As shown in Fig.~\ref{fig:exp}~(b), the state $\rho^n$ overlaps with both fast ($\lambda_3$) and slow ($\lambda_1$) decay modes and hence thermalises slowly, while $\rho^f$ only overlaps with the fast decay mode  (Fig.~\ref{fig:exp}~(b) inset) and hence goes to thermal equilibrium much faster. Thus we experimentally verify the cause behind the existence of observable quantum Mpemba effect in nuclear spins to be the dominance of $\eta_{\pm}^{(2)}$ transitions over $\eta^{(0)}$ transition under dipolar relaxation, as predicted by the theory. After the preparation of $\rho^f$ and $\rho^n$, the nuclear spins are left to evolve without applying any pulse up to $29$ seconds, which extends well beyond their $T_1 = 5.7$ s to ensure the observed Mpemba comes from their natural thermalization process without any bath engineering. Within the limitations of experiments on an open quantum system, we observe remarkable agreement between theoretical prediction and experimental data.

\section{Observing Genuine Mpemba Effect}
It has been proposed recently \cite{PhysRevLett.133.140404} that in order to observe `genuine quantum Mpemba effect' from a thermodynamic perspective, one should use the non equilibrium free energy, which equals the relative entropy $d(\rho_1,\rho_2)=-\text{tr}[\rho_1(\log \rho_1 - \log \rho_2)]$ up to an additive constant, as a metric.  We identify the favorable states for observing the genuine  Mpemba effect to be
$\rho^n = \mathbbm{1}/4 + \epsilon(I_{1z}-I_{2z})/2$ and $\rho^f = \mathbbm{1}/4 - \epsilon(I_{1z}+I_{2z})/2$.  We first prepare the state $\rho^n$, which overlaps with both decay modes $\lambda_1$ and $\lambda_3$. As shown in Fig.~\ref{fig:thseq}~(c), after passing it through a strong dephasor, we let the state to thermalize naturally while measuring $d(\rho^n,\rho^{th})$ with time. The state $\rho^n$ is converted to $\rho^f$ by applying a $\pi$-pulse on the first spin, which overlaps only with the fastest decay mode $\lambda_3$. Even though $\rho^f$ contains higher free energy than $\rho^n$, as shown in Fig~\ref{fig:exp}~(d), under natural thermalization it attains the free energy value of the equilibrium state much faster than $\rho^n$. Thus we  experimentally demonstrate the existence of genuine quantum Mpemba effect in the natural thermalization of nuclear spins.

\section{Dipolar Relaxation at High Temperatures} \label{Mpsec:append}
In the ZQB, the coefficients of the population terms $\proj{ab}$ are $p_{ab}$, and the coefficient of the coherence term $\outpr{01}{10}$ is $c$. The GSKL equation can be written now in the ZQB block as  
\begin{align}
    \partial_t 
\begin{pmatrix}
p_{00} \\
p_{01} \\
p_{10} \\
p_{11} \\
c \\
c^*
\end{pmatrix}=\mathcal{L}_0
\begin{pmatrix}
p_{00} \\
p_{01} \\
p_{10} \\
p_{11} \\
c \\
c^*
\end{pmatrix}
\end{align}
with 
\begin{align}
\mathcal{L}_0 =
\begin{pmatrix}
A & B \\
C & D
\end{pmatrix},
\end{align}
where

\begin{equation}
A =   \nonumber
\end{equation}
\begin{equation}
\scalebox{0.9}{$
\begin{pmatrix}
-\tfrac{1}{8} K_0 (1 - \epsilon) - \tfrac{1}{4} K_0 (1 - 2 \epsilon) & 
\tfrac{1}{16} K_0 (1 + \epsilon) & 
\tfrac{1}{16} K_0 (1 + \epsilon) \\
\tfrac{1}{16} K_0 (1 - \epsilon) & 
-\tfrac{K_0}{24} - \tfrac{1}{16} K_0 (1 + \epsilon) - \tfrac{1}{16} K_0 (1 - \epsilon) & 
\tfrac{K_0}{24} \\
\tfrac{1}{16} K_0 (1 - \epsilon) & 
\tfrac{K_0}{24} & 
-\tfrac{K_0}{24} - \tfrac{1}{16} K_0 (1 + \epsilon) - \tfrac{1}{16} K_0 (1 - \epsilon)
\end{pmatrix}, \nonumber
$}
\end{equation}
\begin{equation}
B = 
\begin{pmatrix}
\tfrac{1}{4} K_0 (1 + 2 \epsilon) & 
\tfrac{1}{16} K_0 (1 + \epsilon) & 
\tfrac{1}{16} K_0 (1 + \epsilon) \\
\tfrac{1}{16} K_0 (1 + \epsilon) & 
-\tfrac{1}{32} K_0 (1 + \epsilon) - \tfrac{1}{32} K_0 (1 - \epsilon) & 
-\tfrac{1}{32} K_0 (1 + \epsilon) - \tfrac{1}{32} K_0 (1 - \epsilon) \\
\tfrac{1}{16} K_0 (1 + \epsilon) & 
-\tfrac{1}{32} K_0 (1 + \epsilon) - \tfrac{1}{32} K_0 (1 - \epsilon) & 
-\tfrac{1}{32} K_0 (1 + \epsilon) - \tfrac{1}{32} K_0 (1 - \epsilon)
\end{pmatrix}, \nonumber
\end{equation}
\begin{equation}
C =
\begin{pmatrix}
\tfrac{1}{4} K_0 (1 - 2 \epsilon) & 
\tfrac{1}{16} K_0 (1 - \epsilon) & 
\tfrac{1}{16} K_0 (1 - \epsilon) \\
\tfrac{1}{16} K_0 (1 - \epsilon) & 
-\tfrac{1}{32} K_0 (1 + \epsilon) - \tfrac{1}{32} K_0 (1 - \epsilon) & 
-\tfrac{1}{32} K_0 (1 + \epsilon) - \tfrac{1}{32} K_0 (1 - \epsilon) \\
\tfrac{1}{16} K_0 (1 - \epsilon) & 
-\tfrac{1}{32} K_0 (1 + \epsilon) - \tfrac{1}{32} K_0 (1 - \epsilon) & 
-\tfrac{1}{32} K_0 (1 + \epsilon) - \tfrac{1}{32} K_0 (1 - \epsilon)
\end{pmatrix}, \nonumber
\end{equation}
\begin{equation}
D = \nonumber
\end{equation}
\begin{equation}
\scalebox{0.87}{$
\begin{pmatrix}
-\tfrac{1}{4} K_0 (1 + 2 \epsilon) - \tfrac{1}{8} K_0 (1 + \epsilon) & 
\tfrac{1}{16} K_0 (1 - \epsilon) & 
\tfrac{1}{16} K_0 (1 - \epsilon) \\
\tfrac{1}{16} K_0 (1 + \epsilon) & 
i \Delta - \tfrac{K_0}{24} - \tfrac{1}{16} K_0 (1 + \epsilon) - \tfrac{1}{16} K_0 (1 - \epsilon) & 
\tfrac{K_0}{24} \\
\tfrac{1}{16} K_0 (1 + \epsilon) & 
\tfrac{K_0}{24} & 
- i \Delta - \tfrac{K_0}{24} - \tfrac{1}{16} K_0 (1 + \epsilon) - \tfrac{1}{16} K_0 (1 - \epsilon)
\end{pmatrix}.  \nonumber
$}
\end{equation}

Now we can easily show that at infinite temperature ($\epsilon=0$), the population and coherence satisfy the following equation 
\begin{align}
\frac{d}{dt}
\begin{pmatrix}
X_{1}(t) \\
X_{2}(t) \\
X_{3}(t)
\end{pmatrix}
=
\begin{pmatrix}
-\tfrac{K_{0}}{4} & \tfrac{K_{0}}{4} & 0 \\
\tfrac{K_{0}}{8} & -\tfrac{K_{0}}{8} & i \Delta \\
0 & i \Delta & -\tfrac{5K_{0}}{24}
\end{pmatrix}
\begin{pmatrix}
X_{1}(t) \\
X_{2}(t) \\
X_{3}(t)
\end{pmatrix}\;,
\end{align}
where $ X_{1}=p_{00}+p_{11}-(p_{01}+p_{10}), X_{2}=c+c^*$ and $X_{3}=c-c^*$.
If the initial density matrix contains only population terms satisfying Eq.~\Eqref{eq:incond} of the main text i.e., $X_{1}(0)=X_{2}(0)=X_{3}(0)=0$, then the coherence terms remain zero throughout the subsequent dynamics. This further reduces the space from $6 \times 6$  to $4 \times 4$. However, since we are working at high but finite temperature, we have to consider up to first order of $\epsilon$. Considering that we can explicitly show that the order of coherence that the system generates will be of the order $ (K_0/\Delta) \epsilon$. This is very a small number as long as $K_0 \ll \Delta$, and so we can neglect this. Therefore, we can safely say even at finite but high temperatures, for initial states containing only population terms satisfying Eq.~\Eqref{eq:incond}, the population and dynamics remain decoupled from the coherence terms :  
\begin{align}
\partial_t 
\begin{pmatrix}
p_{00} \\
p_{01} \\
p_{10} \\
p_{11} 
\end{pmatrix}=\mathcal{L}_p
\begin{pmatrix}
p_{00} \\
p_{01} \\
p_{10} \\
p_{11} 
\end{pmatrix}
\end{align}
with

\begin{equation}
\mathcal{L}_p =   \nonumber
\end{equation}
\begin{equation}
\scalebox{0.7}{$
\begin{bmatrix}
-\tfrac{1}{8} K_0 (1 - \epsilon) - \tfrac{1}{4} K_0 (1 - 2 \epsilon) & 
\tfrac{1}{16} K_0 (1 + \epsilon) & 
\tfrac{1}{16} K_0 (1 + \epsilon) & 
\tfrac{1}{4} K_0 (1 + 2 \epsilon) \\
\tfrac{1}{16} K_0 (1 - \epsilon) & 
-\tfrac{K_0}{24} - \tfrac{1}{16} K_0 (1 + \epsilon) - \tfrac{1}{16} K_0 (1 - \epsilon) & 
\tfrac{K_0}{24} & 
\tfrac{1}{16} K_0 (1 + \epsilon) \\
\tfrac{1}{16} K_0 (1 - \epsilon) & 
\tfrac{K_0}{24} & 
-\tfrac{K_0}{24} - \tfrac{1}{16} K_0 (1 + \epsilon) - \tfrac{1}{16} K_0 (1 - \epsilon) & 
\tfrac{1}{16} K_0 (1 + \epsilon) \\
\tfrac{1}{4} K_0 (1 - 2 \epsilon) & 
\tfrac{1}{16} K_0 (1 - \epsilon) & 
\tfrac{1}{16} K_0 (1 - \epsilon) & 
-\tfrac{1}{4} K_0 (1 + 2 \epsilon) - \tfrac{1}{8} K_0 (1 + \epsilon)
\end{bmatrix} \nonumber
$} 
\end{equation}

We can find the eigenvalues and right eigenvectors of $\mathcal{L}_p$, which are given by
\begin{align*}
v_{0}&= \vec{p}_{th} = \frac{1}{4}
(\, 1 + 2\epsilon,\, 1,\, 1,\, 1 - 2\epsilon \,),~ \lambda_{0}=0;\\
v_{1} &= 
\frac{1}{\sqrt{2}} 
(\, 0,\; -1,\; 1,\; 0 \,),  ~\lambda_{1}=-\frac{5 K_0}{24};\\
v_{2} &= 
\frac{1}{\sqrt{\,4 + \tfrac{8}{3}\epsilon(5 + 4\epsilon)}} 
(\, 1 + \tfrac{16}{3}\epsilon,\; -1 - \tfrac{8}{3}\epsilon,\; -1 - \tfrac{8}{3}\epsilon,\; 1 \,),  \\ &\lambda_{2}=-\frac{K_0}{4};\\
v_{3} &= 
\frac{1}{\sqrt{\tfrac{2}{3}(1-\epsilon)(3-5\epsilon)}} 
(\, -1 + \tfrac{2}{3}\epsilon,\; -\tfrac{\epsilon}{3},\; -\tfrac{\epsilon}{3},\; 1 \,), \\ &\lambda_{3}=-\frac{5 K_0}{8}.
\end{align*}
Similarly, the left eigenvectors of $\mathcal{L}_p$ are given by
\begin{align*}
w_{0} &= 
(\, 1,\; 1,\; 1,\; 1 \,), \\
w_{1} &= 
\frac{1}{\sqrt{2}} 
(\, 0,\; -1,\; 1,\; 0 \,), \\
w_{2} &= 
\frac{1}{\sqrt{\,4 + \tfrac{8}{3}\epsilon(5 + 4\epsilon)}} 
(\, 1 + \tfrac{4}{3}\epsilon,\; -1 - \tfrac{2}{3}\epsilon,\; -1 - \tfrac{2}{3}\epsilon,\; 1 \,), \\
w_{3} &= 
\frac{1}{\sqrt{\tfrac{2}{3}(1-\epsilon)(3-5\epsilon)}} 
(\, -1 + \tfrac{14}{3}\epsilon,\; -\tfrac{\epsilon}{3},\; -\tfrac{\epsilon}{3},\; 1 \,).
\end{align*}
In all the above expressions, terms upto first order of $\epsilon$ is to be considered.
If we now take the initial state to be 
\begin{align}
\{& \tfrac{1}{4} + (1+\cos{2 \theta})\epsilon/2,\; 
\tfrac{1}{4} + (1-\cos{2 \theta})\epsilon/2 \nonumber\\
&\tfrac{1}{4} + (-1+\cos{2 \theta})\epsilon/2,\; 
\tfrac{1}{4} + (-1-\cos{2 \theta})\epsilon/2 \}
\end{align}
then we can show that the population of the density matrix at any time are given by
\[
\begin{aligned}
\rho_{11}(t) &= \tfrac{1}{4} + \Big( -\tfrac{1}{2} + \frac{(2+ \cos{2 \theta})}{2}\, e^{-\tfrac{5 g t}{8}} \Big)\, \epsilon, \\[6pt]
\rho_{44}(t) &= \tfrac{1}{4} + \Big( \tfrac{1}{2} - \frac{(2+ \cos{2 \theta})}{2}\, e^{-\tfrac{5 g t}{8}} \Big)\, \epsilon, \\[6pt]
\rho_{22}(t) &= \tfrac{1}{4} - \frac{1}{2}\big( -1+\cos{2 \theta} \big)\, e^{-\tfrac{5 g t}{24}}\, \epsilon, \\[6pt]
\rho_{33}(t) &= \tfrac{1}{4} + \frac{1}{2}\big( -1+\cos{2 \theta} \big)\, e^{-\tfrac{5 g t}{24}}\, \epsilon.
\end{aligned}
\]

Now $\rho_{11}(t), \rho_{44}(t)$ are decoupled from the other two components and independently satisfy the trace condition. This means the dynamics can be interpreted as two non-interacting qubits, each dissipating through its own bath. Interestingly, while one qubit perceives the bath as being at the same temperature as the original environment, the other effectively experiences the bath as being at infinite temperature.

The complicated two-qubit dynamics can be mapped to a simple effective picture of two non-interacting qubits, each dissipating through its own bath at a different temperature.

\section{Summary and Outlook}
We have investigated the quantum Mpemba effect in the natural thermalization of nuclear spins. By identifying dipolar relaxation as the dominant decoherence process, we derived the conditions under which the effect can emerge. Preparing nuclear spin states that satisfy these conditions, we observed that the state farther from equilibrium relaxes faster than the one closer to equilibrium, thereby establishing the quantum Mpemba effect as an intrinsic feature of natural thermalization, requiring no external bath engineering. In addition, we have provided the experimental demonstration of the genuine quantum Mpemba effect in a natural relaxation process.

As quantum information technologies continue to advance, mechanisms that shorten relaxation times are of critical importance, as they directly reduce idle periods between consecutive runs of quantum devices. Our experiments highlight that the Mpemba effect,long debated in the classical domain, arises naturally in quantum spin systems, which is the hardware of many present day quantum devices. Beyond resolving a fundamental question in nonequilibrium statistical mechanics, these findings open pathways toward exploiting the Mpemba effect as a resource for quantum technologies, where accelerated equilibration could be harnessed for more efficient initialization, error mitigation, and control of large-scale quantum devices.



\chapter{Experimental Localization of Entanglement onto Noninteracting Systems and apparent Violation of Quantum Data Processing Inequality}\markboth{Chapter 5: Localization and Delocalization of Entanglement}{}\label{chap_QDPI}

\begin{center}
``\textit{There is no free lunch in the information theory.}"
\end{center}

\section{Introduction} 
Quantum entanglement \cite{horodecki2009quantum, qian2018entanglement, castelvecchi2020spooky} is undoubtedly one of the most bizarre facets of quantum physics. Over the time, from being a topic of debate about the foundational nature of reality \cite{einstein1935can, reid2009colloquium, bell1964einstein, wehner2006tsirelson},  quantum entanglement has become an important resource \cite{chitambar2019quantum, brandao2013resource} for performing various quantum enabled tasks ranging from computation \cite{shor1999polynomial, grover1996fast, weinstein2001implementation, arute2019quantum, montanaro2016quantum, biamonte2017quantum} to communication \cite{gisin2007quantum, pirandola2020advances, wehner2018quantum} and even sensing \cite{degen2017quantum}. Like a good resource, quantum entanglement can be estimated \cite{PhysRevA.65.032314}, mixed \cite{PhysRevA.82.062316}, distilled \cite{10.5555/2011326.2011330}, concentrated, diluted, swapped, teleported, localized, or even delocalized \cite{pal2021experimental,terhal2003quantum}. In this work, we have focused on the aspect of localization and delocalization of entanglement between two noninteracting systems in a tripartite setting. 

\begin{figure}
\center
\includegraphics[trim=1cm 9cm 0cm 3.5cm,width=8.6cm]{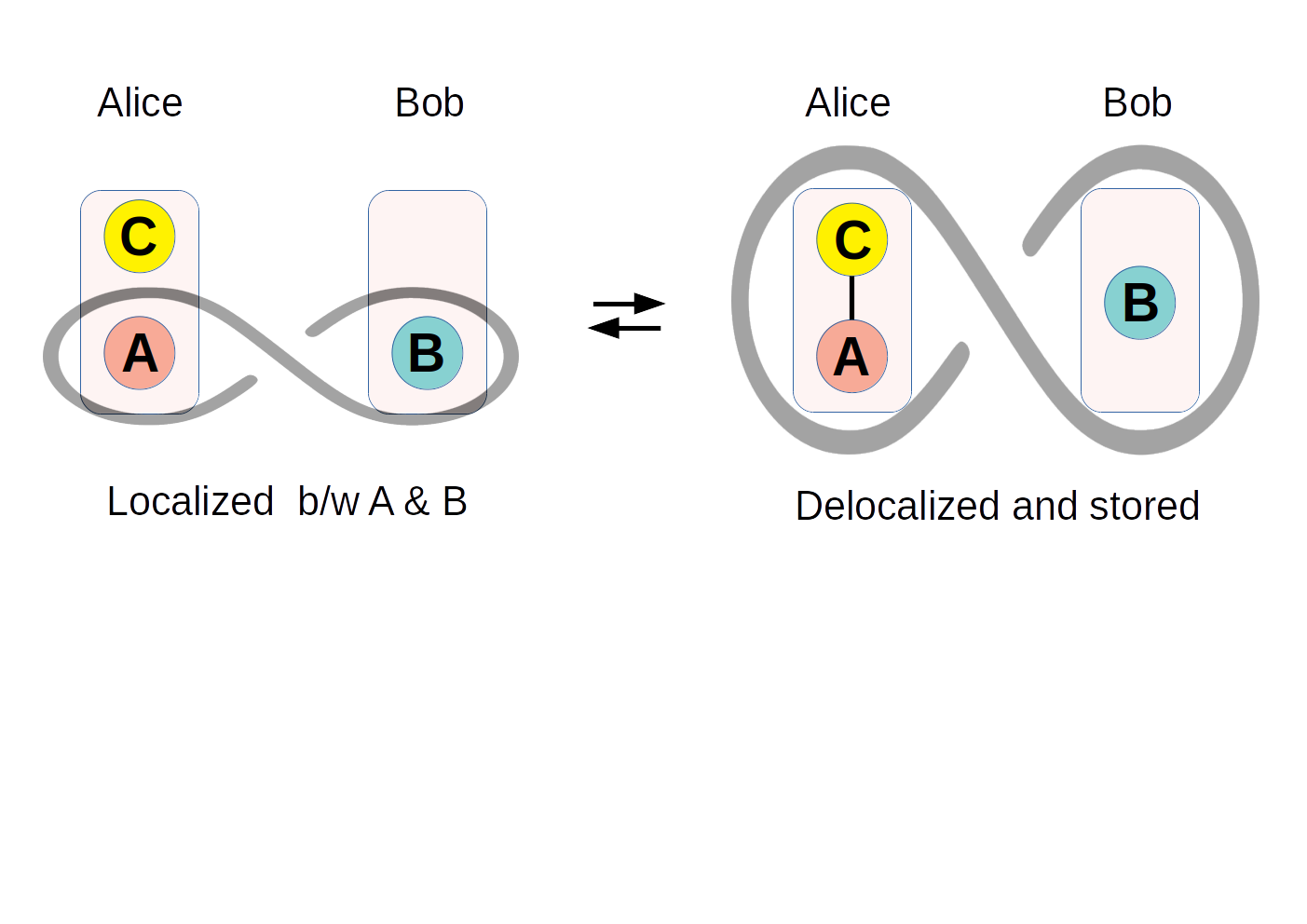}
\caption{ Graphical representation of our protocol where we considered three parties A, B, and C such that only A and C are interacting and B is left alone. Then, assuming the existence of a delocalized entanglement, we study the process by which A and B  get entangled as a result of the interaction between A and C. }
\label{fig:dramatic_representation}
\end{figure}

Consider two parties, Alice and Bob, in possession of systems A and B, respectably, such that there is no entanglement between them, i.e., $E_{\text{A:B}}(\rho_{\text{AB}})=0$, where $E_{X:Y}(\rho)$ is a measure \cite{plenio2006introductionentanglementmeasures} of entanglement between  $X$ and $Y$ when their density operator is $\rho$. Suppose Alice possesses another system C such that there exists global entanglement in the ABC system, which we call delocalized entanglement. Depending on whether C is classically correlated, or entangled, with AB, the delocalized entanglement can be, respectively, either between AC and B, or it can be a genuine multipartite entanglement between CAB. Since Alice only has access to A and C, she can not perform any operation on B.  Given this scenario, we ask how Alice can localize the delocalized entanglement onto A-B bipartition (see Fig. \ref{fig:dramatic_representation}). In other words, given the existence of a delocalized entanglement in the tripartite system, we ask : can A and B get entangled due to A's local interaction with C alone.

We  first consider the case where C is classically  correlated with AB. We show that C and A can interact in such a way that there never grows any quantum correlation between C and AB, yet this seemingly `classical' interaction is enough to localize the entanglement onto A-B bipartition.
We experimentally demonstrate localization of entanglement between A and B via only A-C interaction in nuclear magnetic resonance (NMR) architecture and confirm the `classicality' of the dynamics by observing zero quantum discord between C and AB throughout the process.

Next, we examine the localization of entanglement where C is entangled with A and B. We demonstrate that the delocalized entanglement, which is a genuine multipartite entanglement in this case, can be more robust against dephasing noise than the localized one, as sometimes delocalization can unlock decoherence free subspaces that are not available in lower dimensions. We experimentally validate this robustness using a four qubit quantum register in NMR, considering A as a two qubit system while B and C correspond to single qubits. The experimental demonstration goes against the common intuition that higher dimensional entanglement decoheres more quickly than lower dimensional entanglement.

Finally, note that in both the localization processes, A and B get entangled only due to local interaction of A with C. This seems to violate the quantum data processing inequality (QDPI) \cite{schumacher1996quantum}, indicating that the reduced dynamics during this process can not be completely positive \cite{buscemi2014complete}. However, we show that this violation of QDPI is only apparent and the localization of entanglement can be explained in terms of completely positive reduced dynamics.

The rest of the article is organized as follows. In Sec. \ref{sec:classloc} we consider the localization of entanglement between A and B when C is classically correlated with AB. In Subsec. \ref{subsec:classlocth} we present the formal theory of localization and show how the violation of QDPI and the associated non complete positivity is only apparent by constructing completely positive quantum maps which represent the localization process. In Subsec. \ref{subsec:classlocexp} we present the experimental demonstration of localizing entanglement via only AC interaction while C remains non-discordant with AB throughout. We then present the theory of localization while C is entangled with AB in Sec. \ref{sec:quantloc}. We show how the argument presented for the classically correlated C can still be applied when C is entangled to show that the violation of QDPI still remains only apparent. Finally, we discuss the experimental demonstration of the enhanced robustness of delocalized entanglement in Sec.\ref{sec:delocalize} before concluding in Sec. \ref{sec:concl}. 

\section{Localization of Entanglement from Classical Correlation} \label{sec:classloc}

\subsection{Theory} \label{subsec:classlocth}

Consider A and B to be qudits, each corresponding to a respective $d$-dimensional Hilbert space $\mathbbm{H}_d$. The joint Hilbert space $\mathbbm{H}^{\otimes 2}_d$ of AB is spanned by the generalized Bell basis \cite{Sych_2009,Scholin_2025,Popp_2024}:
\begin{gather}
\Ket{\phi_{mn}} = \left(Z^n X^m \otimes \mathbbm{1} \right) \Ket{\phi_{00}},
\end{gather}
with $X\Ket{j}=\Ket{j+1}$ modulo $d$, and $Z\Ket{j}=w^j \Ket{j}$ with $w=\exp(2\pi i/d)$ for $j\in \{0,1,2,...,d-1\}$ are the Weyl-Heisenberg operators \cite{PhysRevLett.70.1895}. The `seed' state is defined as  $\Ket{\phi_{00}} = (1/\sqrt{d}) \sum_{j=0}^{d-1}\Ket{jj}_{\rm{AB}}$. Note that,  each of these basis states are maximally entangled as measured by, for example, relative entanglement entropy \cite{PhysRevLett.78.2275,PhysRevA.57.1619,PhysRevLett.104.080501} : $E(\proj{\phi_{mn}}) = 1$. Using them, we construct the tripartite states between C, A and B as :
\begin{gather}
\rho^{\text{CAB}} = \sum_{n,m=0}^{d} p_{k}\proj{k}_{\text{C}} \otimes \proj{\phi_{mn}}, \label{eq:qudit_tot}
\\
\text{where}~k=md+n,~\text{and}~\sum_{k=0}^{d^2-1}p_k=1, \nonumber
\end{gather}
such that the quantum discord between C and AB is zero with respect to measurement on both C and AB. Here C is considered as a system having $d^2$ dimensional Hilbert space, which is only classically correlated to AB. Since relative entanglement entropy obeys the `flags condition' \cite{Horodecki2005,PhysRevLett.109.070501}, we can directly compute the delocalized entanglement in state $\rho^{\rm{CAB}}$ of Eq.~\Eqref{eq:qudit_tot} as 
\begin{equation}
E_{\text{CA:B}}(\rho^{\text{CAB}}) = \sum_{n,m=0}^{d-1}p_{k} \, E_{\text{A:B}} (\proj{\phi_{mn}}) = 1.
\end{equation}
The amount of entanglement that is going to be localized between A and B via interaction unitary $U_{\text{CA}}$ will be always be upper bounded by the initial delocalized entanglement, i.e.,
\begin{equation}
 E_{\text{A:B}}(\text{tr}_{\text{C}}[U_{\text{CA}}\,\rho^{\text{CAB}}\,U_{\text{CA}}^{\dagger}]) \leq E_{\text{CA:B}}(\rho^{\text{CAB}}).
\end{equation}
Therefore, the state of Eq.~\Eqref{eq:qudit_tot} contains maximum amount of entanglement that may be possible to localize in A-B bipartition via A-C interaction. Note that for the choice of $p_k = 1/d^2~ \forall k$, the reduced state $\rho^{\text{AB}}=\text{tr}_{\text{C}}(\rho^{\text{CAB}}) = (1/d^2)\sum_{i,j=0}^{d-1} \proj{i}_{\text{A}} \otimes \proj{j}_{\text{B}}$ of AB becomes fully separable and hence contains zero entanglement. In Alice's lab, she can let C and A to interact via the unitary 
\begin{gather}
\mathcal{U}_{\text{CA}}=\sum_{m,n}\proj{k}\otimes (X^{\dagger})^{m} (Z^{\dagger})^n, ~~
\mbox{that transforms} 
\nonumber \\
\rho^{\text{CAB}}
\xrightarrow[]{\mathcal{U}_{\text{CA}} \otimes \mathbbm{1}}
\left(\sum_{k}p_k \proj{k}_C \right) \otimes \proj{\phi_{00}}.   
\label{eq:uca}
\end{gather}
A and B become maximally entangled after the operation since $E_{\text{A:B}}(\proj{\phi_{00}})=1$. Thus the delocalized entanglement gets localized in A-B bipartition due to local interaction between A and C alone.

According to QDPI \cite{schumacher1996quantum}, the total correlation between A and B can not increase due to any local operation on A :
\begin{gather}
I_{\text{A:B}} (\rho^{\text{AB}}) \geq  I_{\text{A:B}}((\Phi_{\text{A}} \otimes \mathcal{I}_{\text{B}})\rho^{\text{AB}}), \label{eq:qdpi} 
\end{gather}
where $I_{\text{A:B}}(\rho^{\text{AB}})=S(\text{tr}_{\text{B}}\rho^{\text{AB}})+S(\text{tr}_{\text{A}}\rho^{\text{AB}})-S(\rho^{\text{AB}})$ is the mutual information quantifying the total correlation between A and B with $S(X)=-\text{tr}(X \log{X})$ being the von Neumann entropy of a system with density operator $X$. $\Phi_{\rm{A}}$ is a completely positive and trace preserving (CPTP) map acting on the subsystem A alone, while B gets acted by the identity map ${\cal I}_{\text{B}}$. Eq.~\Eqref{eq:qdpi} states that any local action on A, which can be described by a CPTP map on A, can not increase the correlation between A and B, which is often quoted as ``\textit{there is no free lunch in information theory}''. However, in the localization process, we saw the entanglement between A and B to increase from zero to maximum, while A only locally interacts  with C and B is left untouched. Thus, there seems to be a violation of Eq.~\Eqref{eq:qdpi}, indicating that the reduced transformation of A must be described by a map which is not completely positive. We show below that this violation of QDPI and the associated non complete positivity is only apparent by focusing on the transformation that the reduced state of AB went through during localization \cite{lidar2019lecture}.
In the following, the localization of entanglement is described by the action of a map $\Phi(\cdot)=\sum_{rs}K_{rs}(\cdot)\,K_{rs}^{\dagger}$, with the Kraus operator $K_{rs}=\outpr{\phi_{00}}{\phi_{rs}}$, on the reduced state of AB as long as the tripartite initial state of CAB is of the form of Eq.~\Eqref{eq:qudit_tot}:
\begin{gather}
\rho^{\text{AB}} = \text{tr}_{\text{C}}\,(\rho^{\text{AB}}) \longrightarrow \widetilde{\rho}^{\text{AB}} = \text{tr}_{\text{C}}\,(\mathcal{U}_{\text{CA}}\,\rho^{\text{CAB}}\mathcal\,{U}^{\dagger}_{\text{CA}}) \nonumber \\
=\text{tr}_{\text{C}}\, \left[ \mathcal{U}_{\text{CA}} \left(\sum_{n,m=0}^{d-1} \, p_k \proj{k} \otimes \proj{\phi_{mn}} \right) \,\mathcal{U}_{\text{CA}}^{\dagger} \right] \nonumber \\
= \sum_{l=0}^{d^2-1}\,\sum_{m,n=0}^{d-1}\Bra{l}\,\mathcal{U}_{\text{CA}}\Ket{k} \left(p_k  \proj{\phi_{mn}} \right)\,\Bra{k}\,\mathcal{U}_{\text{CA}}^{\dagger} \Ket{l} \nonumber \\
= \sum_{l=0}^{d^2-1}\left(\sum_{r,s=0}^{d-1} \Bra{l}\,\mathcal{U}_{\text{CA}}\Ket{q}\,\proj{\phi_{rs}}  \right) \nonumber \\ \left(\sum_{m,n=0}^{d-1}p_k\,\proj{\phi_{mn}} \right) 
\left(\sum_{x,y=0}^{d-1} \Bra{l}\,\mathcal{U}_{\text{CA}}\Ket{w}\,\proj{\phi_{xy}}  \right)^{\dagger},\nonumber\\
~\text{where}~q=rd+s~\text{and}~w=xd+y,\nonumber \\        
= \sum_{r,s=0}^{d-1} \left( \outpr{\phi_{00}}{\phi_{rs}}\right)\,\rho^{\text{AB}}\,\left( \outpr{\phi_{00}}{\phi_{rs}}\right)^{\dagger} \nonumber \\ 
= \sum_{r,s=0}^{d-1}  K_{rs}\,\rho^{\text{AB}}\,K_{rs}^{\dagger} = \Phi(\rho^{AB}),\label{eq:calcul}\\
\text{with}~\sum_{r,s=0}^{d-1}  K_{rs}^{\dagger}\,K_{rs}= \sum_{r,s=0}^{d-1}\proj{\phi_{rs}} = \mathbbm{1}. \nonumber
\end{gather}
 Thus when C is classically correlated with AB, the localization process is completely positive since any map that admits Kraus decomposition is always CPTP. The violation of QDPI was appearing because we were assuming that the local interaction between A and C will induce a local map on A alone : $\Phi_{\text{A}} \otimes {\cal I}_B$. However, as can be seen from Eq.~\Eqref{eq:calcul}, even the local interaction between A and C will induce a global channel $\Phi$ between A and B, which is not decomposable as $\Phi_{\text{A}} \otimes \Phi_{B}$, and hence there is no violation of QDPI. 

\subsection{Experiment} \label{subsec:classlocexp}
\begin{figure}
\begin{center}
(a) \\
\includegraphics[trim = 0cm 5cm 0cm 3cm, width=8.2cm]{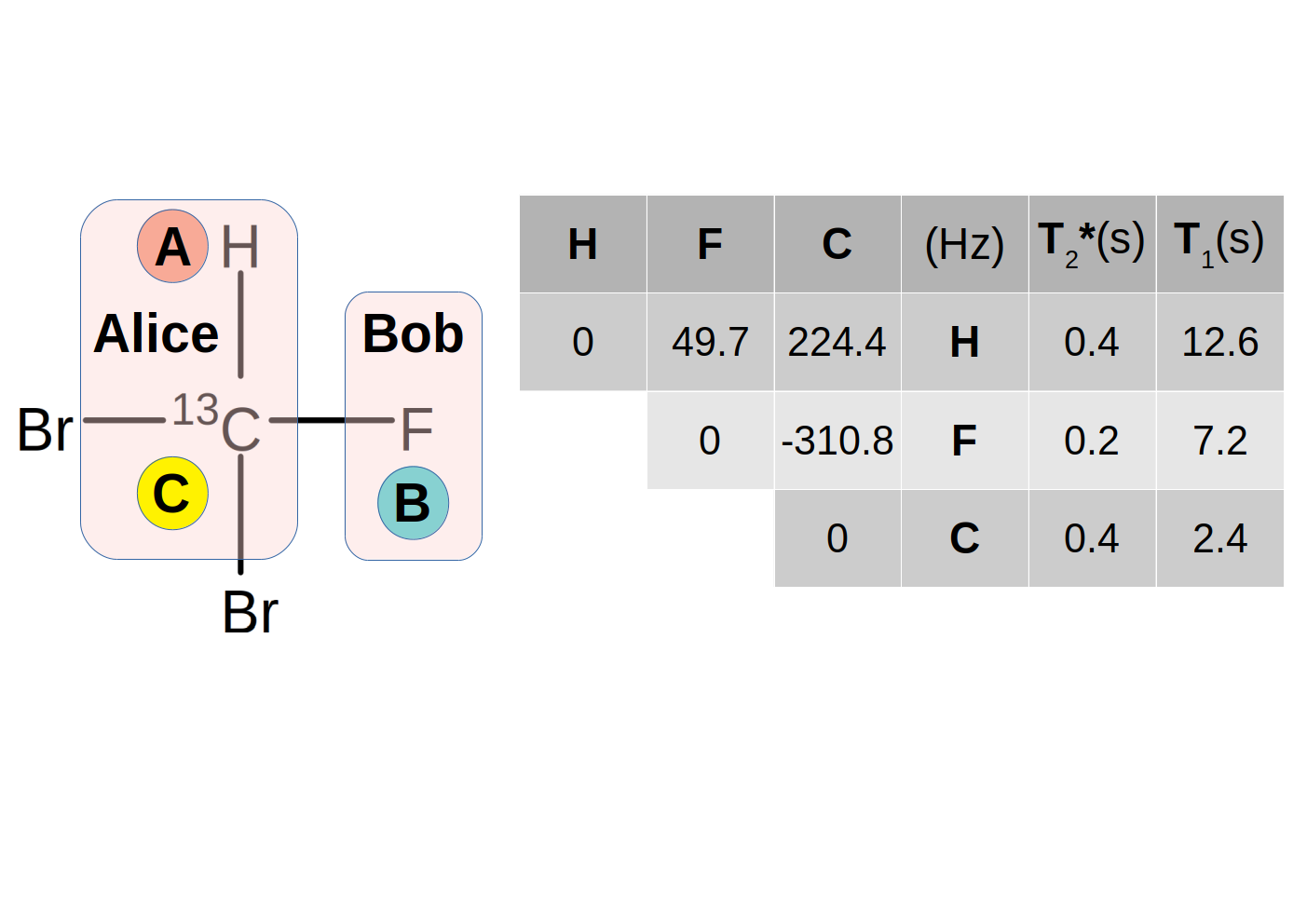} \\
(b) \\
~~\\
\hspace{0.7cm}
\Qcircuit @C=0.85em @R=1.4em {
                                           &            &         &                   & \dstick{\tfrac{1}{2J}}        &     &     &    & \dstick{\tau = \tfrac{l}{10J} }                    &    &\\
\lstick{\text{C}\left(\frac{\mathbbm{1}}{2}\right)}& \qw        & \qw     &\qw                &\ctrl{1}&\qw  & \qw &\qw &\ctrl{1}          &\qw & \meter\\ 
\lstick{\text{A}\left(\proj{0}\right)}            & \gate{H}   & \ctrl{1}&\qw                &\control \qw &\qw  & \qw &\qw &\control \qw          &\qw & \meter\\
\lstick{\text{B}\left(\proj{0}\right)}            & \gate{\pi} & \targ   &\qw                & \qw      &\qw  & \qw &\qw &\qw                 &\qw & \meter 
\gategroup{1}{1}{6}{6}{.7em}{--}  \\
                                           &            & \mbox{Preparation}        & &          &     &     &    &                    &    &\\
                                           &            &         &                   &          &     &     &    &                    &    &\\
} \\
~~\\
(c) \\
\includegraphics[trim = 0cm 0cm 0cm 0cm, clip=true, width=4.2cm]{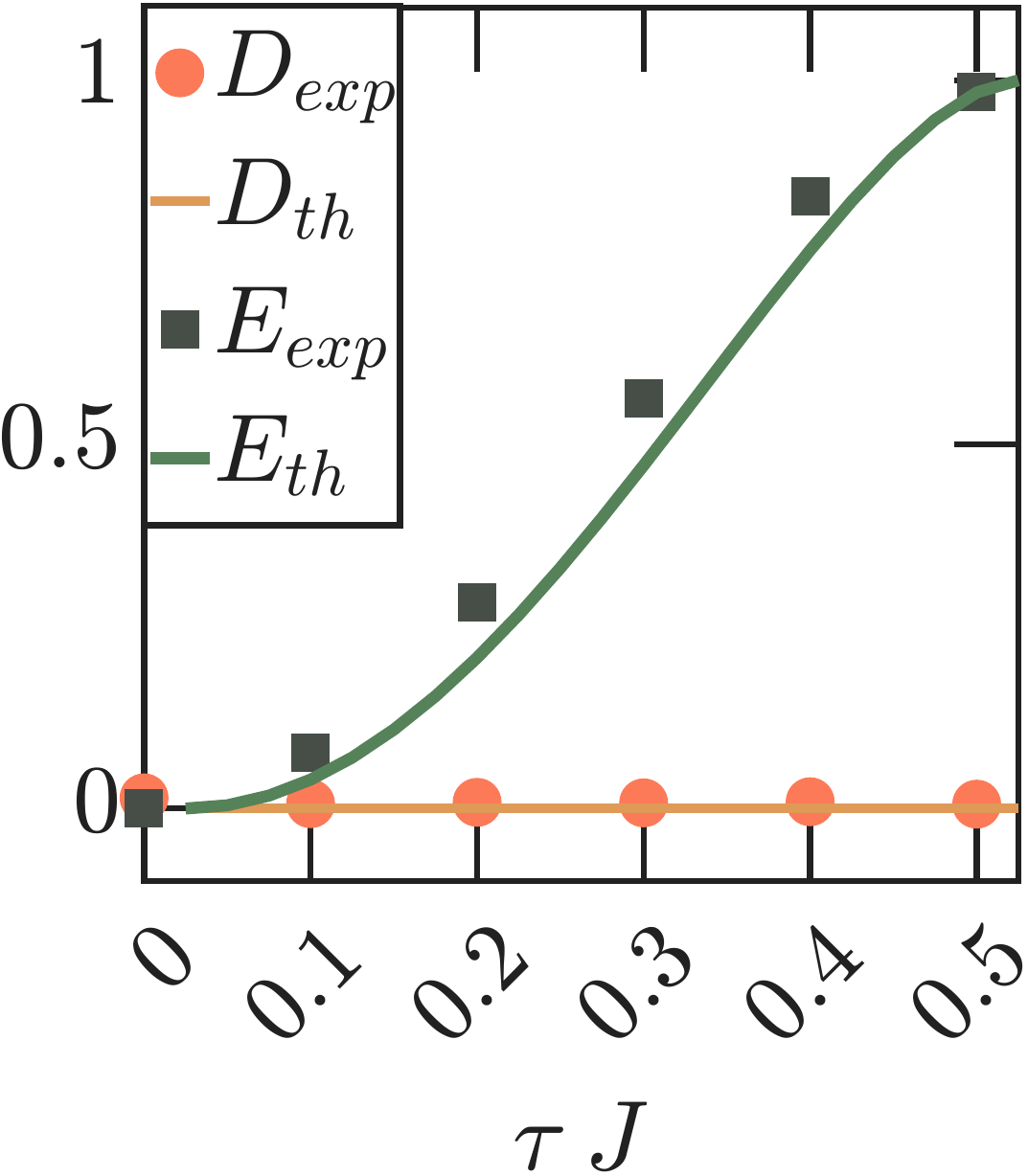}
\caption{(a) Molecular structure of DBFM identifying the three qubits as A, B, and C.  The table contains the coupling constants $J_{ij}$ and relaxation time constants. (b) The quantum circuit for state preparation followed by the localization of entanglement. The $z-z$ interaction is realized by evolving the system under the Hamiltonian $2\pi J I_z^{\text{C}}I_z^{\text{A}}$ for time $\tau$, where $l\in\{0,1,2,3,4,5\}$ stands for the experimental run number. Measurements indicate full three-qubit state-tomography. (c) Experimental results showing entanglement  between A and B, as measured by relative entanglement entropy ($E$), growing with time $\tau$ causing an apparent violation of QDPI. The zero discord $D$ between C and AB highlights the classical nature of C.}
\label{fig:classC_exp}
\end{center}
\end{figure}

We use Dibromofluoromethane-13C (DBFM), dissolved in Acetone-D6, as a three-qubit NMR register and identify $^1$H as A, $^{19}$F as B, and $^{13}$C as C,  as shown in Fig. \ref{fig:classC_exp} (a). 
The internal Hamiltonian of the three spin system in a triply rotating frame  about the Zeeman field with individual Larmor frequencies is given by
\begin{align}
H=2\pi (J_{\text{CA}}I_{z}^{\text{C}}I_{z}^{\text{A}}+J_{\text{CB}}I_{z}^{\text{C}}I_{z}^{\text{B}}+J_{\text{AB}}I_{z}^{\text{A}}I_{z}^{\text{B}}),
\label{eq:internal_hamiltonian_NMR}
\end{align}
where $I_{n}^{\text{K}}$ is the spin angular momentum operator of K along $\hat{n}$. The coupling constants $J$ and the respective relaxation time constants are given in Fig. \ref{fig:classC_exp} (a). At an ambient temperature of 298 K inside a 500 MHz Bruker NMR spectrometer, the three spin system rests in thermal equilibrium state \cite{cavanagh1996protein} $\rho^{th} = \mathbbm{1}/8 + \epsilon (\gamma_{\text{A}}\,I_{z}^{\text{A}}+\gamma_{\text{B}}\,I_{z}^{\text{B}}+\gamma_{\text{C}}\,I_{z}^{\text{C}})$, where `$\gamma$'s are the respective gyromagnetic ratios  and $\epsilon$ is the purity factor \cite{cavanagh1996protein}. Starting from $\rho^{th}$, we prepare the three qubit register in state 
\begin{gather}
\rho_{e}^{\text{CAB}} = \frac{1}{2} \proj{0}_{\text{C}} \otimes \proj{\eta_+} + \frac{1}{2} \proj{1}_{\text{C}} \otimes \proj{\eta_-},
\label{eq:expstate_cl} \\
\text{where}~\Ket{\eta_{\pm}} = \frac{1}{\sqrt{2}} \left(\Ket{01} \pm i \Ket{10} \right). \nonumber
\end{gather}
We first prepare a pseudo-pure state (PPS) between A and B, while destroying the polarization of C using a $\pi/2$ pulse followed by a pulsed field gradient along $\hat{z}$ to get to the state $(\mathbbm{1}/2)_{\text{C}} \otimes \Ket{00}_{\text{AB}}$, which is then fed to the preparation part of the circuit shown in Fig.~\ref{fig:classC_exp}~(b).  

Note that the state of Eq.~\Eqref{eq:expstate_cl} is of the form of Eq.~\Eqref{eq:qudit_tot} with  delocalized entanglement $E_{\text{CA:B}}(\rho^{\text{CAB}}_{e}) = (1/2)E_{\text{A:B}}(\proj{\eta_{+}}) + (1/2)E_{\text{A:B}}(\proj{\eta_{-}}) = 1/2~+~ 1/2=1$. However, the reduced state of AB remains fully separable  $\rho^{\text{AB}}_{e} = \text{tr}_{\text{C}}\,(\rho^{\text{CAB}}_{e}) = (\proj{01} + \proj{10})/2$, containing zero entanglement. 

As shown in Fig.~\ref{fig:classC_exp}~(b), the prepared state is left to evolve under the Hamiltonian of Eq.~\Eqref{eq:internal_hamiltonian_NMR} while we effectively cancel the effects of A-B and B-C interactions with suitable refocusing ($\pi$) pulses. Thus we ensure only C-A interaction to persist. We let C and A to interact for increasing time intervals from $\tau=0$ to $\tau=1/(2J_{\text{CA}})$ in 6 steps, followed by a full three qubit state tomography \cite{nielsen2002quantum} each time. From the experimentally measured tripartite states, the localized entanglement between A and B, quantified by relative entanglement entropy \cite{PhysRevA.82.052336,PhysRevA.91.029901}, is measured as a function of time $\tau$. The results are plotted in Fig.~\ref{fig:classC_exp} (c) confirming the localization of entanglement between A and B due to interaction between A and C alone. Moreover, we also measured the discord between C and AB at each time steps, which again shows zero growth of quantum correlation between them certifying the `classical' nature of the interaction, as mentioned in Sec. \ref{subsec:classlocth}. Within experimental limitations, we observe very good agreement between theoretically predicted values with the measured ones. 

\section{Localization of Entanglement when C is entangled} \label{sec:quantloc}
We now consider the tripartite states of the following structure 
\begin{gather}
\Ket{\psi}_{\text{CAB}} = \frac{1}{d} \sum_{m,n=0}^{d-1} \Ket{k}_{\text{C}} \otimes \Ket{\phi_{mn}},~\text{with}~k=md+n,
\end{gather}
such that A and B contain zero initial entanglement while C is entangled with AB. We can localize the global entanglement between CAB onto A-B bipartition via local interaction between A and C alone. In particular, we observe localization to happen as 
\begin{gather}
\Ket{\psi}_{\text{CAB}} \longrightarrow \Ket{\widetilde{\psi}}_{\text{CAB}} = \mathcal{U}_{\text{CA}}\Ket{\psi}_{\text{CAB}} = \Ket{\chi}_{\text{C}} \otimes \Ket{\phi_{00}}, \\
\text{where}~\Ket{\chi}_{\text{C}} = \frac{1}{\sqrt{d}} \sum_{k=0}^{d^2-1} \Ket{k}, \nonumber
\end{gather}
such that we get A and B to be maximally entangled in the end.  Note that $\mathcal{U}_{\text{CA}}$ is the same unitary used in Eq. \ref{eq:uca}. Considering $\Phi$ to be the map transforming the reduced unentangled state of AB to the entangled one, we observe
\begin{gather}
\Phi(\proj{\phi_{mn}}) = \proj{\phi_{00}},~\forall m,n \nonumber \\
\implies \Phi(\cdot) = \sum_{m,n=0}^{d-1}K_{mn}(\cdot)K_{mn}^{\dagger},~\text{with}~K_{mn} = \outpr{\phi_{00}}{\phi_{mn}}. \nonumber 
\end{gather}
Thus, just like the case with classically correlated C, the localization of entanglement between A and B can again be described by a global CPTP map $\Phi \neq \Phi_{\text{A}} \otimes \Phi_{\text{B}}$. Interestingly, The delocalized entanglement of this kind can sometimes provide enhanced robustness against dephasing noise than localized entanglement, which we experimentally show next. 

\section{Delocalization of entanglement for robust storage} \label{sec:delocalize}
\begin{figure}
\begin{center}
\includegraphics[trim=0cm 4cm 0cm 3cm,width=8cm]{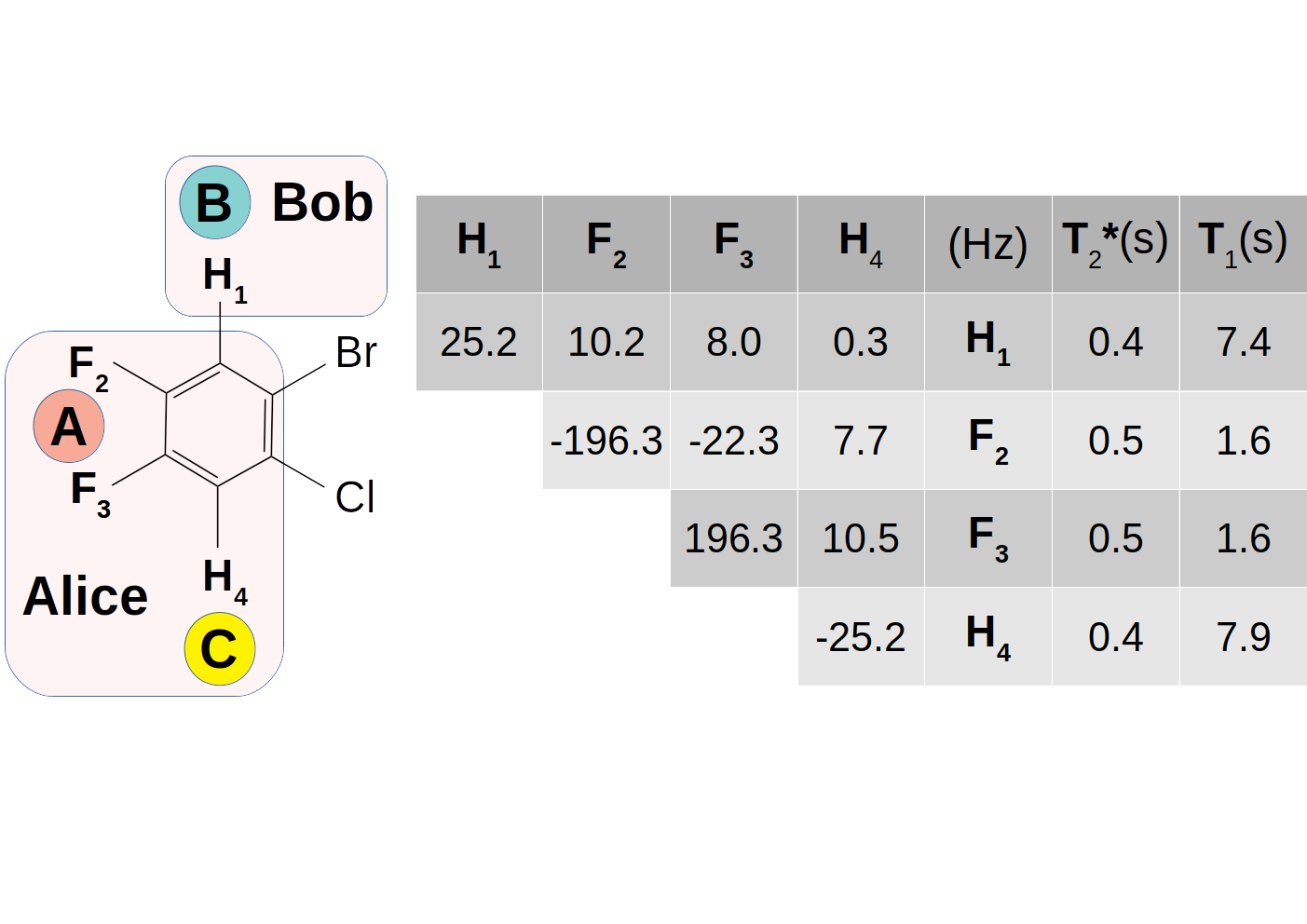} \\ (a) \\
\hspace*{0.5cm}
\Qcircuit @C=0.75em @R=0.85em {
	&        &         &   &          &  &  &   &\mbox{$G_4$} &     &           &    &  &    &              &    &\mbox{$-G_4$}&        &         &\\
	&        &         &             &          &            &          &                &             &     &           &    &   \mbox{store}           &    &              &    &             &        &         &   \\          
	\lstick{F_2}&\gate{H}& \ctrl{1}&\qw          & \ctrl{2} & \qw        & \ctrl{3}          &\qw &\qw          & \qw & \gate{\pi}&\qw &\qw           &\qw &\gate{\pi}    &\qw &\qw          &\ctrl{3}&\ctrl{2} &\meterB{\langle \sigma_x \rangle} \\
	\lstick{F_3}&\qw     & \targ   &\qw          & \qw      & \qw        & \qw              &\qw &\qw          & \qw & \gate{\pi}&\qw &\qw           &\qw &\gate{\pi}    &\qw &\qw          &\qw     &\targ    &\\
	\lstick{H_1}&\qw     & \qw     &\qw          & \targ    & \qw        & \qw             &\qw &\qw          & \qw & \qw       &\qw &\gate{\pi}    &\qw &\qw           &\qw &\qw          &\qw     &\targ    &\\ 
	\lstick{H_4}&\qw     & \qw     &\qw          & \qw      & \qw        & \targ            &\qw &\qw          & \qw & \qw       &\qw &\gate{\pi}    &\qw &\qw           &\qw &\qw  \gategroup{3}{7}{6}{7}{.7em}{.} \gategroup{3}{18}{6}{18}{.7em}{.}  \gategroup{2}{10}{7}{16}{.7em}{--}    &\targ   &  \qw       &     \\        
	&        &         &\cwx[-5]     &          &\cwx[-5]    &          & \cwx[-5]       &\cwx[-5]     &     &           &    &              &    &              &    &\cwx[-5]     &        &         &   \\
	&        &         &\mbox{$G_1$} &          &\mbox{$G_2$}&  &  \mbox{$G_3$}    & &     &           &    & \mbox{$\times$ 2} &    &              &    &&        &         &\\
} \\
~~\\
(b) \\
$
\begin{array}{|c|c|c|c|c|}
\hline 
\mbox{State} & (q_F,q_H) & G_2 & G_3 & \tau_q~ (s) \\
\hline
\begin{array}{c}
\Ket{010}\pm \Ket{101}\\
\Ket{011}\pm \Ket{100}
\end{array}
& (0,\pm 1) & \mbox{OFF} & \mbox{OFF} & 0.25 \\
\hline
\Ket{001} \pm \Ket{110}
& (2,-1) & f_-(G_1)
& \mbox{OFF} & 0.25 \\
\hline
\Ket{000} \pm \Ket{111}
& (2,1) & f_+(G_1)
& \mbox{OFF} & 0.03 \\
\hline
\begin{array}{c}
\Ket{0101}\pm \Ket{1010}\\
\Ket{0110}\pm \Ket{1001}
\end{array}
& (0,0) & \mbox{OFF} & \mbox{ON} & 1.22 \\
\hline
\end{array}
$
~~\\
(c) \\
\includegraphics[trim=0cm 0cm 0cm 0cm,width=8cm]{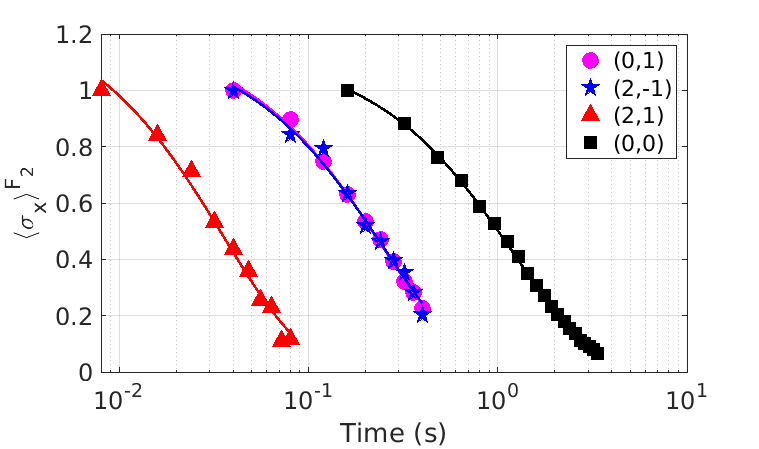} \\
(d)
\end{center}
\caption{(a) Molecular structure BCDF with four qubits as labeled, the table of Hamiltonian parameters, and relaxation time constants. (b) Quantum circuit for selecting various quantum coherences $(q_F,q_H)$ and to test their robustness against the dephasing noise. Here the CNOT gates within dotted boxes are applied only for the $(0,0)$ case.  Finally, each coherence is converted back to $F_2$ coherence $\langle \sigma_x \rangle^{F_2}$ and readout. (c) Gradient strengths and the measured time-constants for various coherences.  Here $f_\pm(G_1) = {-2\gamma_F G_1}/({2\gamma_F \pm \gamma_H})$. (d) Experimentally measured decay of various coherences 
over time, proving that the delocalized entangled state $(0,0)$ being the most robust.
}
\label{fig:deloc}
\end{figure}

We now demonstrate delocalization of entanglement and study its robustness using NMR methods.  For this purpose, we use two $^1$H and two $^{19}$F spin-1/2 nuclei of 1-Bromo-2-chloro-4,5-difluorobenzene (BCDF) dissolved in deuterated dimethyl sulphoxide.
The molecular structure and the labeling of the four qubits are shown in Fig. \ref{fig:deloc} (a). The experiments were performed at an ambient temperature of 300 K on a 500 MHz Bruker NMR spectrometer. Under the weak-coupling approximation, the internal Hamiltonian of the four-spin system in a doubly rotating frame about the Zeeman field of $^1$H and $^{19}$F Larmor frequencies is given by
\begin{align}
H= -2\pi \sum_i  \nu_i I_{z}^i + 2\pi \sum_{i,j}{J_{ij}I_{z}^{i}I_{z}^{j}}.
\label{eq:internal_hamiltonian_FFHH}
\end{align}
The resonance offsets $\nu_i$, the coupling constants $J_{ij}$, and the relaxation time constants are also shown in Fig. \ref{fig:deloc} (a). 

We use the quantum circuit shown in Fig. \ref{fig:deloc} (b) to prepare the following states with localized entanglement
\begin{subequations}\label{eq:ffh}
\begin{align}
\Ket{\psi_{\text{AB}}}^{(0,\pm 1)}& = \frac{1}{\sqrt{2}}(
\Ket{010} \pm \Ket{101}), \frac{1}{\sqrt{2}}(
\Ket{011} \pm \Ket{100})
\label{eq:zq},\\
\Ket{\psi_{\text{AB}}}^{(2,-1)}& = \frac{1}{\sqrt{2}}(\Ket{001} \pm \Ket{110}) \label{eq:zqq}, ~\mbox{and}\\
\Ket{\psi_{\text{AB}}}^{(2,1)}& = \frac{1}{\sqrt{2}} (\Ket{000} \pm \Ket{111}), 
\label{eq:threeq}
\end{align}
\end{subequations}
then delocalize their respective entanglements in a higher dimension, and test their robustness. As indicated in Fig. \ref{fig:deloc} (a), the system A is composed of spins $F_2$ and $F_3$ while the system B composes spin $H_1$.
The superscripts denote coherence orders $(q_F,q_H)$, which indicate the difference in the magnetic quantum numbers in the two components of each coherence for $F$ and $H$
\cite{cavanagh1996protein}.
The overall coherence order is thus $q = q_F+q_H$.  The circuit in Fig. \ref{fig:deloc} (b) uses pulsed field gradients (PFG) to efficiently select the desired coherence order while alleviating the need to prepare pseudopure states in each case.  Notice that all states of Eq.~\Eqref{eq:ffh} are local-unitary equivalent and hence contain same amount of entanglement \cite{horodecki2009quantum}.


We can imagine any one of these states of Eq. \Eqref{eq:ffh} is prepared and distributed at different locations between Alice and Bob. Experimentally, we  realize noninteracting scenario of Alice's and Bob's qubits by refocusing the spin-spin interactions of $F_2$ and $F_3$ with $H_1$ using $\pi$ pulses at suitable intervals as shown in the storage box of Fig. \ref{fig:deloc} (b). 
We have realized quantum dephasing channel using two equal and opposite gradient pulses ($G_4$ and $-G_4$ in Fig:~\ref{fig:deloc}~(b)) and allowing the qubit system to undergo translational diffusion in between them as shown in Fig. \ref{fig:deloc} (b). The PFG strengths and the decay constants of each coherence under depahsing are shown in Fig. \ref{fig:deloc} (c).  In this dephasing dynamics, we have observed the decoherence of all three entangled states of Eq. \Eqref{eq:ffh} and the observed decays are shown in Fig \ref{fig:deloc} (d).  As expected, the two single quantum coherences $(q_F+q_H=\pm 1)$ (Eq. \Eqref{eq:zq} and \Eqref{eq:zqq}) are longer lived compared to the three quantum coherence of Eq. \ref{eq:threeq}.
The question now is, can Alice and Bob increase the lifetime of entanglement within the constraint that A and B can not interact?

Alice can  bring a local ancillary qubit ($H_1$ in our setup) and delocalize the entanglement using a CNOT gate applied on her local qubits (CNOT inside dotted box in Fig. \ref{fig:deloc} (b)). We again emphasize that, in our experiments, suitable refocusing $\pi$ pulses were applied to nullify any interaction between Alice ($F_2$, $F_3$, $H_4$) and Bob ($H_1$) throughout the delocalization to mimic the setup that they are at different locations. The state now becomes 
\begin{align}
\Ket{\psi_{AB}}&\otimes\Ket{\psi_{C}}
=\frac{1}{\sqrt{2}}(\Ket{010} \pm \Ket{101}) \nonumber \\
& \xrightarrow{\mathrm{C_{F_2}NOT_{H_1}}}
\Ket{\psi_{ABC}}^{(0,0)} = \frac{1}{\sqrt{2}}(
\Ket{0101} \pm \Ket{1010}).
\label{eq:ffhh}
\end{align}
As shown by Fig. \ref{fig:deloc} (d), this state is remarkably more robust to dephasing than any of the localized ones.  The decay time constant (see Fig. \ref{fig:deloc} (c)) improved by a factor of almost five compared to single-quantum coherence and by a factor of forty compared to the three quantum coherence.

We thus experimentally verified that certain kind of entangled states (such as in Eq. \ref{eq:ffh}) are not robust against decoherence.  However, using local operations alone, they can be delocalized, which can drastically enhance the robustness of those states against decoherence. Thus, delocalization can be used to store entanglement before localising it and performing quantum tasks. \\
\section{Summary and Conclusions}
\label{sec:concl}
Entanglement is a central resource for achieving quantum advantage in emerging technologies, making it essential to understand how it can be localized onto a specific bipartition starting from a delocalized, or shared, state. We investigated this problem in a tripartite (ABC) system under two distinct scenarios: when the third party (C) is only classically correlated with AB, and when C is entangled with AB. In both cases, the initial states are chosen such that the delocalized entanglement is present in the system—either as entanglement between CA and B (in the classical-correlation case), or as a genuine multipartite entanglement across CAB (in the quantum-correlation case). Crucially, however, the reduced state of AB is fully separable and contains no entanglement.

We then examined how this delocalized entanglement can be localized onto the AB bipartition through local interaction between A and C alone. At first sight, the localization process appears to violate the quantum data-processing inequality, suggesting that the reduced dynamics of AB is non–completely positive (non-CP). We showed, however, that this violation is only apparent: the process can be rigorously described within the framework of a completely positive and trace-preserving (CPTP) quantum channel. We experimentally demonstrated this localization mechanism using a three-qubit NMR register.

Looking ahead, one of the foremost challenges in scaling quantum devices is the long-term preservation of entanglement. Earlier, Vidal et al. \cite{vidal1999robustness} had argued for enhanced robustness when entanglement is delocalized across larger Hilbert spaces. Inspired by this idea, though employing a different mechanism, we explored the delocalization of entanglement by extending it onto a local ancillary system. Using a four-qubit NMR register, we experimentally observed a striking enhancement in robustness, with significantly prolonged storage time of delocalized entanglement by up to a factor of 40.

In summary, we presented a combined theoretical and experimental study of both localization and delocalization of entanglement between remote parties. Our results resolve interesting conceptual puzzles, introduce experimentally viable protocols for manipulating entanglement, and open new directions for sharing, preserving, and harnessing entanglement in scalable quantum technologies.

\chapter{Summary and Future Directions}\markboth{Chapter 6: Summary and Future Directions}{}\label{chap_QDPI}

Finally, I will briefly point out some of the key findings that came out of the research work that has been reported in this thesis so far. It has been known previously that for a two-level system to surpass the temporal Tsirelson's bound (TTB) of Leggett-Garg inequality (LGI), either one needs to consider non-unitary parity-time $\mathcal{PT}$-symmetric evolution, or invoke a higher-dimensional ancillary system on which generalized measurement has to be performed. We showed that it is possible to violate TTB within two-level systems under unitary maps themselves. We experimentally confirm this violation and show that it can even approach the algebraic maximum of the LGI parameter. In the case of the Mpemba effect in the classical systems, it is still under debate how natural and ubiquitous the phenomenon is. Our experiments showed decisively that in the quantum realm, the Mpemba effect is very much a natural phenomenon, and in fact, it can be directly observed during natural thermalization of nuclear spins without any bath engineering. On a more application side, the determination of Lee-Yang zeros is a problem of crucial importance in condensed matter and many-body physics. We proposed a method of doing so for a general asymmetric system using the tools of quantum simulation and experimentally demonstrated it. Finally, with the advancement of quantum technologies, storing and manipulating the entanglement would be of prime importance. We did a detailed study of its localization and delocalization, subject to local actions without violating any physical principles like complete positivity or quantum data processing inequality.   

I believe the research reported in this thesis can pave the way for various future works. In the case of determining the algebraic variety of Lee-Yang zeros, we found that the structure of the co-amoeba depends very sensitively on the values of the coupling constants between the system qubits with the probe qubit. If the coupling constants are altered by even $0.01~\%$, the experimental data points get scattered all over the co-amoeba. This hyper-sensitivity can be used to build quantum sensors, and thus it can lead to new research in the area of quantum metrology. On the other hand, the numerical and analytical signature of enhanced robustness gained by superposition of unitary operators can provoke future experimental explorations, which could eventually lead to the application of superposed unitary operators in faithful and robust quantum control tasks. Our successful experimental realization of the superposition of unitary in the nuclear spin system can lead to the experimental realization of superposition between much more general quantum channels, which can unravel some very interesting physics from the perspective of open systems.     
\fancyhf{}
\fancyhead[LO]{Closing Remarks and Outlook}
\fancyhead[RE]{Closing Remarks and Outlook}
\fancyhead[LE,RO]{\thepage}

\newpage
\thispagestyle{empty}  
\mbox{}  

\clearpage  

\fancyhf{}
\fancyhead[LO]{Appendix}
\fancyhead[RE]{Appendix}
\fancyhead[LE,RO]{\thepage}

 


\fancyhf{}
\fancyhead[LO]{\nouppercase{\leftmark}}
\fancyhead[RE]{\nouppercase{\rightmark}}
\fancyhead[LE,RO]{\thepage}



\clearpage
\addcontentsline{toc}{chapter}{Bibliography}
\printbibliography

\end{document}